\documentclass[usenatbib]{mnras}

\usepackage[dvipsnames]{xcolor}
\usepackage{graphicx}
\usepackage{amstext}
\usepackage{amsmath}
\usepackage{natbib}
\usepackage{url}
\usepackage{setspace}
\usepackage{tabularx}
\usepackage{mathtools}
\usepackage[colorinlistoftodos]{todonotes}
\usepackage{booktabs}
\usepackage{rotating}
\usepackage{multirow}

\def \aj {{\rm {AJ}}}
\def \araa {{\rm {ARA\&A}}}
\def \apj {{\rm {ApJ}}}

\def \apjs {{\rm {ApJS}}}
\def \apjl {{\rm {ApJL}}}

\def \aap {{\rm {A\&A}}}
\def \aapr {{\rm {A\&AR}}}

\def \mnras {{\rm {MNRAS}}}

\def \pasj {{\rm {PASJ}}}

\def \dir {} 

\def	 \ntwoh {{\rm N$_2$H$^+$}}
\def	 \ntwohoz {{\rm N$_2$H$^+$\,($1-0$)}}
\def     \ntwod {{\rm N$_2$D$^+$}}
\def	 \ceo	{{\rm C$^{18}$O}}
\def	 \ceooz	{{\rm C$^{18}$O\,($1-0$)}}

\def	 \tco	{{\rm $^{13}$CO}}
\def	 \tcooz	{{\rm $^{13}$CO\,($1-0$)}}

\def 	 \tco {{\rm $^{13}$CO}}

\def	 \ntwod {{\rm N$_2$D$^+$}}

\def 	 \kms {{\rm \,km\,s$^{-1}$}}
\def     \sol {{\rm M$_\odot$}}
\def     \arcsec {{\rm $^{\prime\prime}$}}

\def     \micron{{\rm \,$\mu$m}}
\def     \ltsimm{\mathrel{\spose{\lower 3pt\hbox{$\sim$}}\raise 2.0pt\hbox{$<$}}}
\def     \gtsimm{\mathrel{\spose{\lower 3pt\hbox{$\sim$}}\raise 2.0pt\hbox{$>$}}}

\newcommand{\change}[1]{\textcolor{black}{#1}}

\begin{document}

\title[Similar kinematics within two infrared dark clouds]
	{Similar complex kinematics within two massive, filamentary infrared dark clouds\thanks{Based on observations carried out with the IRAM 30m Telescope. IRAM is supported by INSU/CNRS (France), MPG (Germany) and IGN (Spain).}
	} 
 
\author[Barnes, Henshaw, Caselli, Jim\'{e}nez-Serra, Tan, Fontani, Pon, Ragan]
    {A . T. Barnes$^{1,2,3}$\thanks{E-mail: a.t.barnes@2014.ljmu.ac.uk},
    J. D. Henshaw$^{4}$,
    P. Caselli$^{3}$,
    I. Jim\'{e}nez-Serra$^{5}$,
    J. C. Tan$^{6}$, \and 
    F. Fontani$^{7}$,
    A. Pon$^{8}$, and 
    S. Ragan$^{9}$
    \\ {$^{1}$Astrophysics Research Institute, Liverpool John Moores University, 146 Brownlow Hill, Liverpool L3 5RF, UK }
    \\ {$^{2}$School of Physics and Astronomy, University of Leeds, LS2 9JT, Leeds, UK}
    \\ {$^{3}$Max Plank Institute for Extraterrestrial Physics (MPE), Giessenbachstrasse 1, 85748 Garching, Germany }
    \\ {$^{4}$Max Plank Institute for Astronomy (MPIA), Konigstuhl 17, 69117 Heidelberg, Germany}
    \\ {$^{5}$Queen Mary University of London, Astronomy Unit, Mile End Road, London E1 4NS, UK}
    \\ {$^{6}$Department of Astronomy, University of Florida, Gainesville, FL 32611, USA }
    \\ {$^{7}$INAF - Osservatorio Astrofisico di Arcetri, L.go E. Fermi 5, I-50125, Firenze, Italy}
    \\ {$^{8}$Department of Physics and Astronomy, The University of Western Ontario, 1151 Richmond Street, London, N6A 3K7, Canada}
    \\ {$^{9}$School of Physics \& Astronomy, Cardiff University, Queen's building, The parade, Cardiff, CF24 3AA, UK}
    }
    
\date{Accepted 2018 January 17. Received 2018 January 11; in original form 2017 November 27.}
\pagerange{\pageref{firstpage}--\pageref{lastpage}} 
\pubyear{2017}

\maketitle
\label{firstpage}

\begin{abstract}\label{abstract}
Infrared dark clouds (IRDCs) are thought to be potential hosts of the elusive early phases of high-mass star formation. Here we conduct an in-depth kinematic analysis of one such IRDC, G034.43+00.24 (Cloud F), using high sensitivity and high spectral resolution IRAM-30m \ntwohoz\ and \ceooz\ observations. To disentangle the complex velocity structure within this cloud we use Gaussian decomposition and hierarchical clustering algorithms. We find that \change{four} distinct coherent velocity components are present within Cloud F. The properties of these components are compared to those found in a similar IRDC, G035.39-00.33 (Cloud H). We find that the components in both clouds have: high densities (inferred by their identification in \ntwoh), trans-to-supersonic non-thermal velocity dispersions with Mach numbers of $\sim$\,$1.5-4$, a separation in velocity of $\sim$\,3\kms, and a mean red-shift of $\sim$\,0.3\,\kms\ between the \ntwoh\ (dense gas) and \ceo\ emission (envelope gas). The latter of these could suggest that these clouds share a common formation scenario. We investigate the kinematics of the larger-scale Cloud F structures, using lower-density-tracing \tcooz\ observations. A good correspondence is found between the components identified in the IRAM-30m observations and the most prominent component in the \tco\ data. We find that the IRDC Cloud F is only a small part of a much larger structure, which appears to be an inter-arm filament of the Milky Way. 
\end{abstract}

\begin{keywords}
stars: formation --- stars: massive --- ISM: clouds --- ISM: individual (G034.43+00.24) --- ISM: molecules.
\end{keywords}
\section{Introduction}\label{sec:introduction}

Young stars, particularly the most massive, are of great astrophysical importance. The huge amounts of energy and momentum that they inject into the interstellar medium have a significant effect on the evolution of their host galaxy. Yet, despite ongoing efforts, the formation process of massive stars is not fully understood. To gain such an understanding, one needs to study the initial conditions under which they form, before protostellar feedback removes information (e.g. kinematic and chemical) of the environment in which the earliest stages of star formation occur. Therefore, observations of quiescent star-forming regions have to be made in order to study the initial conditions of high-mass star, and stellar cluster, formation. This necessitates the identification of molecular clouds with sufficient mass and density, which currently exhibit a low star formation activity. 

Infrared dark clouds (IRDCs) are a group of molecular clouds that were first identified in the mid-nineties as promising astro-laboratories to study the initial conditions of high-mass star formation. The Infrared Space Observatory ({\it ISO}; 15\,\micron; \citealp{perault_1996}) and the Midcourse Space Experiment ({\it MSX}; 7 to 25\,\micron; \citealp{egan_1998}) were used to initially discover IRDCs, and identified them as regions of strong mid-infrared extinction against the background Galactic emission. More recent works have shown that IRDCs are cold ($<$\,20\,K; \citealp{pillai_2006, ragan_2011}), are massive ($\sim$\,10$^{3-5}$\,\sol; \citealp{rathborne_2006, longmore_2012, kainulainen_2013}), have large column densities ({\it N}(H$_2$)\,$\sim$\,$10^{22-25}$\,cm$^{-2}$; \citealp{egan_1998, carey_1998, simon_2006a, vasyunina_2009}), and have high volume densities ({\it n}(H$_2$)\,$\sim$\,10$^{3-5}$cm$^{-3}$ ; e.g. \citealp{peretto_2010, hernandez_2011, butler_2012}). Of particular importance, IRDCs have been shown to have large reservoirs of relatively pristine gas, which has not been influenced by star formation, as inferred from their chemical composition (e.g. \citealp{miettinen_2011, gerner_2015, barnes_2016, kong_2016}).

Although the physical and chemical properties of IRDCs have been well studied (the extinction or continuum dust morphology, dust and gas masses, dust and gas temperatures, and levels of molecular depletion) only recently have dedicated studies of their complex kinematic structure been attempted. Observations of molecular line transitions have shown that molecular clouds, even with relatively simple extinction morphologies, can contain a complex network of velocity components \citep{henshaw_2014, hacar_2013, hacar_2017b}. However, reliably disentangling such structures is difficult (e.g. \citealp{jimnez-serra_2010, jimnez-serra_2014, devine_2011, henshaw_2013, henshaw_2014, tan_2013, pon_2016b, kong_2017, zamora-aviles_2017}), yet doing so is key to understanding the role of molecular cloud structure and evolution in star formation (e.g. \citealp{ragan_2006, devine_2011, rygl_2013, kirk_2013, tackenberg_2014}).

\begin{figure}
\centering
\includegraphics[trim = 42mm 3mm 50mm 4mm, clip, angle=0, width= 1.01\columnwidth]{\dir 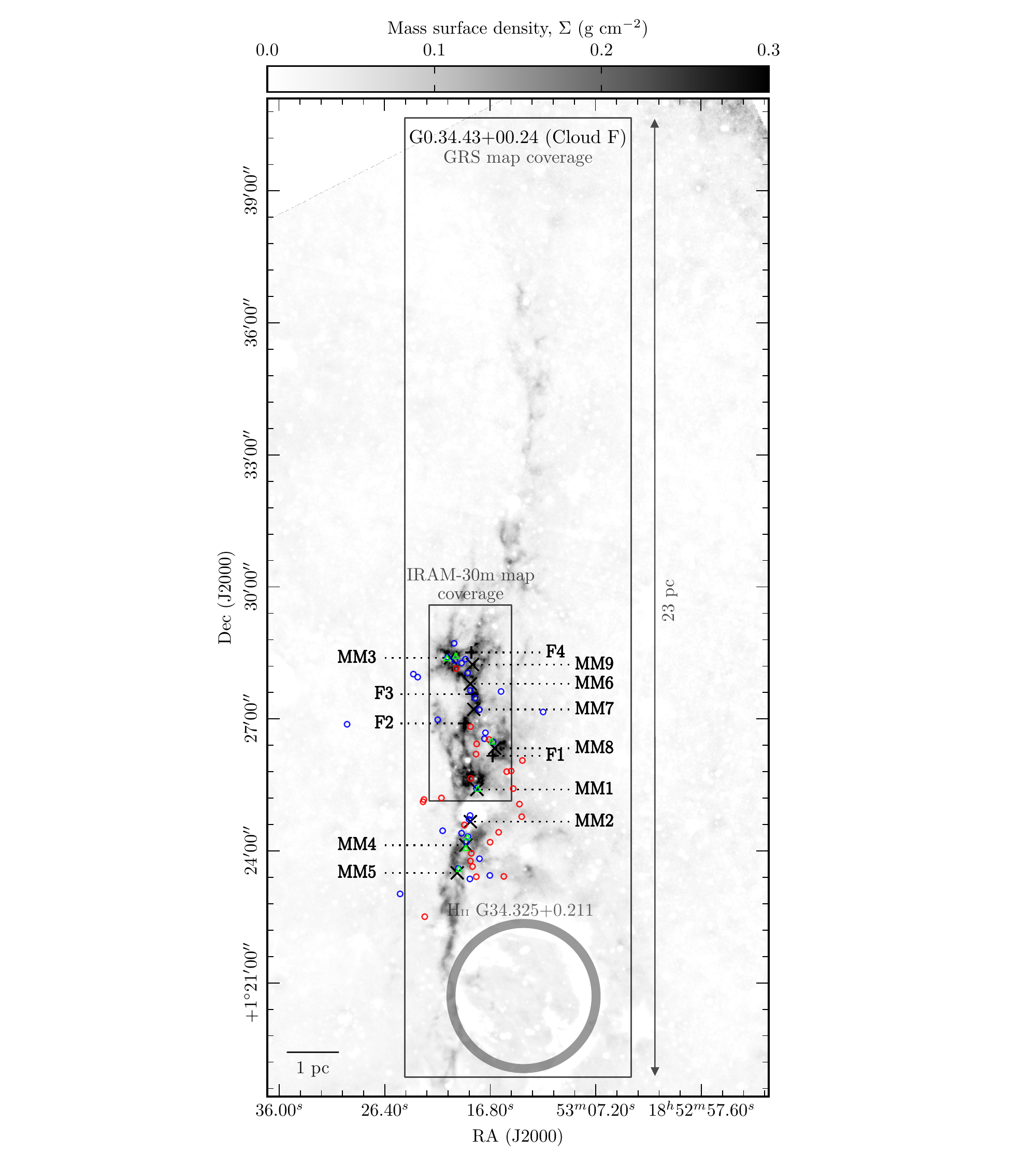} \vspace{-6mm}
\caption{Shown in greyscale is the high-resolution, high-dynamic-range mass surface density map of the IRDC G034.43+00.24, produced by combining the dust extinction at the near- and far-infrared wavelengths \citep{kainulainen_2013}. The black rectangles show the coverage of the Galactic Ring Survey \citep[][see Appendix\,\ref{Appendix C}]{jackson_2006} and IRAM-30m observations. Shown with $+$ and $\times$ symbols and labeled are the positions of the ``core'' regions identified by \citet[][with F prefix]{butler_2012} and those from \citet[][with MM prefix]{rathborne_2006}, respectively. Shown as coloured circles are the young stellar objects candidates, identified by their spectral energy distribution in the {\it Spitzer} bands \citep{shepherd_2007}: ``good'' and ``poor'' (i.e. those with a poor or no spectral energy distribution fit) detections are shown in blue and red, respectively. Sources with extended, enhanced 4.5\micron\ emission, or ``green fuzzies'', are plotted as green triangles \citep{chambers_2009}. Shown in the lower right of the map is the approximate shell of the \ion{H}{II} region G34.325+0.211, see \citealp{xu_j_2016} for a discussion of its influence on the IRDC.} 
\label{cloudf_msd}
\end{figure}

This work will focus on the IRDC G034.43+00.24 (henceforth Cloud F), which was first identified by \citet{miralles_1994} as an unresolved elongated structure in NH$_3$ emission (a tracer of cold, dense gas) to the north of the bright IRAS source 1807+0121 (see Figure\,\ref{cloudf_msd}). Further investigation of this region was, however, delayed until the advent of higher resolution infrared telescopes, such as the {\it MSX}, which \citet{simon_2006a} used to identify Cloud F, along with 10,930 other candidate IRDCs, as having an extended structure silhouetted against diffuse background emission. \citet{simon_2006b} then investigated the global properties of the clouds from the \citet{simon_2006a} sample which resided within the Galactic Ring Survey's coverage (a survey of \tcooz\ molecular line emission), and were especially extended (major axis $>$1.$^\prime$53) and had a strong average extinction contrast against the background (with [background\,-\,image]/background\,$>$\,0.25). Using 1.2 mm continuum observations, \citet{rathborne_2006} then investigated the core properties within 38 of these clouds (the positions of these cores within Cloud F are shown in Figure\,\ref{cloudf_msd}), selecting those which had kinematic distance estimates \citep{simon_2006b}. \citet{butler_2009, butler_2012} and \citet{kainulainen_2013} studied the core properties within 10 of the \citet{rathborne_2006} sample IRDCs, which were relatively nearby, massive, dark, and showed relatively simple surrounding diffuse emission (positions shown in Figure\,\ref{cloudf_msd}). These maps highlighted Cloud F in particular (along with G035.39-00.33; see section\,\ref{comparison_to_cloudh}), as having a complex filamentary morphology containing several massive cores, and a large amount of dense, quiescent gas (also see \citealp{fontani_2011} and \citealp{kong_2017} for chemical studies towards the quiescent gas within this cloud). The stringent selection process, through several datasets, summarised here, has singled out Cloud F as an ideal candidate in which to investigate the initial conditions of massive star formation. Table\,\ref{cloud_props} presents the physical properties of interest for Cloud F (determined within the area mapped by the IRAM-30m observations), and Figure\,\ref{cloudf_msd} shows the mass surface density map across the cloud region \citep{kainulainen_2013}.

To investigate the kinematic structures on various scales within Cloud F, we use emission from the \ceooz\ and \ntwohoz\ molecular line transitions. Assuming that the \ceooz\ line is thermalised and optically thin, this should trace the more extended gas, as its critical density is comparable to the average volume density expected within IRDCs ($\sim$\,$10^{3-4}$\,cm$^{-3}$ when observed at scales of $\sim$\,0.5\,pc; e.g. \citealp{henshaw_2013}). The \ntwohoz\ transition has a significantly higher critical density ($\sim$\,$10^{4-5}$\,cm$^{-3}$), and therefore is expected to trace the higher density regions. In nearby low-mass star-forming regions \ntwohoz\ typically traces dense cores (e.g. \citealp{caselli_2002, andre_2007, friesen_2010}), however, given the significantly higher volume densities seen within some IRDCs, this line is found to be extended (e.g \citealp{tackenberg_2014, henshaw_2013, henshaw_2014}). 

We note, however, that the transitions from the \ntwoh\ molecule contain hyperfine structure, which can complicate the analysis of kinematically complex regions, where the hyperfine components can be merged to form one broad component (e.g in the case where the line width is larger than the separation of the components). Unlike its higher J-transitions, however, \ntwohoz\ has a hyperfine component (the F${_1}$, F = 0,1 $\rightarrow$ 1,2 transition) which is ``isolated'' by $>$\,7\kms\ from the main group (i.e. those with a separation of $\sim$\,1\kms; \citealp{caselli_1995}), and is, therefore, unlikely to merge given the typical line properties observed within IRDCs (e.g. with line-widths of $\sim$\,1\kms; \citealp{henshaw_2013}). As all the analysis presented in this work will be conducted on the isolated hyperfine component of \ntwohoz\, henceforth, unless otherwise stated, when mentioning the \ntwohoz\ transition we are referring to this hyperfine component. To do so, we will centre on the frequency of the isolated hyperfine component from \citet{pagani_2009}. We note, however, that slightly different frequencies for the isolated hyperfine component are available in the literature  \change{($93176.2637 - 93176.2650$\,MHz; \citealp{caselli_1995, cazzoli_2012})},\footnote{See \url{https://www.astro.uni-koeln.de/cdms}} yet changing to these will only shift the centroid velocity by \change{($3.7 - 4.1$)\,$\times$\,10$^{-2}$\,\kms}. As this variation is below the spectral resolution of $\sim$\,6\,$\times$\,10$^{-2}$\,\kms\ of the \ntwohoz\ observations used throughout this work (see Table\,\ref{parameters}), we do not expect this to significantly affect the results presented throughout this work (particular importance for section\,\ref{shift}).

This paper is structured in the following manner. The details of the IRAM-30m observations towards Cloud F can be found in Section\,\ref{sec:observations}. The results are presented in Section\,\ref{sec:results}. The analysis of the kinematic structure is then given in Section\,\ref{Analysis} and is discussed in Section\,\ref{Discussion}. Here we compare the structures identified to those identified in a similar IRDC (G035.39-00.33), and the larger scale structures identified from \tcooz\ observations. The conclusions of this work are given in Section\,\ref{Conclusions}. A discussion of kinematics with reference to previous analyses of Cloud F, and calculation of the CO depletion within the cloud are given in Appendices\,\ref{Appendix D} and \ref{Appendix B}. In Appendix\,\ref{Appendix E} we briefly discuss the physical interpretation of the structures identified from the molecular lines observations presented in this work. The analysis of the IRAM-30m Cloud H observations and of the \change{Galactic Ring Survey \citep[GRS;][]{jackson_2006}} observations are presented in Appendices\,\ref{Appendix A} and \ref{Appendix C}. 

\begin{table}
\caption{Cloud properties within the IRAM-30m mapped region, shown in Figure\,\ref{cloudf_msd}. See section\,\ref{comparison_to_cloudh} for comparison to the IRDC G035.39-00.33 (or Cloud H; \citealp{butler_2009}).  \vspace{0.2cm}}
\centering
\begin{tabular}{c c c}
\hline
Cloud property & Cloud F & Cloud H \\
within IRAM-30 map & (G034.43+00.24)  & (G035.39-00.33) \\
\hline
Distance, $d$ (kpc) $^a$              &  3.7\,$\pm$\,0.6 & 2.9\,$\pm$\,0.4  \\
Map size, $R$ (pc)   $^b$           & 3.4\,$\pm$\,0.5 & 2.3\,$\pm$\,0.3 \\ 
Aspect ratio, $A_0$         & 2.4 & 2.6 \\ 
$\Sigma$ (g\,cm$^{-2}$) $^c$     & 0.10\,$\pm$\,0.03 & 0.09\,$\pm$\,0.03 \\
$f_{\rm D}$ $^d$                  & 1.1$\,\pm$\,0.6 & 2.8\,$\pm$\,1.4 \\
Mass, $M$ (\sol) $^e$                    & 4700\,$\pm$\,1400 & 1700\,$\pm$\,500 \\

$T$ (K) $^f$                           & $\sim$\,17 & $\sim$\,13 \\
$m$ (\sol\,pc$^{-1}$) $^g$               & 1400\,$\pm$\,400 & 740\,$\pm$\,200 \\ 
\hline
\end{tabular}
\begin{minipage}{\columnwidth}\vspace{0.1cm}
$a$: Near kinematic distance to the sources \citep{simon_2006b, roman-duval_2009}. See section\,\ref{grs_comparison} for further discussion of the source distance.\\
$b$: Calculated from the mean value of the Ra and Dec range at the assumed source distance.\\
$c$: Average mass surface density \citep{kainulainen_2013}. \\
$d$: CO depletion factor presented in Appendix\,\ref{Appendix C} compared to the value measured by \citet{hernandez_2012a}. \\
$e$: Masses calculated for the region covered by the IRAM-30m observations. Total cloud masses calculated by \citep{butler_2012} are 4,460 and 13,340\,\sol\ for Clouds F and H, respectively.\\
$f$: \citet{dirienzo_2015, pon_2016a, sokolov_2017} \\
$g$: The mass per unit length can be given as $m = M / R$.
\end{minipage}

\label{cloud_props}
\end{table}
\section{Observations}\label{sec:observations}

The \ceooz\ and \ntwohoz\ observations towards Cloud F were obtained using the \change{Institute for Radio Astronomy in the Millimeter Range} 30-m telescope (IRAM-30m) on Pico Veleta, Spain, over the 27$^{th}$ - 28$^{th}$ July 2012.\footnote{Project code: 025-12} The data cubes were produced from On-The-Fly (OTF) mapping, covering an area of $\sim$\,104\arcsec\,$\times$\,240\arcsec\ (corresponding to 2pc\,$\times$\,4.8pc, at the source distance of 3.7 kpc; \citealp{simon_2006b}), using central reference coordinates of RA\,(J2000)\,=\,18$^h$53$^m$19$^s$, Dec\,(J2000)\,=\,01$^{\circ}$27$'$21${''}$,\footnote{In Galatic coordinates {\it l} = 34.441$^{\circ}$, {\it b} = 0.247$^{\circ}$.} which is shown on Figure\,\ref{cloudf_msd}. These observations were carried out using the EMIR receivers. The VErsatile Spectrometer Assembly (VESPA) provided spectral resolutions of $\sim$ 20 - 80 kHz. 

The {\sc gildas}\footnote{see \url{https://www.iram.fr/IRAMFR/GILDAS/}} packages {\sc class} and {\sc mapping} were used to reduce and post-process the data. This included subtracting a single-order polynomial function to produce a flat baseline and convolving the OTF-data with a Gaussian kernel, thereby increasing the signal-to-noise ratio and allowing us to resample the data onto a regularly spaced grid. All the intensities were converted from units of antenna temperature, $T^{*}_{A}$, to main-beam brightness temperature, $T_{MB}$, using the beam and forward efficiencies shown in Table\,\ref{parameters}. The native angular resolution of the IRAM-30m antenna at the frequency of the \ceooz\ and \ntwohoz\ transitions are $\sim$\,23\arcsec\ and 26\arcsec, respectively. Both data sets are smoothed to achieve an effective angular resolution of $\sim$\,28\arcsec, with a pixel spacing of 14\arcsec, to allow comparison (corresponding to a spatial resolution of $\sim$\,0.5\,pc at the source distance of $\sim\,3.7\,$kpc; \citealp{simon_2006b}).

\begin{table}
\caption{Observational parameters.\vspace{0.2cm}}
\centering
\begin{tabular}{c c c c}
\hline
\begin{tabular}[x]{@{}c@{}}Observational\\ parameter \end{tabular}   & \ntwohoz\ & \ceooz\ \\
\hline
Frequency (MHz) & 93176.7637 $^{a}$ & 109782.1780 $^{b}$ \\ 
HPBW (\arcsec) $^{c}$ & 26 & 23 \\
\begin{tabular}[x]{@{}c@{}}Velocity Resolution\\  (km s$^{-1}$) \end{tabular} & 6.28 $\times$ 10$^{-2}$ & 5.33 $\times$ 10$^{-2}$ & \\
Beam Efficiency & 0.81 & 0.78 \\
Forward Efficiency & 0.95 & 0.94\\
{\it rms} (K) & 0.13 & 0.15 \\
\hline
\end{tabular}
\begin{minipage}{\columnwidth}\vspace{0.1cm}
$a$: Frequency of main hyperfine component, the isolated component \ntwoh\ (J, F${_1}$, F = 1,0,1 $\rightarrow$ 0,1,2) has a frequency of 93176.2522 MHz \citep{pagani_2009}. \\
$b$: \citet{cazzoli_2003} \\
$c$: Calculated as $\theta\mathrm{_{HPBW}}\,=\,1.16\,\lambda\,/\,D$, where $\lambda$ and $D$ are the wavelength and telescope diameter, respectively (see \url{http://www.iram.es/IRAMES/mainWiki/Iram30mEfficiencies}).   \\
\end{minipage}
\label{parameters}
\end{table}
\section{Results}\label{sec:results}

\subsection{Moment analysis}\label{subsection:moment analysis}

\begin{figure}
\centering
\includegraphics[trim = 3mm 3mm 3mm 3mm, clip,angle=0,width=1\columnwidth]{\dir 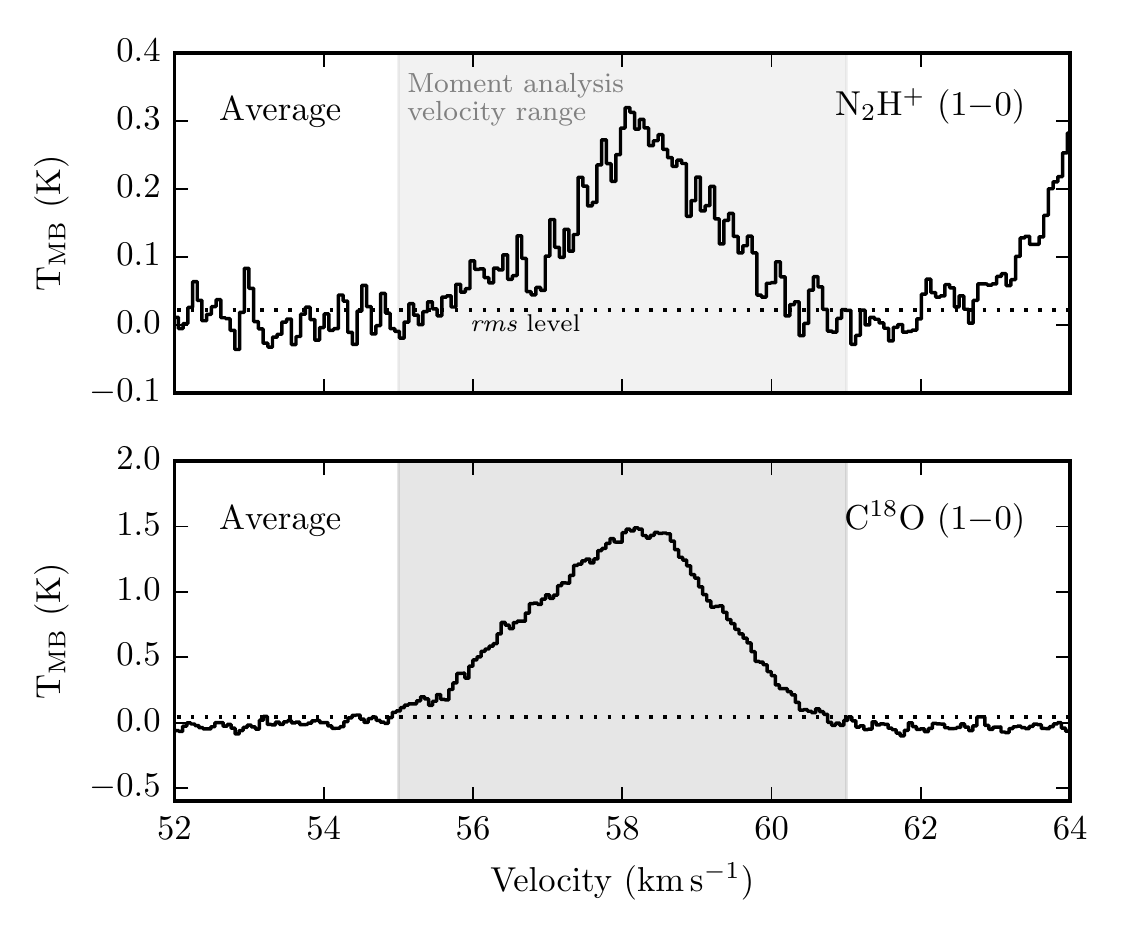}  \vspace{-3mm}
\caption{Shown are the average spectrum of the \ntwohoz\ transition (upper) and the \ceooz\ transition (lower) across the mapped region of Cloud F. The horizontal dotted line represents the {\it rms} level on the average spectrum of $\sim$\,0.02\,K and $\sim$\,0.04\,K for \ntwohoz\ and \ceooz, respectively. Note, these values are different to the average of the {\it rms} within individual positions, which is given in Table\,\ref{parameters}. The shaded region shows the velocity range used for the moment map analysis (see Figure\,\ref{moment_maps_cloudf}).} 
\label{spec_ave_cloudf}
\end{figure}

To gain an initial insight into the intensity distribution and kinematics of the molecular line emission we conduct a moment analysis, using the {\sc spectral cube} package for {\sc python}.\footnote{\url{https://spectral-cube.readthedocs.io/en/latest/}.} This analysis has been carried out for a velocity range of $55 - 61$\,\kms\ for both lines, which was chosen to best incorporate all the significant emission from Cloud F identified in the spectrum averaged across the whole mapped area, shown in Figure\,\ref{spec_ave_cloudf}. The average uncertainty on the integrated intensity towards each position for \ntwohoz\ and \ceooz\ are $\sigma \sim$\,0.08\,K\,\kms\ and $\sigma \sim$\,0.09\,K\,\kms, respectively (the {\it rms} is shown in Table\,\ref{parameters}, and the uncertainty has been calculated following \citealp{caselli_2002a}). The pixels below a 3\,$\sigma$ threshold have been masked after the moment analysis procedure. 

\begin{figure*}
\centering
\includegraphics[trim = 3mm 4mm 3mm 2mm, clip,angle=0,width=1\textwidth]{\dir 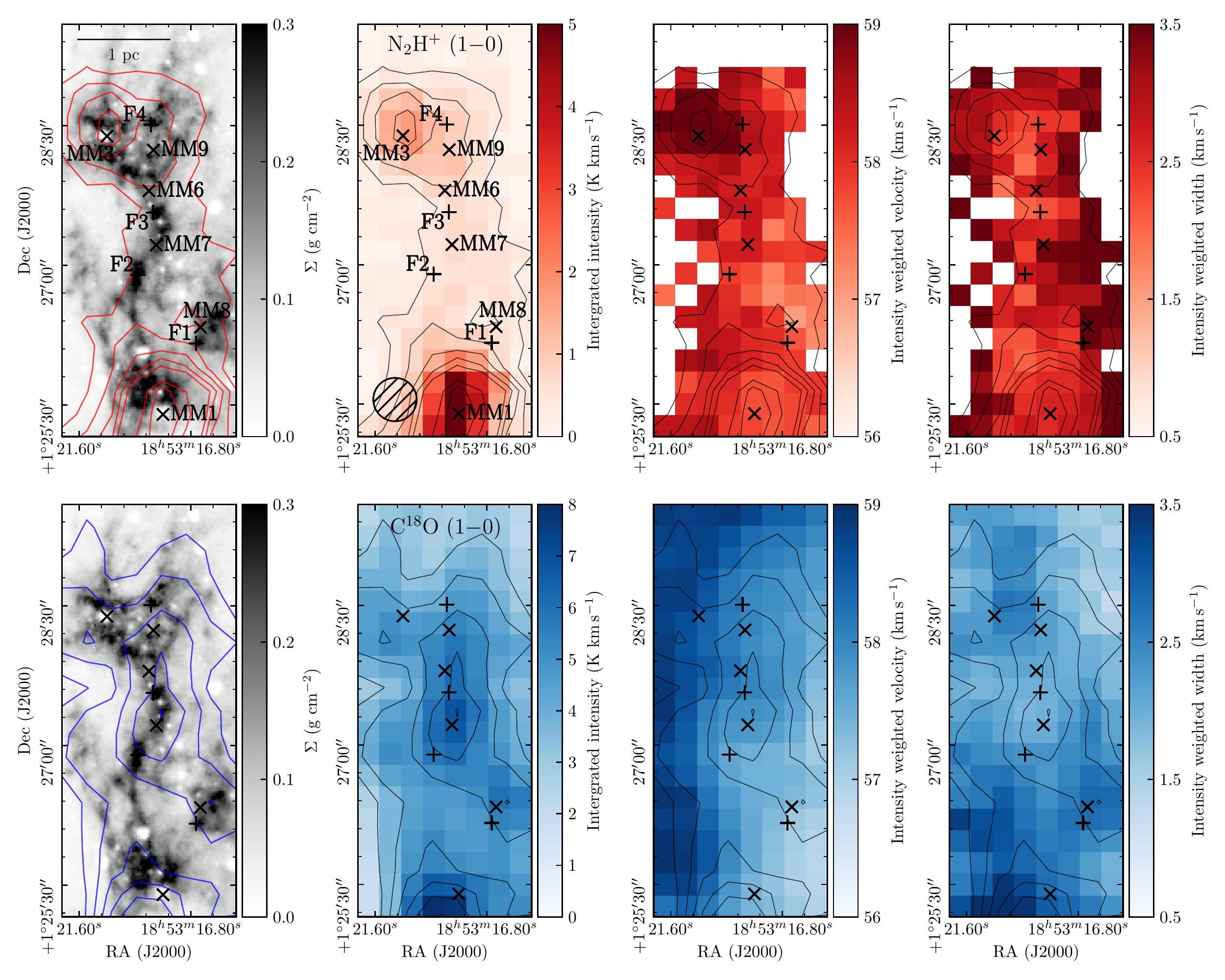}
\caption{Moment map analysis of the Cloud F \ntwoh\,($1 - 0$) (upper row) and \ceo\,($1 - 0$) (lower row) observations. Shown in greyscale in the first column is the mass surface density map of Cloud F, determined from near and mid- infrared extinction \citep{kainulainen_2013}. Shown with $+$ and $\times$ symbols are the positions of the ``core'' regions identified by \citet{butler_2012} and those from \citet{rathborne_2006}, respectively, which are labeled in both the upper left and upper centre-left panels. Shown in the second, third and fourth columns are the integrated intensities (0$^\mathrm{th}$ order moment), intensity weighted velocity field (1$^\mathrm{st}$ order moment), and intensity weighted line width (2$^\mathrm{nd}$ order moment) maps. Overlaid as red (first column) and black (second, third and fourth columns) contours in the upper row is the integrated intensity of \ntwoh\,($1 - 0$), in steps of $\{ 5, 10, 15, 20, 25, 45, 55\}\,\sigma$; where $\sigma \sim$\,0.08\,K\,\kms. Overlaid as blue (first column) and black (second, third and fourth columns) contours in the lower row is the integrated intensity of \ceo\,($1 - 0$), in steps of $\{ 40, 50, 60, 70, 80\}\,\sigma$; where $\sigma \sim$\,0.09\,K\,\kms. The moment analysis has been performed above 3$\sigma$ for all transitions. Shown in the lower left corner of the second upper panel is the smoothed angular beam size.} 
\label{moment_maps_cloudf}
\end{figure*}

The results of the moment analysis towards Cloud F are presented in Figure\,\ref{moment_maps_cloudf}. Shown in greyscale in the first column is the mass surface density map determined from extinction in the near infrared \citep{kainulainen_2013}. Shown in the second, third and fourth columns are the integrated intensities (0$^\mathrm{th}$ order moment), intensity weighted velocity field (1$^\mathrm{st}$ order moment), and intensity weighted line width (2$^\mathrm{nd}$ order moment) maps, respectively. For reference, contours of the integrated intensity are overlaid on each panel, and the positions of the \citet{rathborne_2006} and \citet{butler_2012} core regions are plotted on the mass surface density map. 

This analysis shows that both the \ntwohoz\ and \ceooz\ emission is extended across the length of the IRDC, where only a few ($<$\,10 per cent) of the pixels do not meet the 3\,$\sigma$ integrated intensity threshold. The \ntwohoz\ emission traces the mass surface density map morphology relatively well, with peaks towards the MM3 and MM1 core regions. The \ceooz\ emission also traces the mass surface density morphology, albeit to a lesser extent than the \ntwohoz\ emission, peaking at the position of the MM7 core, to the west of the F1 and MM8 regions, to the south-east of the MM3 region, and towards the MM1 region. A likely cause of the different spatial distributions of the \ceooz\ and \ntwohoz\ emission is that \ceo\ traces the extended envelope material, whereas \ntwoh\ is expected to trace the dense gas, which follows the continuum cores and mass surface density distribution. Furthermore, towards these densest regions, unlike \ntwoh, \ceo\ can suffer from freeze-out (see Appendix\,\ref{Appendix B}). 


The intensity weighted velocity field maps for both transitions show an increasing velocity from the west to east (right to left on Figure\,\ref{moment_maps_cloudf}). The total difference of velocity across the mapped region is $\sim$\,$2-3$\,\kms, which corresponds to a gradient of the order $\sim$\,$0.5 - 0.7$\,\kms\,pc$^{-1}$ for the approximate distance diagonally across the mapped region of $3-4$\,pc, at the assumed source distance (see Table\,\ref{cloud_props}). 

The intensity-weighted line width maps show different morphologies. The \ceooz\ shows the largest values of the intensity weighted line width towards the MM1 region and the south-east corner of the mapped region ($\sim$\,3.5\,\kms), with peaks towards the MM3 region ($\sim$\,2.7\,\kms), and towards the peak in integrated intensity towards the west of the F1 and MM8 regions ($\sim$\,2.9\,\kms). The \ntwohoz\ emission shows narrower line widths towards the centre of the cloud, with values of $\sim$\,2.5\,\kms\ towards the MM1 region, and $\sim$\,2\,\kms\ towards the MM3 region.

\subsection{Channel map analysis}\label{subsection:channel maps}

\begin{figure*}
\centering
\includegraphics [trim = 2mm 3mm 2mm 2mm, clip,angle=0,width=1\textwidth]{\dir 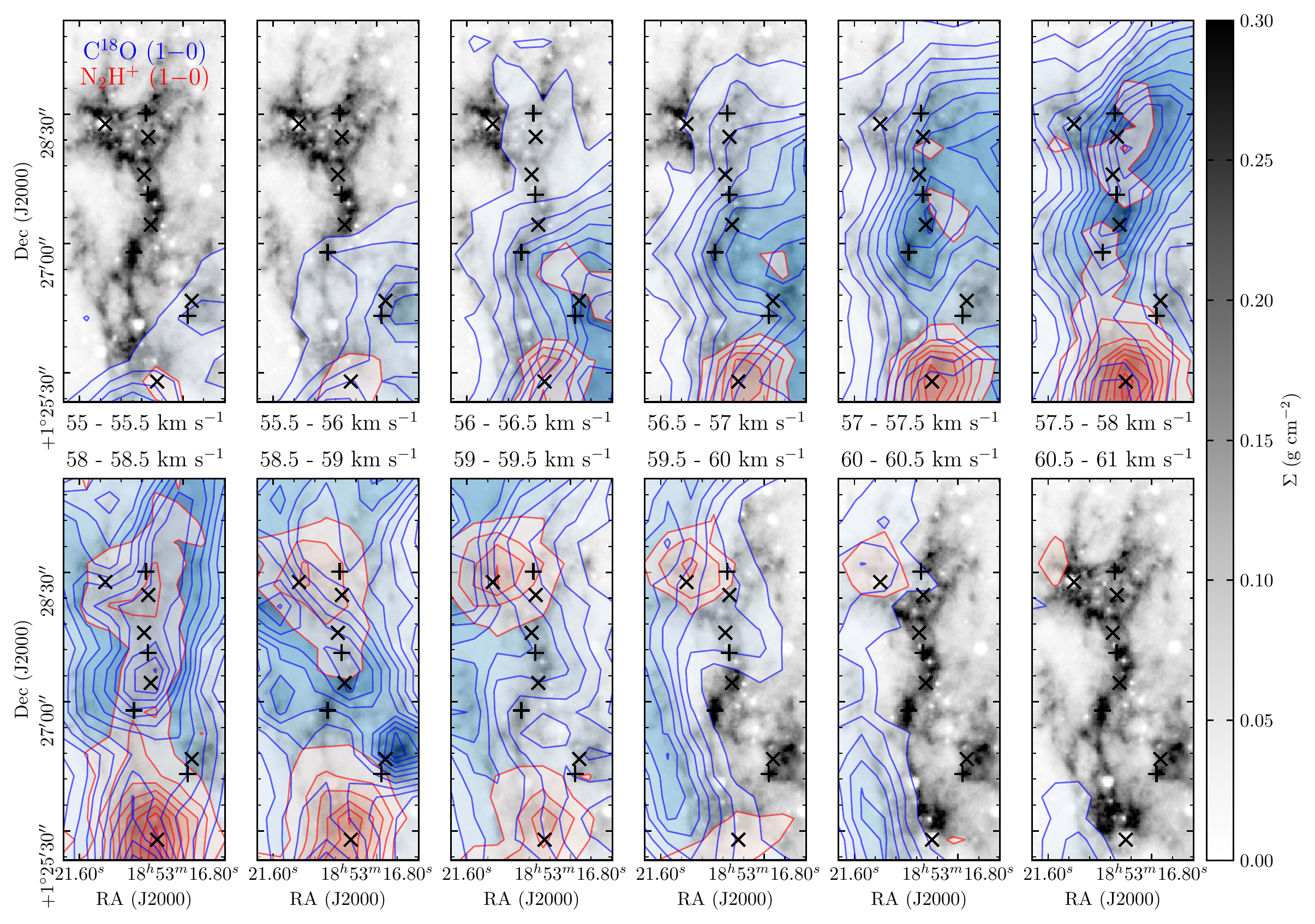}
\caption{Cloud F channel maps of \ntwohoz\ and \ceooz\ shown in red and blue filled contours, which begin at 5\,$\sigma$, and increase in steps of 5\,$\sigma$, where $\sigma\,\sim$\,0.023\,K\,\kms\ and  $\sigma\,\sim$\,0.024\,K\,\kms, respectively. The intensities are integrated from $55-61$\,\kms\ in steps of 0.5\,\kms, as shown below or above each map. Each map is overlaid on the mass surface density map of \citet{kainulainen_2013}. Shown with $+$ and $\times$ symbols are the positions of the ``core'' regions identified by \citet{butler_2012} and those from \citet{rathborne_2006}, respectively.}
\label{channel_maps_cloudf}
\end{figure*}

To investigate the velocity gradients identified in the \ntwoh\ and \ceo\ moment map analysis, the emission from these transitions has been integrated across subsets of the total velocity range used to create the moment maps (referred to as channel maps). We integrate the \ntwohoz\ and \ceooz\ transitions from $55-61$\,\kms\ in steps of 0.5\,\kms\ (which corresponds to approximately 10 channels for both lines). Figure\,\ref{channel_maps_cloudf} shows contours of the integrated intensity in these steps for \ntwohoz\ (in red) and \ceooz\ (in blue), overlaid on the mass surface density map \citep{kainulainen_2013}. 

The channel maps show a complex morphology, where both lines appear to peak towards the south for the majority of the velocity range, with several local maxima appearing at different velocities towards the north of the cloud. These suggest that the velocity gradients identified in the moment map analysis are not continuous, but rather that they are due to distinct peaks in velocity across the map, which, when averaged, mimic a smoothly varying centroid velocity. Identifying velocity structures by arbitrarily separating these maxima can be, however, dependent on the applied spatial and/or velocity boundaries.

\begin{figure*}
\centering
\includegraphics[trim = 4.5cm 10mm 6.3cm 9mm, clip,angle=0,width=1\columnwidth]{\dir 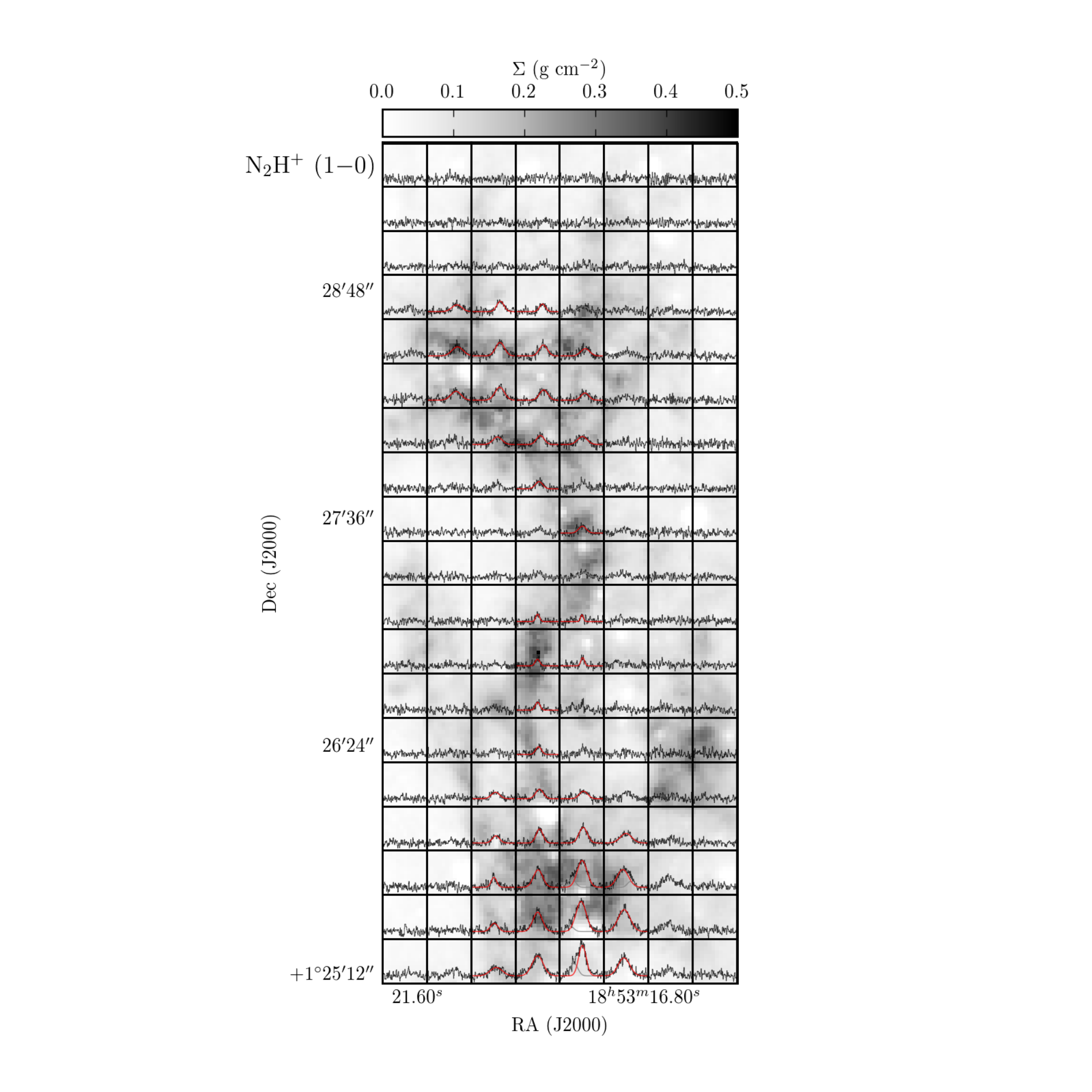}
\includegraphics[trim = 4.5cm 10mm 6.3cm 9mm, clip,angle=0,width=1\columnwidth]{\dir 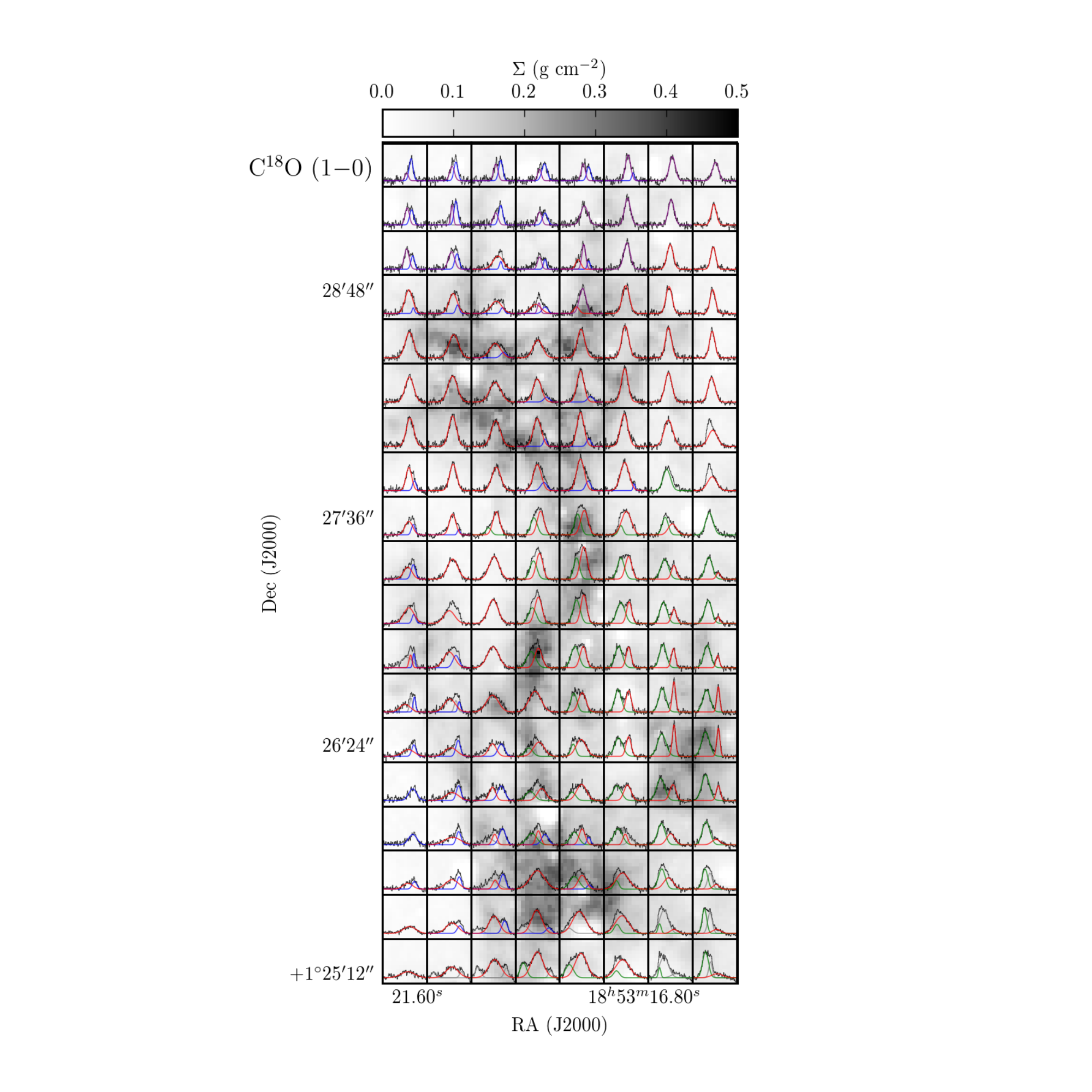}
\caption{The shown are the spectra of the \ntwohoz\ transition (left) and the \ceooz\ transition (right) across the mapped region of Cloud F. The velocity ranges are 54 to 62\kms, and the intensity ranges are -0.5 to 2.5\,K for \ntwoh and -0.5 to 3.5\,K for \ceo. Overlaid on each spectrum are the results of the line fitting ({\sc scouse}) and clustering ({\sc acorns}) routines, which are discussed in section\,\ref{subsection:spectral line fitting and linking}. The colours of these profiles represent the various velocity component associations \change{given in Table\,\ref{velocity component parameters}}. The background greyscale is the mass surface density map, determined from near and mid-infrared extinction \citep{kainulainen_2013}.} 
\label{spec_plot_cloudf}
\end{figure*}

\section{Analysis}\label{Analysis}

To determine if multiple velocity components are present across Cloud F, as the various intensity peaks in the channel map analysis would seem to suggest, we check the individual \ntwohoz\ and \ceooz\ spectra. The two panels of Figure\,\ref{spec_plot_cloudf} show the spectra at each position across the cloud. Multiple distinct velocity components can indeed be clearly identified, predominately in the \ceooz\ emission, at several positions across the cloud. A result which is not evident from the average spectra, shown in Figure\,\ref{spec_ave_cloudf}. A more reliable method to separate these components than is possible with moment or channel maps is, therefore, required to accurately analyse the kinematics within this complex IRDC. \change{In this section, we use use a semi-automated gaussian fitting algorithm and automated hierarchical clustering algorithm. These have been chosen such that the identified coherent velocity structures can be tested for robustness against a range of input parameters, within both the fitting and clustering algorithms. This method ensures that the structures are both reliable and reproducible. Importantly, in section\,\ref{comparison_to_cloudh} we investigate an apparently similar IRDC to Cloud H, for which we use this same method to identify the coherent velocity structures, allowing for a systematic comparison of their kinematic properties.}    
 
\subsection{Spectral line fitting and velocity coherent features}\label{subsection:spectral line fitting and linking}

\begin{figure*}
\centering
\includegraphics[trim = 46mm 33mm 48mm 34mm, clip, angle=0, width=1\columnwidth]{\dir 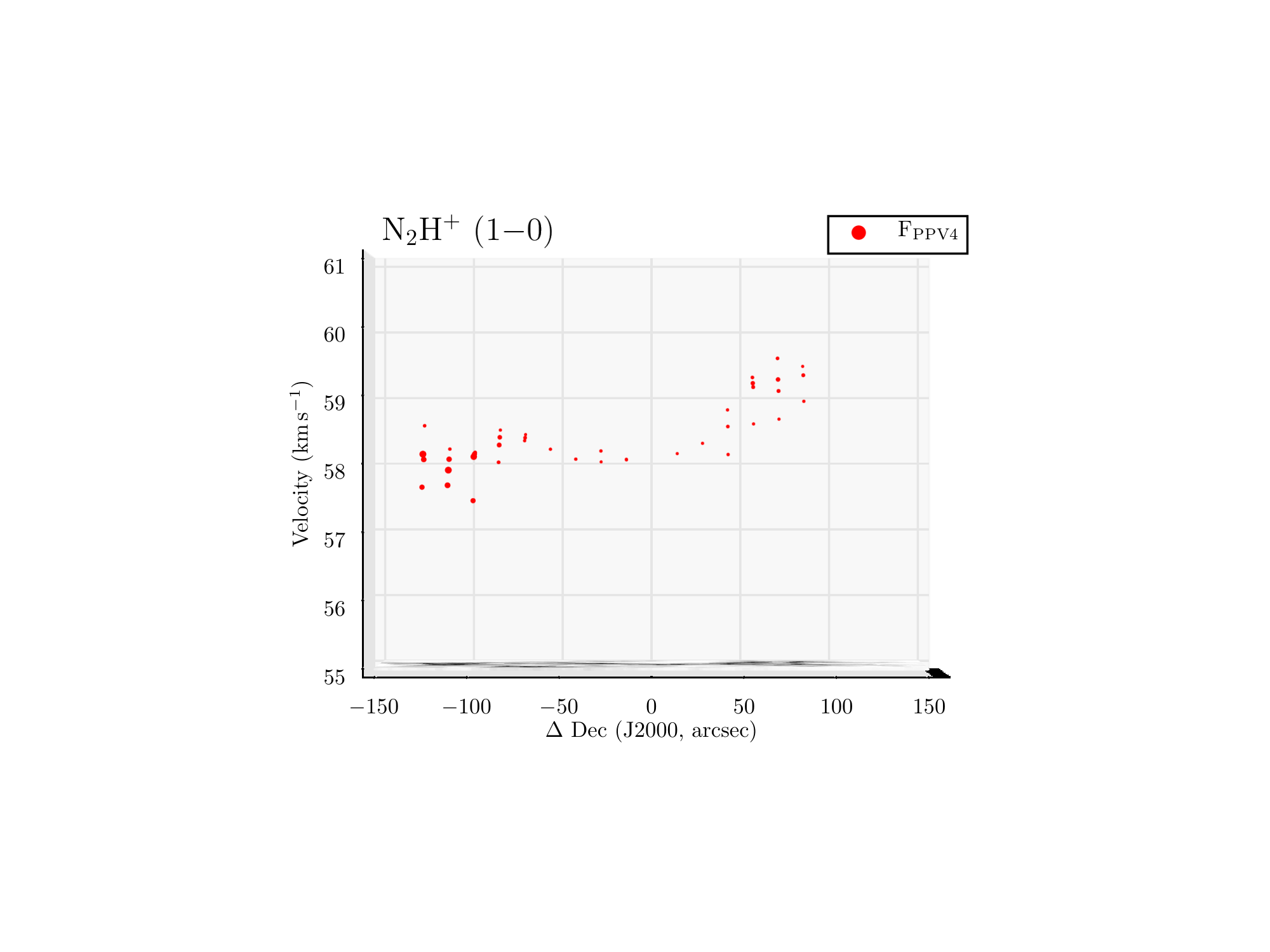} 
\includegraphics[trim = 46mm 33mm 48mm 34mm, clip, angle=0, width=1 \columnwidth]{\dir 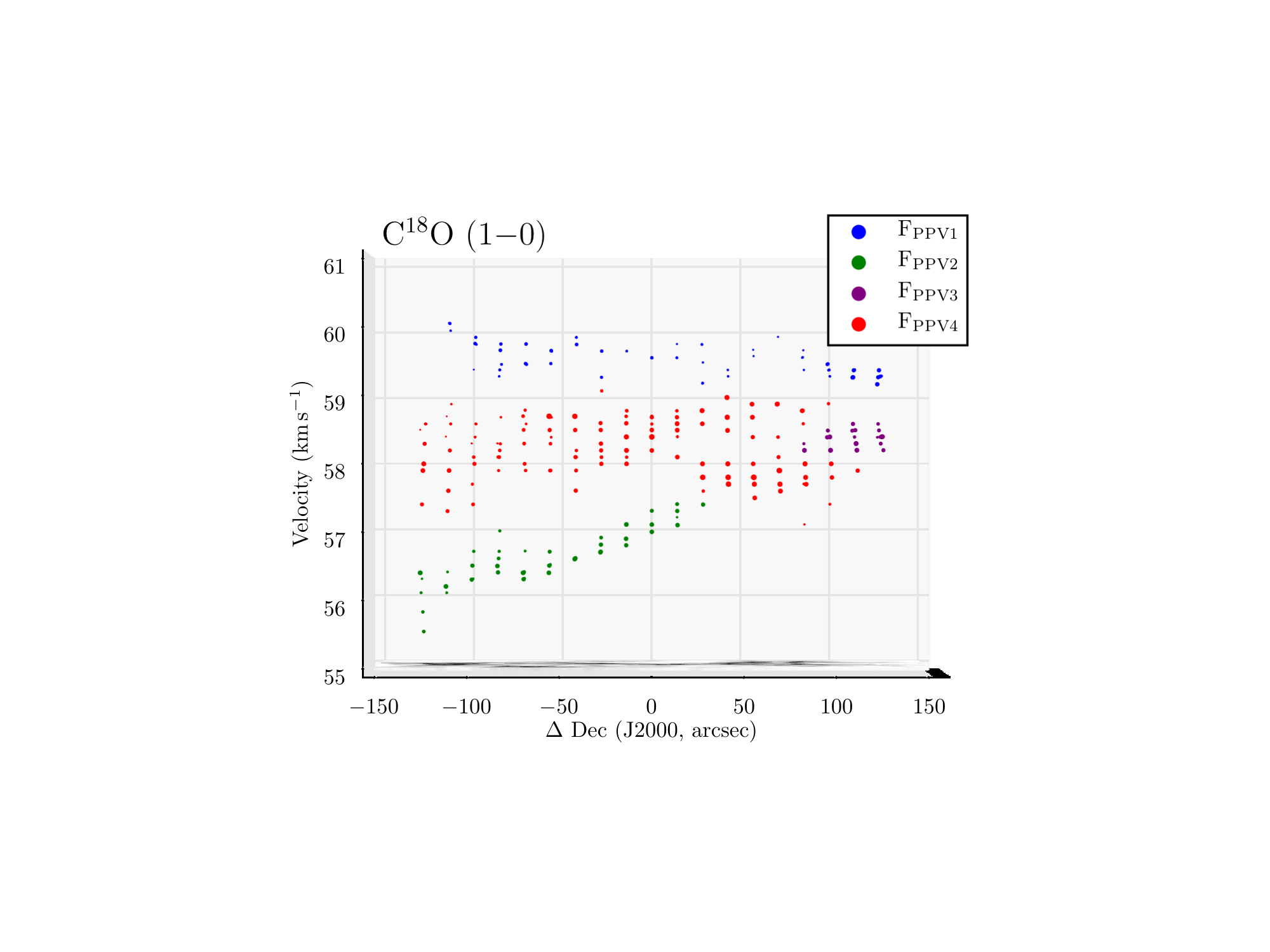}

\includegraphics[trim = 86mm 33mm 30mm 34mm, clip, angle=0, height=0.7\columnwidth]{\dir 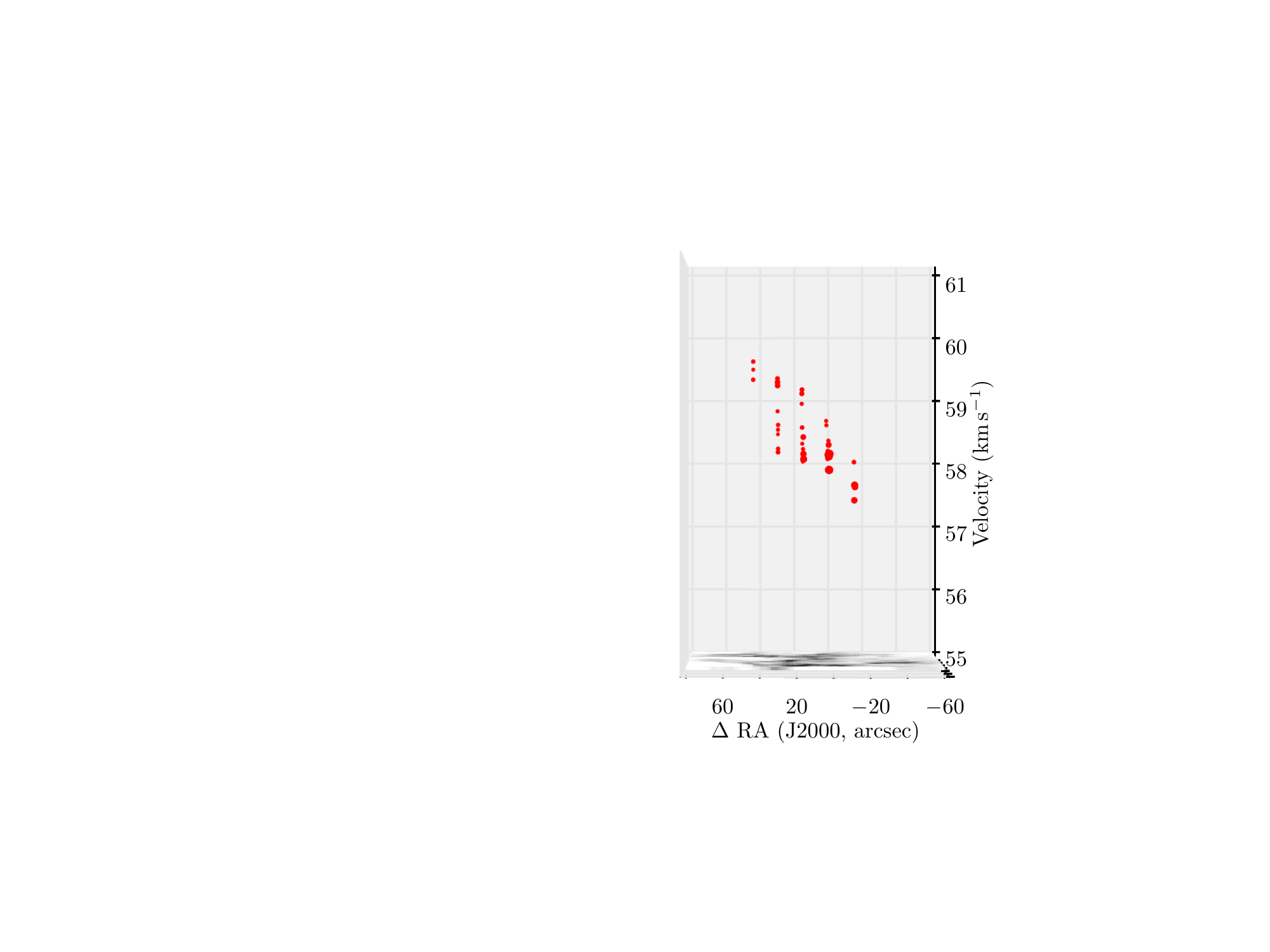} 
\includegraphics[trim = 86mm 33mm 30mm 34mm, clip, angle=0, height=0.7\columnwidth]{\dir 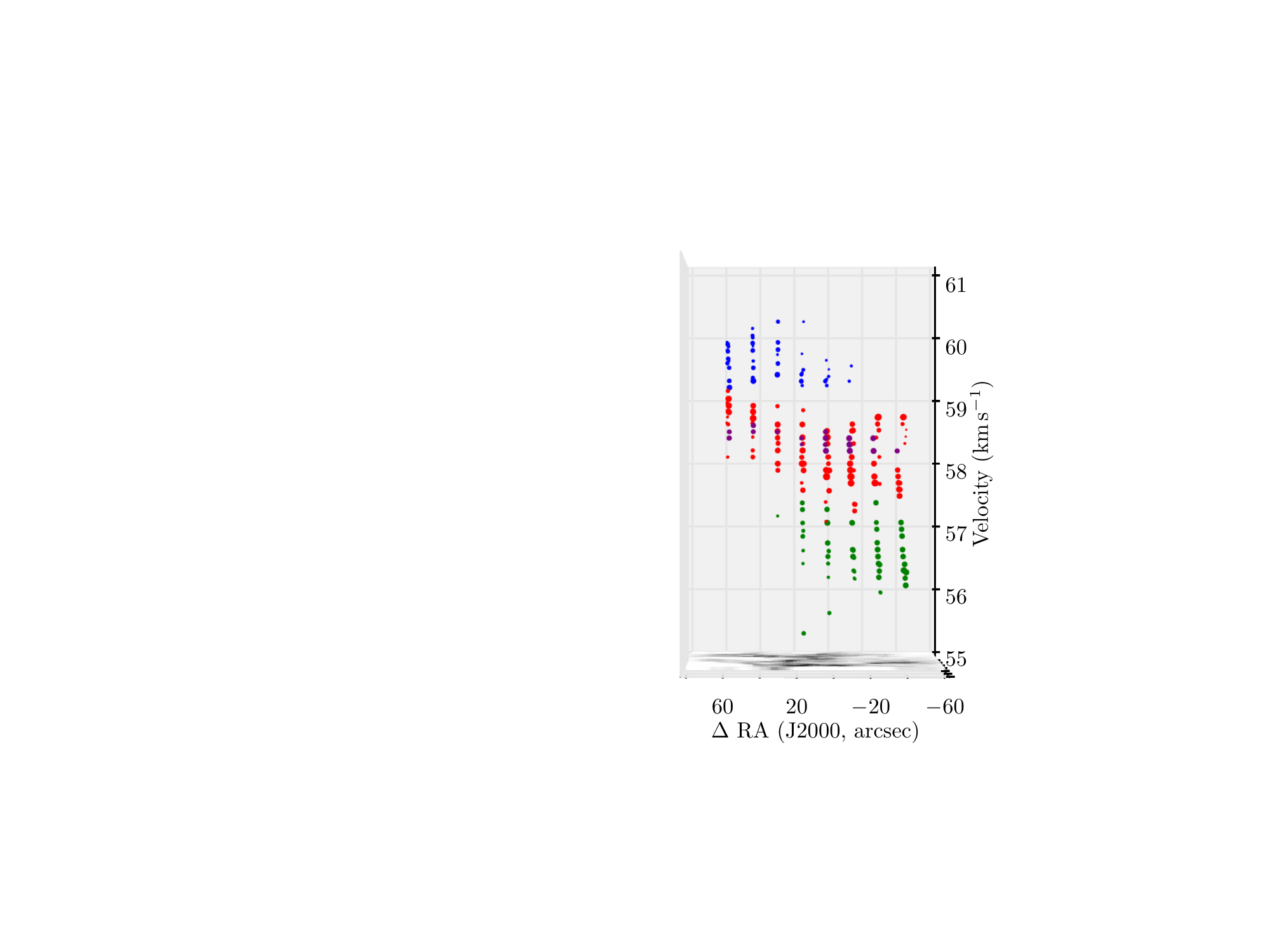}

\includegraphics[trim = 53mm 12mm 25mm 42mm, clip,angle=0, clip,angle=0,width=1.03\columnwidth]{\dir 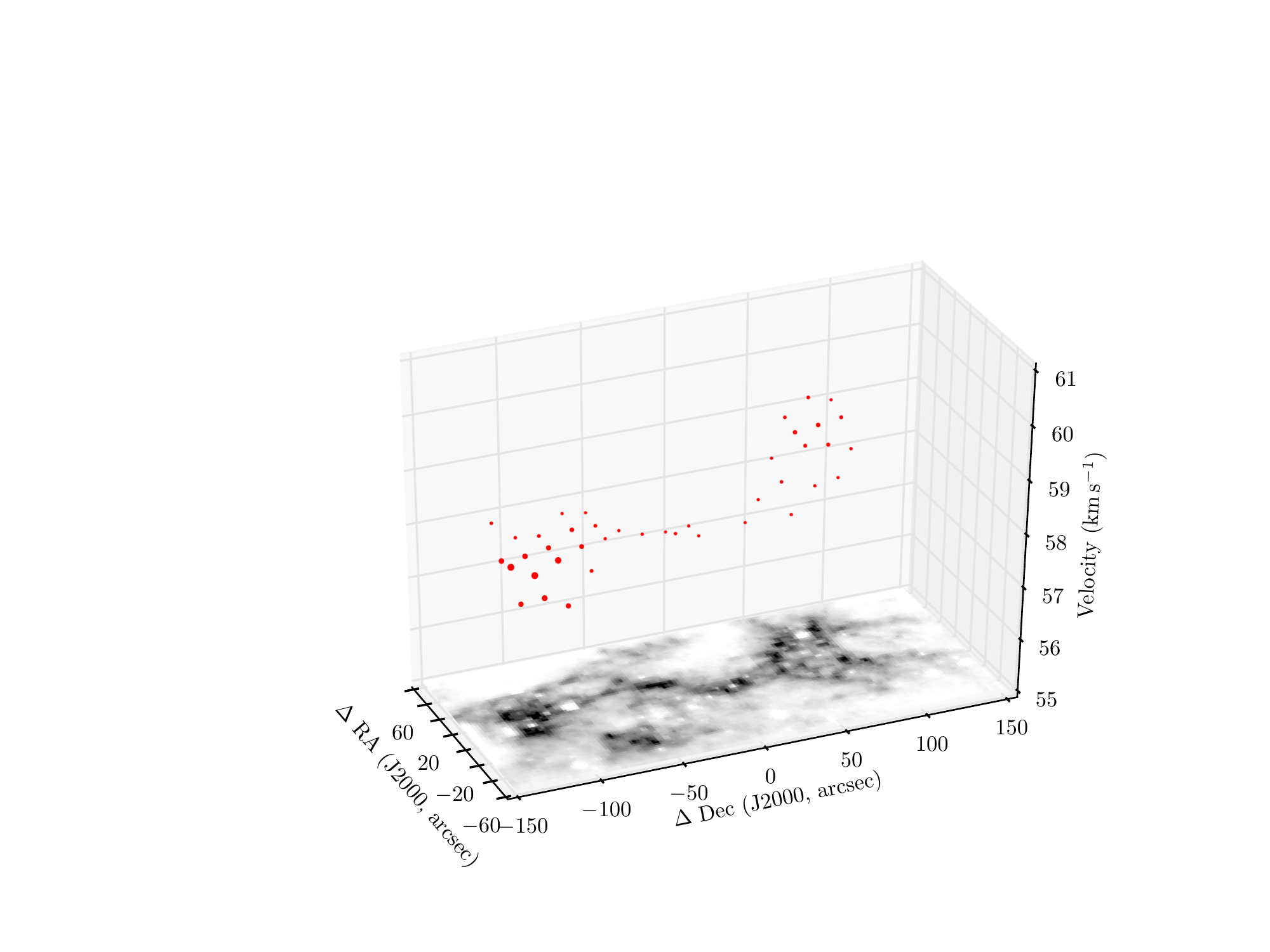} 
\includegraphics[trim = 53mm 12mm 25mm 42mm, clip,angle=0, clip,angle=0,width=1.03\columnwidth]{\dir 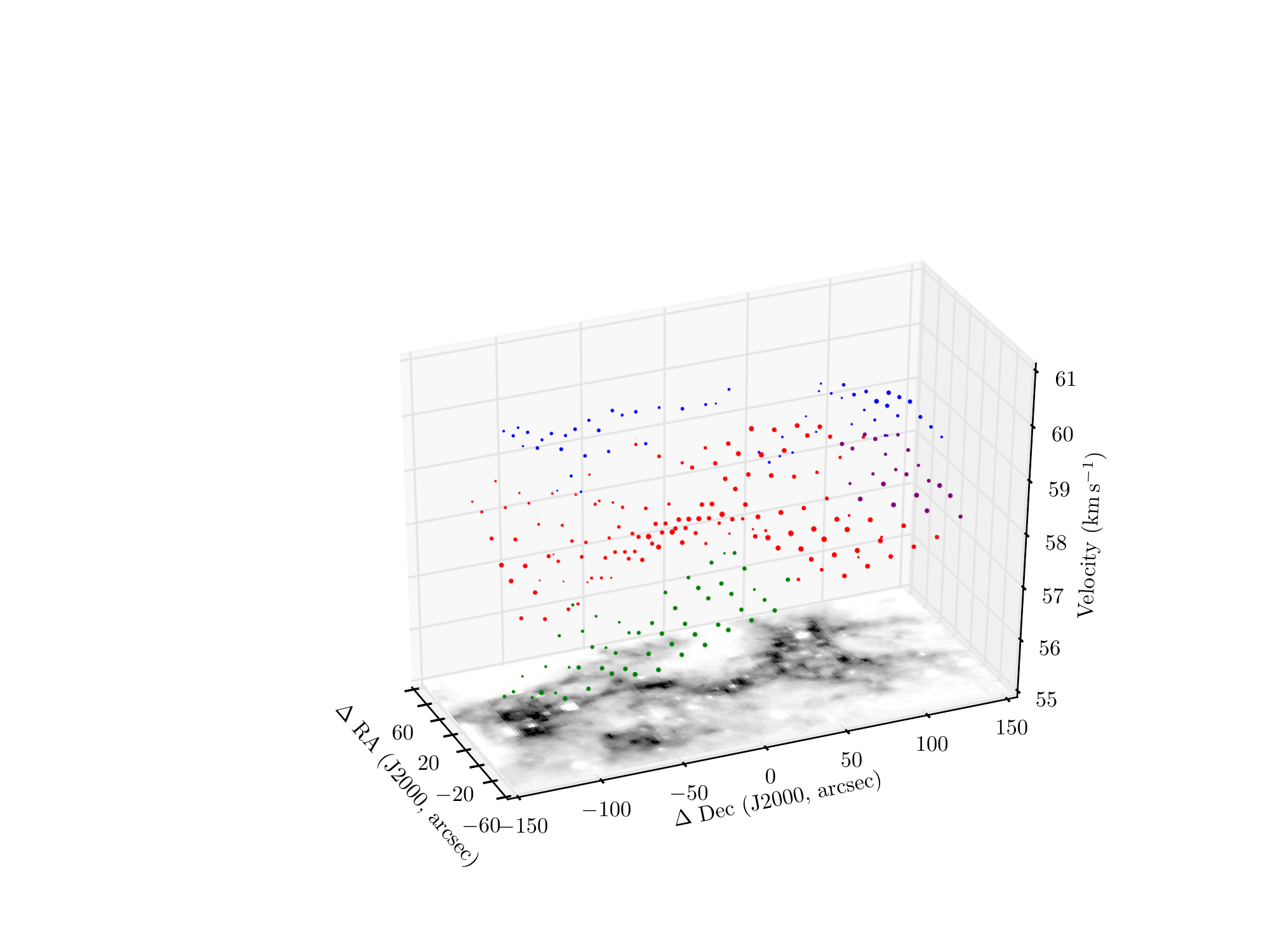}
\caption{Displayed in each panel is the position-position-velocity diagram of Cloud F, shown at three viewing angles for comparison. The left and right panels show \ntwohoz\ and \ceooz\ results, respectively. The colour of each point represents its association to one of the coherent velocity components, F$_{\rm PPV1}$ in blue, F$_{\rm PPV2}$ in green, F$_{\rm PPV3}$ in purple, and F$_{\rm PPV4}$ in red. The size of each point represents its relative peak intensity. The mass surface density map of \citet{kainulainen_2013} is shown on the base of each plot. Note that the coordinate offsets of these plots are relative to the centre of the mapped region: RA\,(J2000)\,=\,18$^h$53$^m$19$^s$, Dec\,(J2000)\,=\,01$^{\circ}$27$'$21${''}$ ({\it l} = 34.441$^{\circ}$, {\it b} = 0.247$^{\circ}$).}
\label{ppv_plots_cloudf}
\end{figure*}

\begin{table*}
\caption{Parameters of the velocity components identified in the IRAM-30m observations towards Cloud F (F$_{\rm PPV}$, upper rows) and Cloud H (H$_{\rm PPV}$, lower rows). Shown are the molecules used to identify the components, and for each component: the name with the colour used for each Figure in parentheses, the total number of points, the average centroid velocity, the average line width, the velocity gradient and the angle of this gradient with respect to East of North. When the uncertainty on the velocity gradient is larger than or equal to the calculated velocity gradient, the velocity gradient angle is unconstrained, and therefore not shown.}
\centering
\begin{tabular}{ccccccccc}
\hline
Line & Component & $\#$ points & Centroid velocity & Line width & Velocity gradient  & Gradient angle \\ 
& (colour) & & ($V_0$) \kms &  ($\Delta \upsilon$)  \kms & ($\nabla v$) \kms\,pc$^{-1}$ & ($\theta_{\nabla v}$) degrees \\
\hline
\ceooz \\
& F$_{\rm PPV1}$ \textcolor{blue}{(blue)} & 54 & 59.56 \,$\pm$\, 0.24 & 0.96 \,$\pm$\, 0.31 & 0.12 \,$\pm$\ 0.03 & -38.89 \,$\pm$\, 14.81 \\
& F$_{\rm PPV2}$ \textcolor{OliveGreen}{(green)} & 53 & 56.68 \,$\pm$\, 0.40 & 1.48 \,$\pm$\, 0.31 & 0.25 \,$\pm$\, 0.08 & -31.07 \,$\pm$\, 23.86 \\
& F$_{\rm PPV3}$ \textcolor{purple}{(purple)} & 22 & 58.39 \,$\pm$\, 0.12 & 1.04 \,$\pm$\, 0.27 & 0.16 \,$\pm$\, 0.05 & -83.77 \,$\pm$\, 4.32 \\
& F$_{\rm PPV4}$ \textcolor{red}{(red)} & 128 & 58.26 \,$\pm$\, 0.43 & 1.75 \,$\pm$\, 0.60 & 0.28 \,$\pm$\, 0.07 & -85.83 \,$\pm$\, 5.93 \\
\ntwohoz \\
& F$_{\rm PPV4}$ \textcolor{red}{(red)} & 41 & 58.44 \,$\pm$\, 0.51 & 1.75 \,$\pm$\, 0.50 & 0.75 \,$\pm$\, 0.15 & 70.20 \,$\pm$\, 3.20 \\
\\
\ceooz \\
& H$_{\rm PPV1}$ \textcolor{orange}{(orange)} & 20 & 46.12 \,$\pm$\, 0.11 & 0.47 \,$\pm$\, 0.15 & 0.12 \,$\pm$\, 0.06 & 76.08 \,$\pm$\, 12.41 \\
& H$_{\rm PPV2}$ \textcolor{purple}{(purple)} & 27 & 46.61 \,$\pm$\, 0.20 & 1.39 \,$\pm$\, 0.41 & 0.38 \,$\pm$\, 0.11 & 61.15 \,$\pm$\, 9.66 \\
& H$_{\rm PPV3}$ \textcolor{OliveGreen}{(green)} & 26 & 43.67 \,$\pm$\, 0.16 & 1.33 \,$\pm$\, 0.38 & 0.29 \,$\pm$\, 0.09 & -87.80 \,$\pm$\, 7.77 \\
& H$_{\rm PPV4a}$ \textcolor{red}{(red)} & 21 & 45.07 \,$\pm$\, 0.06 & 1.48 \,$\pm$\, 0.34 & 0.01 \,$\pm$\, 0.02 & \dots \\
& H$_{\rm PPV4b}$ \textcolor{blue}{(blue)} & 32 & 45.52 \,$\pm$\, 0.25 & 1.35 \,$\pm$\, 0.47 & 0.33 \,$\pm$\, 0.10 & -56.27 \,$\pm$\, 11.72 \\
\ntwohoz \\ 
& H$_{\rm PPV2}$ \textcolor{purple}{(purple)} & 14 & 46.88 \,$\pm$\, 0.10 & 0.74 \,$\pm$\, 0.20 & 0.15 \,$\pm$\, 0.08 & -29.98 \,$\pm$\, 29.89 \\
& H$_{\rm PPV4a}$ \textcolor{red}{(red)} & 38 & 45.42 \,$\pm$\, 0.10 & 1.26 \,$\pm$\, 0.16 & 0.02 \,$\pm$\, 0.02 & \dots \\
& H$_{\rm PPV4b}$ \textcolor{blue}{(blue)} & 27 & 45.99 \,$\pm$\, 0.09 & 1.02 \,$\pm$\, 0.35 & 0.10 \,$\pm$\, 0.05 & -77.40 \,$\pm$\, 10.68 \\

\hline 
\end{tabular}
\label{velocity component parameters}
\end{table*}

To separate the velocity components, we fit Gaussian profiles to the spectra across the cloud using the Semi-automated multi-COmponent Universal Spectral-line fitting Engine ({\sc scouse}; \citealp{henshaw_2016}).\footnote{\change{Written in the {\sc idl} programming language. See \url{https://github.com/jdhenshaw/SCOUSE} for more details.}} \change{{\sc scouse} has been chosen over manually fitting each individual spectrum (a total of $\sim$300 for both Cloud F maps), as this algorithm was specially produced to efficiently and systematically fit a large number of spectra.} To do so, {\sc scouse} works in several steps. Firstly, the map is split into regions (referred to as ``spectral averaging areas'', SAA), within which the data are spatially averaged. For each SAA spectra, the user is instructed to fit the appropriate number of Gaussian components. The individual spectra contained within the SAAs are then automatically fitted using the parameters from the Gaussian fits of their SAA within given tolerance limits on the peak intensity, line centroid velocity, line width and separation between components. As a final step, the results are checked for anomalies, which can be re-fitted if required. 

\change{As discussed in \citealp{henshaw_2016}, the size of the SAA selected will be somewhat data dependent. A size of 30\arcsec\ represented the maximum size for which the spatially averaged spectrum was a good representation of the line profiles of its composite spectra. Changing the SAA size will not significantly affect the final best-fitting solutions across the cloud. Rather, decreasing the size of the SAA will result in an increase in the number of spectral that require manual fitting during the SAA fitting stage (and a reduction of fits that need to be corrected during the later stages). Alternatively increasing the size of the SAA will have the opposite effect. An SAA radius of 30\arcsec\ for the Cloud F data, such that each SAA contained four to six spectra, therefore, represented the most efficient choice.} The tolerance limits were set such that each fit had to have a peak intensity of at least three times the {\it rms}, a centroid velocity similar to an SAA fit to within three times the value of the velocity dispersion and a line width to within a factor of two. To be considered a multi-component fit, the components had to be separated by a factor of two times the line width. These parameters gave reasonable fits across the cloud for both \ntwohoz\ and \ceooz, where the mean residual across all positions after fitting was $<3$ {\it rms} and only $\sim$\,10\,per cent of the spectra required manual checking. The results of {\sc scouse} are over-plotted on the spectrum at each position in Figure\,\ref{spec_plot_cloudf}.

\change{To identify coherent velocity features within our decomposed data, we use Agglomerative Clustering for ORganising Nested Structures ({\sc acorns}; Henshaw et al. in prep).\footnote{\change{Written in the {\sc python} programming language, soon available at \url{https://github.com/jdhenshaw/acorns}.}} A complete description of the algorithm and the process will be presented in Henshaw et al. (in prep), however, the key details are included below. }

\change{{\sc acorns} is specifically designed to work on decomposed spectroscopic data, i.e. the output of {\sc scouse} (or an equivalent algorithm). The algorithm follows the principles of hierarchical agglomerative clustering and searches for a hierarchical system of clusters within the decomposed dataset. In agglomerative clustering, each data point begins its life as a `cluster'. Clusters grow by merging with other clusters provided they satisfy a number of conditions which are supplied by the user (see below). As the algorithm progresses, hierarchies develop. These hierarchical clusters, as in many areas of research, can be visualised graphically as a dendrogram.}

\change{The merging process is governed by a series of conditions controlled by the user. These conditions are known as linkage criteria, and can refer to, for example, a euclidean distance between two clusters or an absolute difference in a variable of the users choice (e.g. velocity or velocity dispersion). If two adjacent clusters satisfy the linkage criteria, they will be merged. In this work, for two data points to be linked, we require several that criteria are satisfied. Namely, adjacent clusters must: i) have a Euclidean separation which less than a beam size; ii) have an absolute velocity difference less than twice the spectral resolution of the data; iii) have an absolute difference in velocity dispersion which is less than the thermal velocity dispersion of the gas (estimated to be cs=0.23 km/s at 17K, from Table 1, given a mean molecular weight of 2.33 a.m.u).\footnote{The observed velocity dispersion is given as $\sigma_{\rm obs}$\,=\,$\Delta \upsilon$\,(8\,ln\,(2))${^{-0.5}}$, where $\Delta \upsilon$ is the observed full width half maximum (or line width). Here and throughout this work, we use the classical value of the abundance to be consistent with the previous IRDC analyses (e.g. \citealp{henshaw_2013, jimnez-serra_2014}), and not the value obtained when accounting for heavier elements (2.37; see \citealp{kauffmann_2008}). Taking the latter would not significantly affect the results of this work.} We consider these criteria to be fairly strict and representative of the limitations of our data (i.e. our spatial and spectral resolution). Once this initial robust hierarchy has been established (i.e. all possible links satisfying these criteria have been exhausted), {\sc acorns} then allows the user to relax these conditions in order to further develop the hierarchy. This can be performed in several ways, in incremental stages, both interactively and non-interactively, or in a single step. In this study we relaxed the conditions in a single step, however, we conducted a parameter space study in order to establish the set of relaxation parameters which produced the most robust hierarchy. These optimal parameters were chosen when a hierarchy appeared most persistent across an area of the parameter space (i.e. when the hierarchy did not significantly change for a range of relaxation parameters), whilst being comparable for both the \ceo\ and \ntwoh\ data (assuming that \ntwoh\ traces similar, or a least the densest, components traced by \ceo). This was achieved with relaxation factors of 2.5, 1.75, and 0.75, for the spatial separation, centroid velocity and line width, respectively, for both transitions.}

Figure\,\ref{ppv_plots_cloudf} shows the position-position-velocity diagram for the above analysis, where the centroid velocities of the Gaussian profiles at each position have been plotted, and the colours correspond to the identified components (shown in the legend). Four coherent velocity components have been identified in the \ceooz\ emission from Cloud F: F$_{\rm PPV1}$, F$_{\rm PPV2}$, F$_{\rm PPV3}$ and F$_{\rm PPV4}$, which are shown in blue, green, purple, and red, respectively, on Figures\,\ref{spec_plot_cloudf} and \ref{ppv_plots_cloudf}. We find that the components F$_{\rm PPV1}$ and F$_{\rm PPV4}$ are extended across the length of the cloud, whereas F$_{\rm PPV2}$ and F$_{\rm PPV3}$ are limited to the southern and northern portions of the cloud, respectively. The component F$_{\rm PPV4}$ is also identified in the \ntwohoz\ transition emission. The basic properties of these components are given in Table\,\ref{velocity component parameters}, and they are analysed in the following sections. 

\subsection{Velocity gradients}\label{velocity gradients}

As previously shown in the moment map analysis, Cloud F appears to have a smooth velocity gradient increasing in velocity from for west to east (see Figure\,\ref{moment_maps_cloudf}). However, in the kinematic structure identified from the Gaussian decomposition, we do not see such a smooth gradient, instead, we observe the velocity components at distinct velocities across the cloud. We find that the F$_{\rm PPV1}$ is at a high velocity ($\sim$\,60\,\kms) on the east of the mapped region, and the F$_{\rm PPV2}$ is at a low velocity ($\sim$\,57\,\kms) on the west of the mapped region (see Figure\,\ref{ppv_plots_cloudf}). When averaged with F$_{\rm PPV4}$, as in the case of the moment map analysis, these would mimic a smooth velocity gradient across the cloud. Rather than a primarily west to east velocity gradient, the identified components show large-scale velocity gradients running along the south-north axis of the cloud. Here we determine the magnitude and angle of the larger scale gradients across the cloud following the analysis of \citet{goodman_1993}, who assume that the line centroid velocities can be represented by the linear function,
\begin{equation}
V_{LSR} = V_0 + A\,\Delta{\rm RA} + B\,\Delta{\rm Dec},
\end{equation}
where $\Delta{\rm RA}$ and $\Delta{\rm Dec}$ are the offset right ascension and declination in radians, $V_0$ is the average velocity of the velocity component, $V_{LSR}$ is the centroid velocity of the Gaussian profile fit at each position, and $A$ and $B$ are solved for using the non-linear least squares optimisation routine {\sc scipy.optimize.curve\_fit} in {\sc python}. The velocity gradient, $\nabla v$, can be calculated with, 
\begin{equation}
\nabla v = \frac{(A^2 + B^2)^{0.5}}{d},
\end{equation}
where $d$ is the source distance. The angle of the gradient, $\theta_{\nabla v}$, can be determined from, 
\begin{equation}
\theta_{\nabla v} = {\rm tan}^{-1} \Big( \frac{A}{B} \Big).
\end{equation}
The magnitudes and angles, with respect to East of North, of the velocity gradients for each velocity component, are given in Table\,\ref{velocity component parameters}. We find velocity gradients across the cloud in the range of 0.12-0.75\,\kms\,pc$^{-1}$, with an average over all components of $\sim$\,0.3\,\kms\,pc$^{-1}$, which is lower than the range determined from the moment map analysis $\sim$\,$0.5 - 0.7$\,\kms\,pc$^{-1}$. This is due to the fact that here we are analysing the gradients of the individual components, rather than the gradient produced by the separation of the components when they are averaged. These gradients are discussed further in appendix\,\ref{Appendix D}.

\subsection{Different \ntwohoz\ and \ceooz\ centroid velocities}\label{shift}

\begin{figure*}
\centering

\includegraphics[trim = 20mm 0mm 15mm 3.5mm, clip,angle=0,width=0.63\columnwidth]{\dir 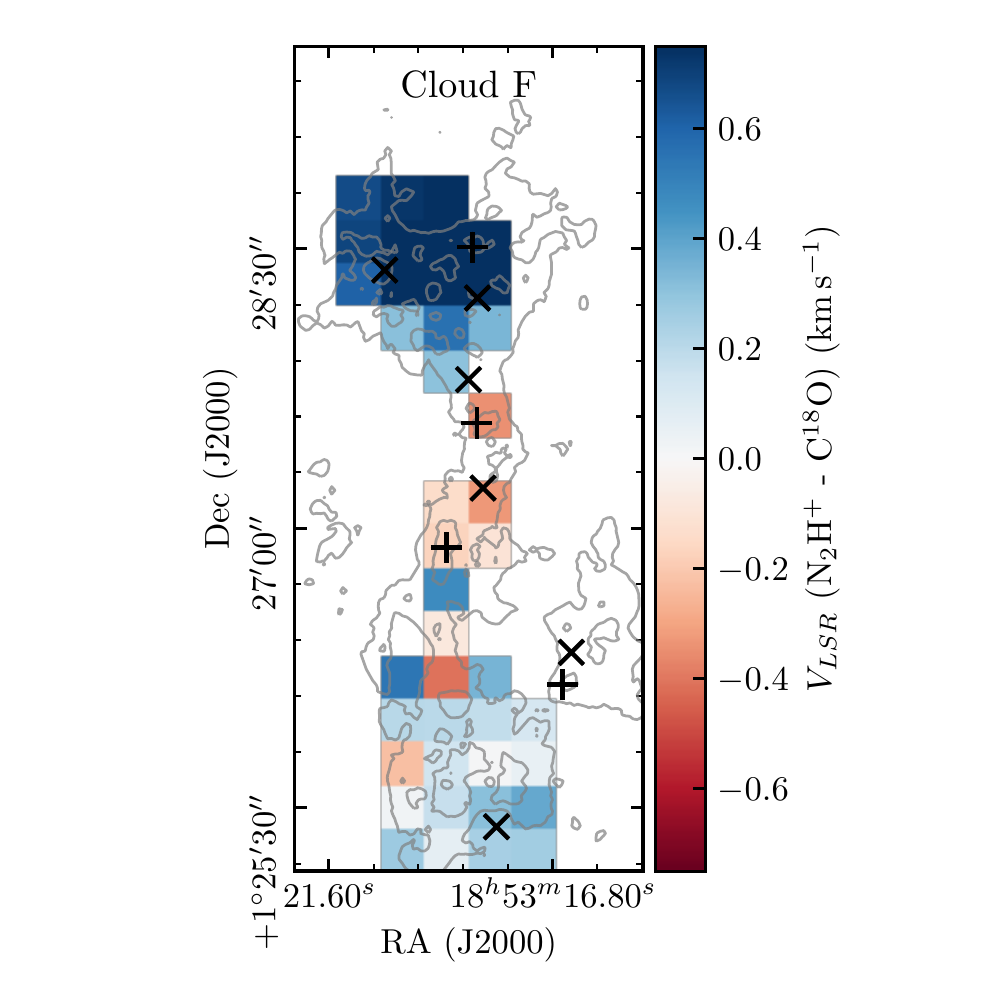}
\includegraphics[trim = 20mm 0mm 15mm 3.5mm, clip,angle=0,width=0.63\columnwidth]{\dir 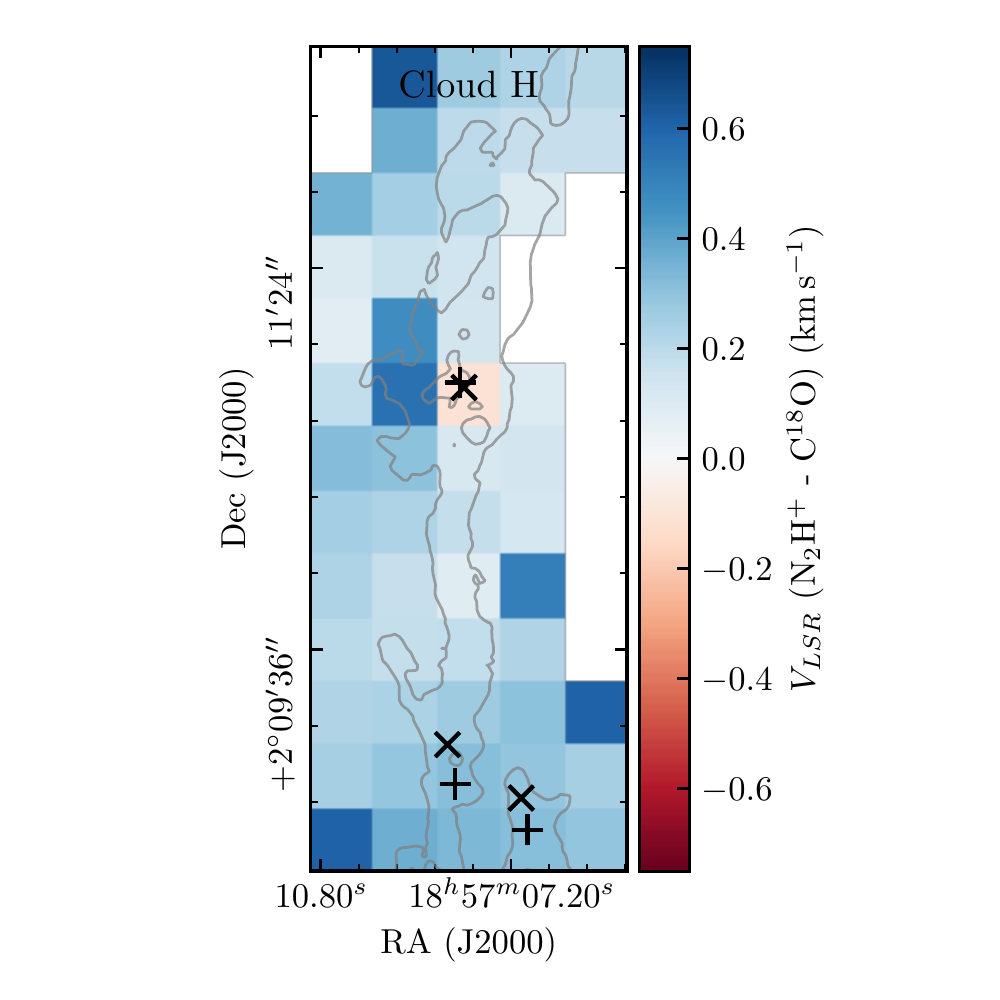}
\includegraphics[trim = 0mm -5mm 0mm 0mm, clip,angle=0,width=0.54\columnwidth]{\dir 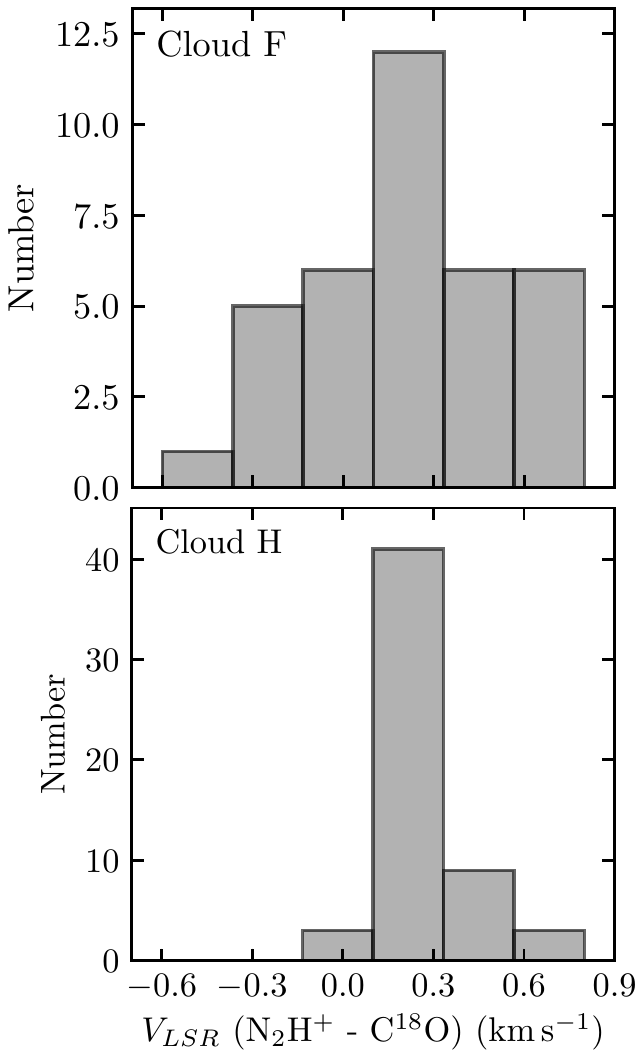}
\caption{The left and centre panels are maps of the difference in the centroid velocity, $V_{LSR}$\,(N$_2$H$^{+}$\,-\,C$^{18}$O), of Clouds F and H (see section\,\ref{comparison_to_cloudh}), respectively. Over-plotted as $+$ and $\times$ symbols are the positions of the ``core'' regions identified by \citet{butler_2012} and those from \citet{rathborne_2006}, respectively. Shown in the right panels are histograms of this difference, where the upper is for the Cloud F map, and the lower for the Cloud H map. For reference, the clouds are labeled at the top of each plot.}
\label{velo_shift_cloudf}
\end{figure*}

Comparing the distribution of different molecular species, both spatially and in velocity, within star-forming regions, can provide clues of their formation scenarios (e.g \citealp{henshaw_2013}). To do so within Cloud F, we compare the centroid velocity at each position of the component identified in the \ntwohoz\ and \ceooz, F$_{\rm PPV4}$. The centroid velocity difference, $V_{LSR}$\,(N$_2$H$^{+}$\,-\,C$^{18}$O), map and histogram are presented in Figure\,\ref{velo_shift_cloudf}. We find that the average difference in velocity is $+0.32\,\pm\,0.06$\,\kms.\footnote{Uncertainty given is the standard error on the mean, where the standard deviation is $\pm\,0.40$\,\kms.} 

A velocity shift between two tracers may, however, be produced when comparing the velocity from emission which is not tracing the same gas; or in other words, here we choose to compare the emission from the F$_{\rm PPV4}$ component, as this is seen in both \ceo\ and \ntwoh, however, if some of this emission is actually from different component which has been wrongly assigned, then an artificial velocity shift would be produced. Such an effect could be plausible towards the north of the cloud, where the \ntwohoz\ emission appears to originate at a velocity in-between the F$_{\rm PPV1}$ and F$_{\rm PPV4}$ components (see Figure\,\ref{ppv_plots_cloudf}). To investigate this, we temporarily attribute the emission above Dec\,(J2000)\,=\,01$^{\circ}$27$'$36${''}$ to the F$_{\rm PPV1}$ component, and re-determine the velocity shift between the \ntwohoz\ and \ceooz. Doing so, we find an average value towards this northern region is $-0.67$\,$\pm$\,0.10\,\kms; a negative shift which is a factor of two larger in magnitude than that found when assigning this region to the F$_{\rm PPV4}$ component. Demonstrating that there is a significant shift in velocity between the \ntwohoz\ and \ceooz\ emission regardless of which component the \ntwohoz\ emission within the northern region is assigned, and this result is persistent, regardless of which \ntwoh\ and \ceo\ components are compared. We continue with the assumption that components closer in velocity are the most likely to be physically associated, hence keep the original assignment of all the \ntwohoz\ emission to the F$_{\rm PPV4}$ component, which provides a smaller, yet still significant, average velocity shift ($+0.32\,\pm\,0.06$\,\kms).

\subsection{Velocity dispersions}\label{velocity dispersions}

\begin{figure*}
\centering
\includegraphics[trim = 3mm 4.2mm 3.5mm 3.5mm, clip,angle=0,height=0.205\textheight]{\dir 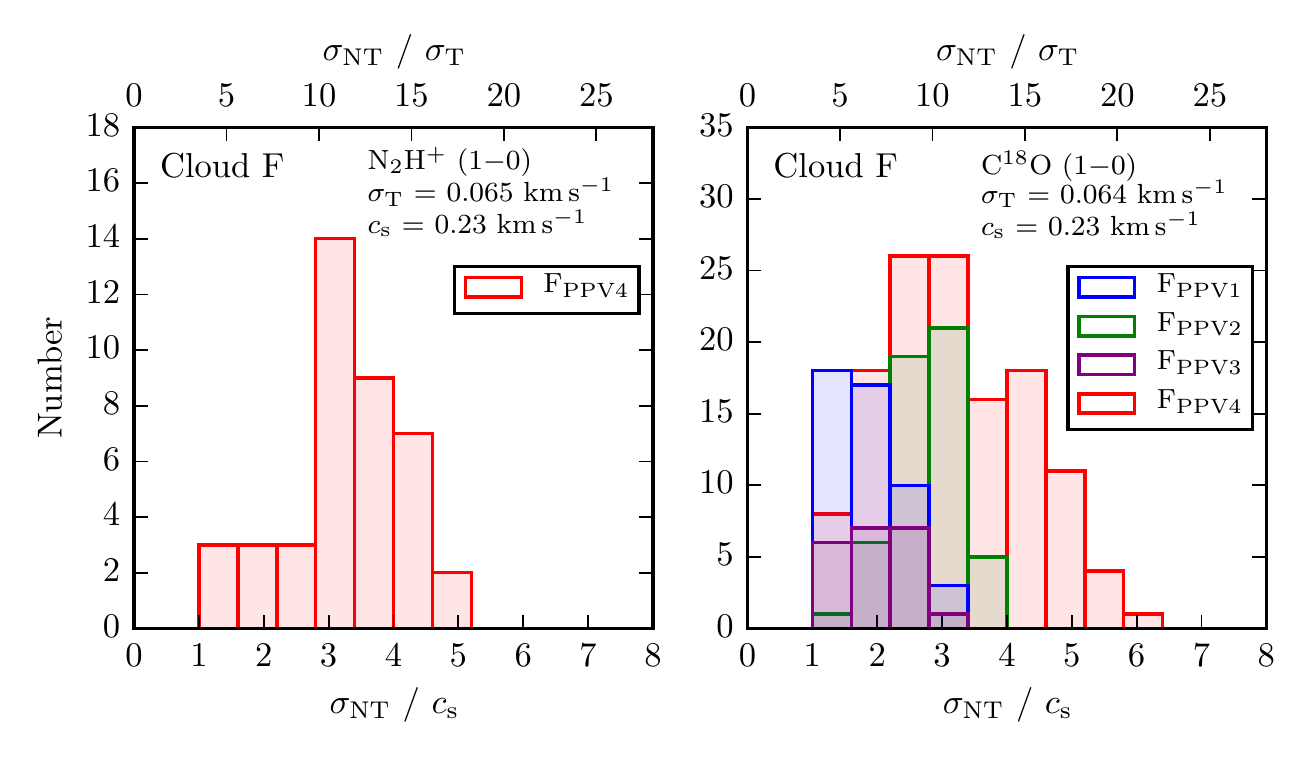}
\includegraphics[trim = 8mm 4.2mm 3.5mm 3.5mm, clip,angle=0,height=0.205\textheight]{\dir 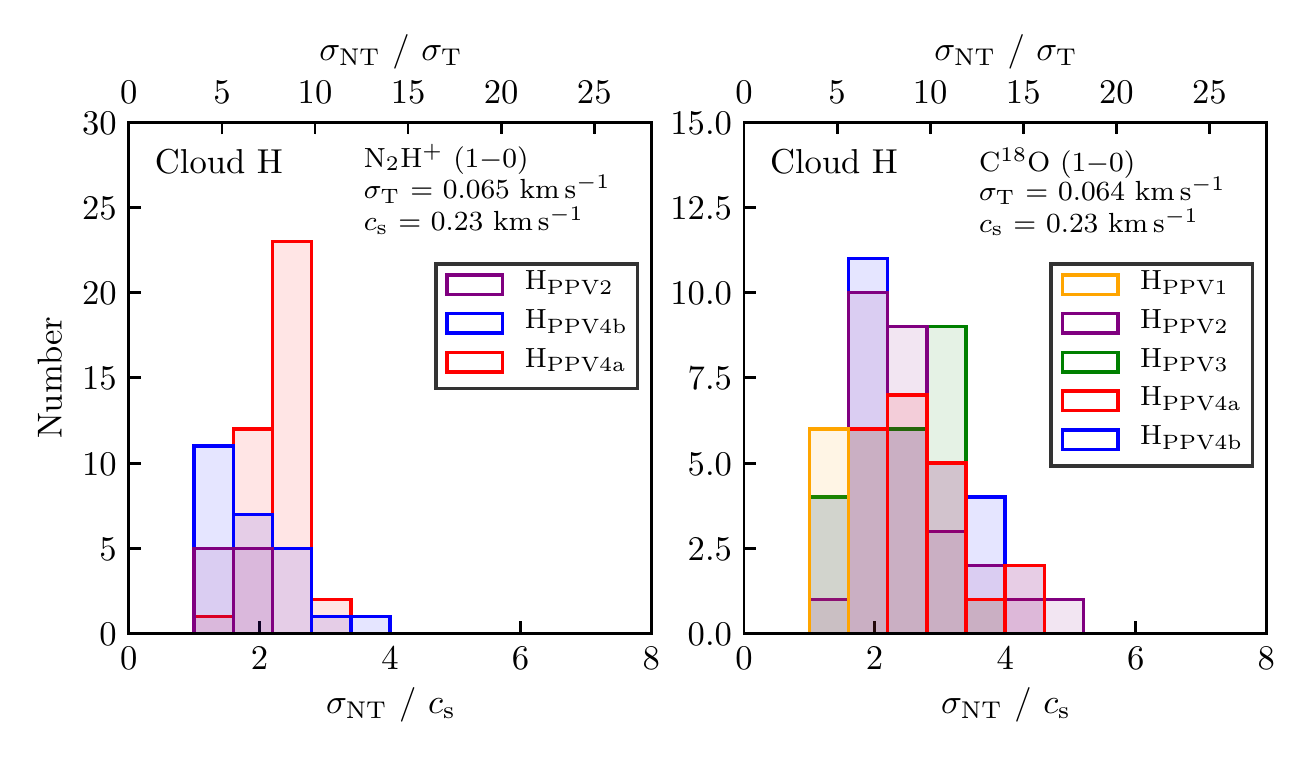}
\caption{Cloud F (left two panels) and Cloud H (right two panels) histograms of the non-thermal contribution to the velocity dispersion over the gas sound speed (lower axis on each plot; c$_\mathrm{s}$ = 0.23 \kms\ at 17\,K, given a mean molecular weight of 2.33 a.m.u). The upper axis labels on each plot show the non-thermal contribution to the velocity dispersion over the thermal contribution to the velocity dispersion, where $\sigma_\mathrm{T}$ = 0.065 \kms\ and 0.066 \kms, for \ceo \ (30 a.m.u) and \ntwoh \ (29 a.m.u), respectively. Colours represent the various velocity components (see Figures  \,\ref{ppv_plots_cloudf} and \,\ref{ppv_plots_cloudh}).}
\label{velo_disp}
\end{figure*}

In order to study the non-thermal motions of the gas from the observed line widths we use the expression from \citet{myers_1983},
\begin{equation}
\sigma _\mathrm{NT} = \sqrt{\sigma _\mathrm{obs}^{2} - \sigma _\mathrm{T}^{2}} = \sqrt{\frac{\Delta\upsilon^{2}}{8\mathrm{ln}(2)} - \frac{k_{B}T_{kin}}{m_\mathrm{obs}}},
\end{equation}
\noindent where $\sigma _\mathrm{NT}$, $\sigma _\mathrm{obs}$, and $\sigma _\mathrm{T}$, are the non-thermal, the observed, and the thermal velocity dispersion, respectively. $\Delta\upsilon$ is the observed full width half maximum (or line width), $k_{B}$ is the Boltzmann constant, $T_{kin}$ is the kinetic temperature of the gas and $m_\mathrm{obs}$ refers to the mass of the observed molecule (30 a.m.u for \ceo; 29 a.m.u for \ntwoh). We assume a constant kinetic gas temperature of 17\,K (e.g. \citealp{dirienzo_2015, pon_2016a}), which gives thermal dispersion contributions of 0.064\,\kms\ and 0.065\,\kms, for \ceo \ and \ntwoh, respectively.\footnote{Given the weak dependence on the temperature, varying between the expected limits within IRDCs does not affect the results presented here.}


Figure\,\ref{velo_disp} shows the non-thermal component of the velocity dispersion compared to both the gas sound speed (lower axis) and thermal component of the dispersion (upper axis) at each position within Cloud F. Comparison to the gas sound speed has been made assuming a temperature of 17\,K and a mean molecular weight of 2.33 a.m.u (c$_\mathrm{s}$ = 0.23 \kms). We find that the velocity dispersions (or Mach numbers $\mathcal M\,=\,\sigma _\mathrm{NT} / c_s$) averaged over all velocity components are 0.75\,$\pm$\,0.03\,\kms\ ($\mathcal M$\,=\,3.2\,$\pm$\,0.14) and 0.63\,$\pm$\,0.02\,\kms\ ($\mathcal M$\,=\,2.70\,$\pm$\,0.07) for \ntwohoz\ and \ceooz, respectively. To link these observed non-thermal motions to the turbulent motions (i.e. with corresponding Mach number) within the cloud, we make the assumption that no velocity gradients or substructure are present with size scales less than the beam size. This may, however, not be the case for these IRAM-30m observations presented here given the large physical beam size ($\sim$\,0.5\,pc; cf. \citealp{henshaw_2013} and \citealp{henshaw_2014}), hence the values of the turbulent velocity dispersion calculated here most likely represent upper-limits on the true turbulent motions within the cloud. We find that all the velocity components have non-thermal velocity dispersions factors of several larger than the gas sound speed, which suggests that their internal turbulent motions are (at most) only moderately supersonic over scales of $\sim$\,0.5\,pc (traced by these observations). 

We note that the calculated average \ntwoh\ velocity dispersion is larger than the average \ceo\ velocity dispersion, which is not typically expected if the \ntwoh\ is tracing the denser, more compact regions within the cloud, unless the dense gas is associated with embedded young stellar objects. However, when comparing the velocity dispersions for the F$_{\rm PPV4}$ component only, we find comparable values (see Table\,\ref{velocity component parameters}). This suggests that both molecular lines are tracing similar gas within this component.    

\section{Discussion}\label{Discussion}

The kinematic analysis of Cloud F has unveiled a complex structure, consisting of several extended, coherent velocity components. Previous studies of this cloud have shown that it contains a distribution of both quiescent and active star-forming regions, which are discussed in relation to the kinematic structure in appendix\,\ref{Appendix D}, with a particular focus on the \citet{rathborne_2006} and \citet{butler_2012} shown on Figure\,\ref{cloudf_msd}. In the following sections, the kinematic structure of Cloud F is compared to that within similar IRDC, and to the larger scale gas kinematics surrounding the cloud.

\subsection{Cloud F in the context of other massive star-forming regions: comparison to Cloud H}\label{comparison_to_cloudh}

Galactic Plane surveys, undertaken with infrared space-based telescopes (most recently {\it Spitzer} and {\it Herschel}), have shown that filamentary structures appear to be ubiquitous throughout the interstellar medium (e.g. \citealp{molinari_2016}). Recently, kinematic analysis of molecular line emission has shown that coherent structures, believed to be velocity space representation of these filamentary structures, are equally common, appearing in both low- and high-mass star-forming regions (e.g. \citealp{hacar_2013, henshaw_2014, hacar_2016a, henshaw_2016d}). However, despite these structures being morphologically and qualitatively similar, their physical properties may be very different. Currently, direct comparisons of the structure and its properties within massive star-forming regions are lacking. To address this, in this section, we discuss how the properties determined for Cloud F compare to a similar IRDC, Cloud H (G035.39-00.33; \citealp{butler_2009}; see Table\,\ref{cloud_props}), with the aim of highlighting which kinematic properties are shared between these, and potentially other, massive star-forming regions. 

The analysis of Cloud H's kinematic structure from the IRAM-30m observations has been carried out by \citet{henshaw_2013}. Here we smooth these data to the same spatial resolution of the Cloud F observations ($\sim$\,0.5\,pc), such that structure identification is not biased to a spatial scale, given the hierarchical nature of the interstellar medium. Furthermore, we use the same analysis tools to determine the kinematic structure ({\sc scouse} and {\sc acorns}), such that the systematic comparison is possible (see Appendix\,\ref{Appendix A}), as it has been recently suggested that the results from different structure finding algorithms can vary (e.g. \citealp{chira_2017}). We find that Cloud H contains a complex structure, harbouring several coherent velocity components, seen in both the \ntwohoz\ and \ceooz\ emission, which are in agreement with the results previously found by \citet{henshaw_2013}. 

Similar to Cloud F, from the measured line width, we have determined the non-thermal velocity dispersions within Cloud H. We find Mach numbers of ${\mathcal M}$\,=\,2.0\,$\pm$\,0.07 and ${\mathcal M}$\,=\,2.28\,$\pm$\,0.08 using \ntwohoz\ and \ceooz\ (averaged over all velocity components), respectively; histograms of these results are shown next to those for Cloud F in Figure\,\ref{velo_disp}. Therefore, we find that the non-thermal contributions are factors of $2-3$ larger than the sound speed of the gas within both clouds, which sets an upper limit on the turbulent motions being mildly supersonic over the examined physical scale (i.e. the smoothed spatial resolution of $\sim$\,0.5\,pc). 

We investigate the velocity distribution of the components by comparing the separation of the lowest and highest velocity component within both clouds. We use the components seen in \ceooz\ emission, and find a line-of-slight difference for Cloud F (F$_{\rm PPV1}$ and F$_{\rm PPV2}$) of 2.9\,$\pm$\,0.5\,\kms, and for Cloud H (H$_{\rm PPV2}$ and H$_{\rm PPV3}$) of 2.9\,$\pm$\,0.3\,\kms. Assuming a simple three-dimensional morphology, this result could indicate that both clouds have kinematic structures which are interacting at a speed of up to 5\kms; accounting for a factor of $\sqrt{3}$, assuming the velocity in the plan of the sky is equivalent in all directions. 

A filament merging scenario is in agreement with the wide-spread narrow SiO emission observed within Cloud H, generated by the sputtering of dust grains within large-scale C-shocks \citep[][requiring a shock velocity of $\sim$\,12\kms]{jimnez-serra_2010}. It is possible that a similar scenario of filament merging is causing the velocity difference within Cloud F, and indeed Cosentino et al. (accepted) also found similarly wide-spread SiO emission throughout this cloud. However, it is difficult to determine if this emission is due to a merging scenario or higher level star-formation within Cloud F, which would have affected the chemistry of the molecular gas as a result of stellar feedback (e.g. towards the F4, or MM3, region). \change{The elevated level of star formation within Cloud F suggest that it is at a later evolutionary stage than Cloud H}, which has previously been suggested to be $\sim$\,3\,Myr old \citep{henshaw_2013, jimnez-serra_2014, barnes_2016}.

We also compare the velocity separation between the coherent velocity components identified in the \ntwohoz\ and \ceooz\ emission. When doing so within Cloud F we found a significant positive systematic offset of $V_{LSR}$\,(N$_2$H$^{+}$\,-\,C$^{18}$O) = $+ 0.32\,\pm\,0.03$\,\kms. To conduct a similar analysis within Cloud H, we compare the H$_{\rm PPV4a}$ and H$_{\rm PPV4b}$ structures. These components have been identified simultaneously at three positions within the \ntwohoz\ map (see Figure\,\ref{spec_plot_cloudh}). At these positions, we average the centroid velocity of the components in \ntwohoz\ and compare this velocity to the component seen in the \ceooz\ emission (H$_{\rm PPV4}$).\footnote{These positions are towards the complex H6 regions \citep{henshaw_2014, henshaw_2016d}.} We note that omitting these positions, which make up only $\sim$\,5\,per cent of the total positions used for this comparison, would not significantly affect the result presented here. The centroid velocity difference map and histogram for Cloud H is presented with the Cloud F results in Figure\,\ref{velo_shift_cloudf}. We find an average velocity shift of $+\,0.26\,\pm\,0.02$\,\kms,\footnote{Uncertainty given is the standard error on the mean, where the standard deviation is $\pm\,0.14$\,\kms.} which is in agreement with the value found by \citet{henshaw_2013}. It is intriguing that both clouds share such a similar positive velocity shift between \ntwohoz\ and \ceooz, given below are several possible scenarios its formation.

It has been previously proposed that the velocity difference within Cloud H is a result of a filament merging, whereby higher velocity filaments (H$_{\rm PPV1}$ and H$_{\rm PPV2}$, from this work) are merging with a lower velocity, less massive filament (H$_{\rm PPV3}$), increasing the density of an intermediate velocity filaments (H$_{\rm PPV4a}$ and H$_{\rm PPV4b}$). Given the majority of the mass within Cloud H is situated at a higher velocity, the densest gas forming within the intermediate velocity filament, as traced by \ntwoh, is pushed to a higher velocity with respect to its envelope material traced by \ceo, also formed by the merging process \citep{henshaw_2013, jimnez-serra_2014, henshaw_2014}. Indeed, simulations have shown that certain lines-of-sight through density fluctuations and varying velocity fields within collapsing clouds can cause significant velocity difference between molecular tracers \citep{smith_2013, bailey_2015}. As previously discussed, such a scenario is plausible for Cloud F, and would also be in agreement with the observed velocity difference between the components. We note, however, a common physical mechanism driving this interaction is not determinable from the data presented here (e.g. cloud-cloud merging or global gravitational collapse).

A second explanation, proposed by \citet{zhang_2017}, is that velocity shifts between low and high-density tracers could be a signature of gas which is both expanding and contracting within the core regions of the cloud. This is based on the assumption that the higher critical density molecules, in their case HCO$^+$ emission, trace the inner dynamics of a core, while lower critical density molecules, \ceo, trace the outer, envelope dynamics. These authors find blue-shifted and red-shifted profiles of the high and low-density tracers, respectively, towards a sample of cores, which they suggest shows the different core layers are moving in opposing directions; a scenario of ``envelope expansion with core collapse'' (e.g. \citealp{lou_2011}). Indeed, these authors proposed such a scenario for a core region within Clouds H (H6/MM7; \citealp{rathborne_2006, butler_2012}). We suggest that this could, in theory, be applied to explain the velocity shifts observed across both clouds, yet this would require the effect being extrapolated over larger scales.

In summary, here we have shown that two massive, morphologically and qualitatively similar IRDCs share several kinematic properties, hinting at similar internal gas conditions. An intriguing result given that they are drawn from a different of Galactic environments and their various internal physical processes (e.g. level of star formation). It would be interesting to examine a larger sample of clouds to determine if these properties are inherent to the wider IRDC population. 

\subsection{Connection between IRDC scales and GMC scales for Cloud F}\label{grs_comparison}

\begin{figure*}
\centering
\includegraphics[trim = 45mm 27mm 50mm 37mm, clip,angle=0, clip,angle=0,width=0.70\textwidth]{\dir 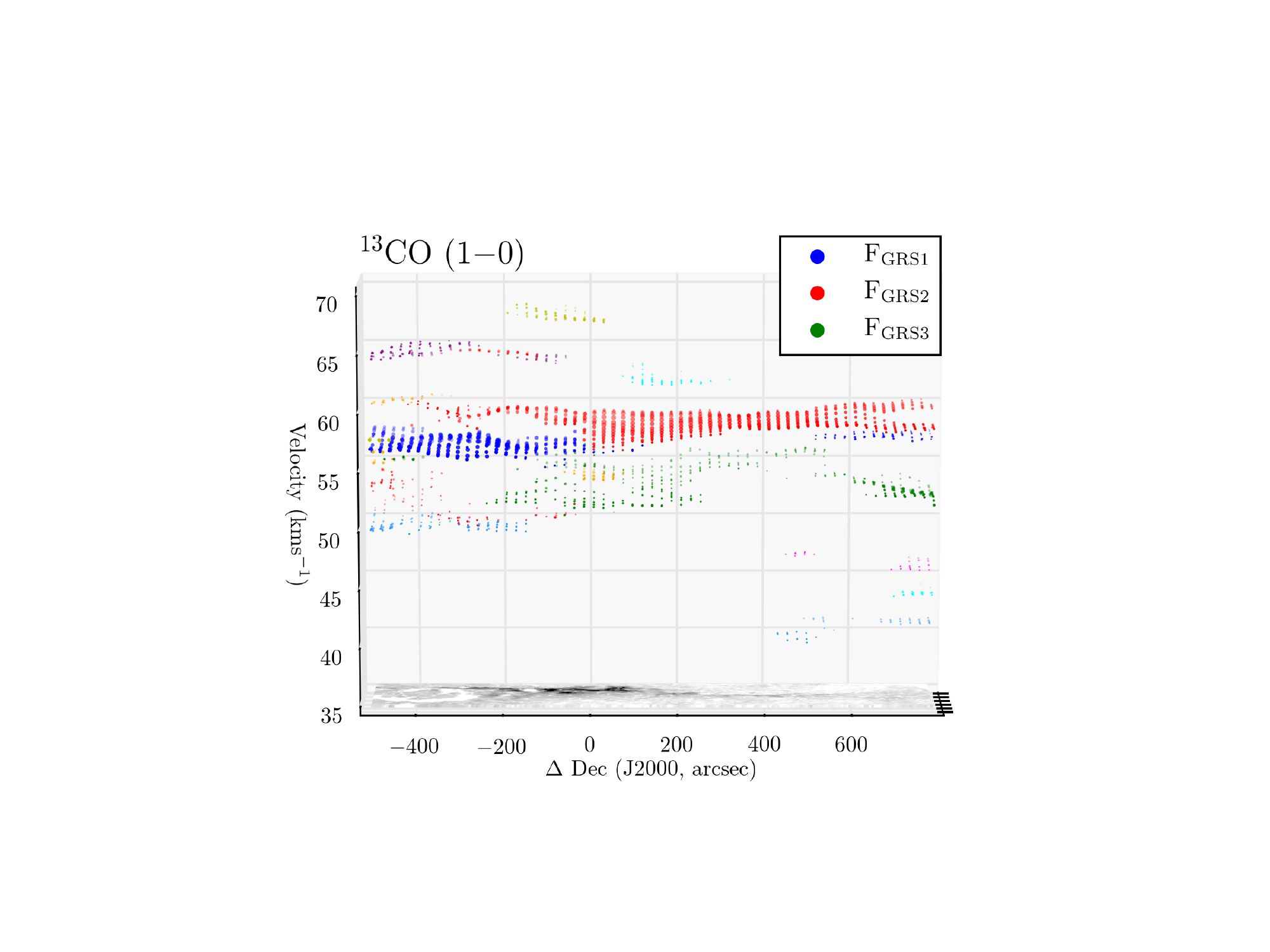} 
\includegraphics[trim = 45mm 27mm 50mm 37mm, clip,angle=0, clip,angle=0,width=0.70\textwidth]{\dir 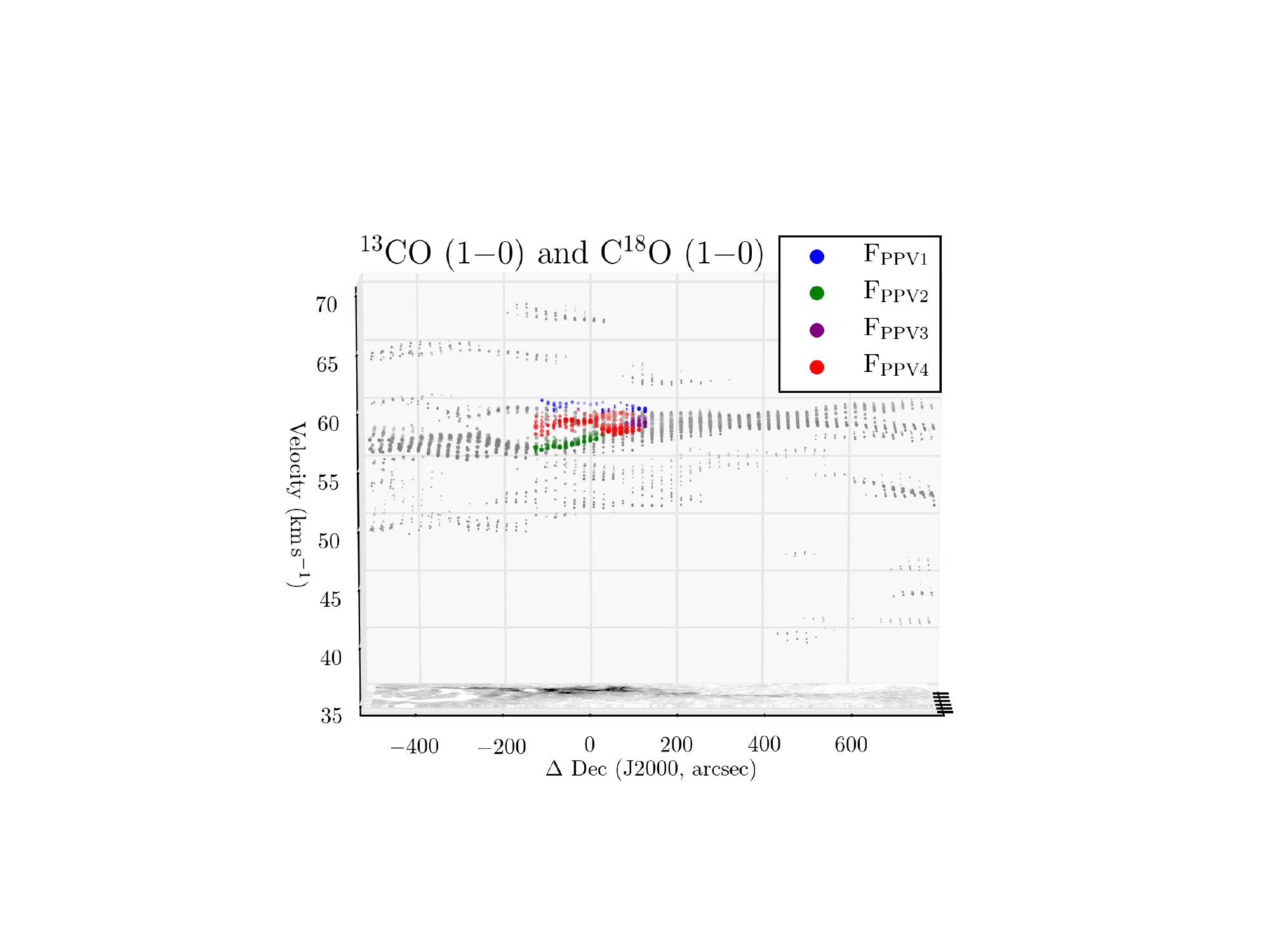}
\caption{Shown are position-position-velocity diagrams covering the large-scale region surrounding Cloud F. The \change{upper} panel displays the Gaussian decomposition results for the \tcooz\ GRS observations, where the colour of each point represents its association with a coherent velocity component (see Appendix\,\ref{Appendix C}). The three most extended components are shown in the legend in the upper right of the panel. The \change{lower} panel displays the same position-position-velocity diagram with the GRS observations shown in grey, overlaid with the structures determined from the IRAM-30m \ceooz\ observations shown in colours identical to Figure\,\ref{ppv_plots_cloudf} (see legend in upper right of panel). The size of each point represents its relative peak intensity. The mass surface density map of \citet{kainulainen_2013} is shown on the base of each plot. Note, the coordinate offsets of these plots are relative to the centre of the mapped region: \change{RA\,(J2000)\,=\,18$^h$53$^m$19$^s$, Dec\,(J2000)\,=\,01$^{\circ}$27$'$21${''}$ ({\it l} = 34.441$^{\circ}$, {\it b} = 0.247$^{\circ}$).}}
\label{ppv_plots_cloudf-grs}
\end{figure*}

As previously mentioned, the interstellar medium is hierarchically structured, with massive star-forming regions hosting a complex sub-structure through various scales (e.g. filaments to cores). However, these regions are by no means at the top of this hierarchy, rather they are believed to be only a small part of larger, Galactic scale structures, which typically have masses and spatial extents one to two orders of magnitude larger than IRDCs (e.g. \citealp{ragan_2014, hernandez_2015, zucker_2015}). Indeed, several works have already studied the larger scale environment surrounding both Clouds F and H (as defined here by the IRAM-30m coverage). \citet{hernandez_2015} find that the larger scale structures which host Clouds F and H share many similar kinematic properties, such as velocity dispersions ($\sim\,3-5$\,\kms), velocity gradients ($\sim$\,few 0.1\,\kms\,pc$^{-1}$) and virial parameters ($\sim$\,unity), and \citet{ragan_2014} showed how these IRDCs could be part of ``Giant Molecular Filaments'' (henceforth, GMFs) structures, which stretch over hundreds of parsecs and have masses of $\sim$\,10$^{5-6}$\,\sol. 

Using the same analysis tools used on the IRAM-30m observations, we determine the kinematic structure of the region surrounding Cloud F (see appendix\,\ref{Appendix C}). The results of this analysis are presented in the upper panel of Figure\,\ref{ppv_plots_cloudf-grs}, which shows the position-position-velocity diagram with each point coloured to the identified velocity components given in Table\,\ref{grs velocity component parameters} (features of interest are shown in the figure legend). The most extended and prominent of these structures, F$_{\rm GRS2}$, is coherent over nearly the entire mapped region, a projected distance of $>$\,20\,pc at the assumed source distance of 3.7\,kpc \citep{simon_2006b}. 

The lower panel of Figure\,\ref{ppv_plots_cloudf-grs} shows the structures identified from the IRAM-30m \ceooz\ observations overlaid on those from the \tcooz\ GRS observations. We find that the F$_{\rm PPV4}$ structure appears to coincide spatially and in velocity with the F$_{\rm GRS2}$ towards the north of the IRAM-30m mapped region. However, towards the south of the IRAM-30m map, towards the MM1 region, the F$_{\rm PPV4}$ component appears at a velocity in-between the F$_{\rm GRS1}$ and F$_{\rm GRS2}$ components, which suggests that the splitting of these two GRS components is an optical depth effect in the \tco\ observations, which would make sense as this is one the densest regions within the cloud. Furthermore, as discussed in appendix\,\ref{Appendix D}, this optically thick emission profile can be linked to the infall motions previously identified towards this region \citep{ramesh_1997, sanhueza_2010}. 

We find that F$_{\rm PPV4}$ and F$_{\rm PPV2}$ appear to trace F$_{\rm GRS1}$ and F$_{\rm GRS2}$ on the west side of the mapped region, towards the MM8/F1 region. However, given the spatial resolution of GRS observations ($\sim$\,0.8\,pc), it is difficult to distinguish the transition to the optically thin regime when inspecting the spectra from the MM1 region to the F1/MM8 regions, hence the \tco\ observations towards this region may also be optically thick. Towards the F4/MM3 region, the F$_{\rm PPV1}$ component does not appear to have any associated component in the GRS observations. It is possible that this component has blended in the GRS data, which seems feasible given the broad line width of the F$_{\rm GRS2}$ component within this region ($\sim$\,2\kms), and the close proximity in velocity to the F$_{\rm PPV4}$ component ($\sim$\,1\,\kms). Despite the caveats discussed here, the brightest and most extended structures in the GRS observations appear to correspond to the structures in the IRAM-30m, indicating that Cloud F could be the central, densest part of this larger scale structure. 

\subsubsection{Cloud F as part of a massive inter-arm filament}

\citet{ragan_2014} identified a structure within the spatial coverage and velocity range (50--60\,\kms) of the F$_{\rm GRS1}$/F$_{\rm GRS2}$ components as the Giant Molecular Filament 38.1-32.4a, which has the largest mass ($\sim$\,10$^{6}$\,\sol) and projected length ($\sim$\,200\,pc) in their sample. It was suggested that this structure resides between the near and far Sagittarius arm, hence could be classified as an ``inter-arm cloud'' (e.g. \citealp{zucker_2015}). However, the recent Bayesian distance estimator from \citet{reid_2016}, which takes into account the kinematic distance, displacement from the plane, and proximity to individual parallax and the probability of residing within a spiral arm (i.e. as priors), places this source in the far Sagittarius arm, at a distance of 10.6\,$\pm$\,0.3\,kpc. This is in disagreement with the kinematic distance analysis from \citet{roman-duval_2009}, which places the cloud at the near distance of 3.7\,$\pm$\,0.8\,kpc based on the absorption of the background HI emission towards this region, the near-infrared extinction distance of $\sim$\,3\,kpc \citep{foster_2012}, and the parallax distance of 1.56$^{+0.12}_{-11}$\,kpc (\citealp{kurayama_2011}; see \citealp{foster_2012} and \citealp{foster_2014} for discussion of potential issues with this measurement).

We adjust the weighting on near/far kinematic distance within the \citet{reid_2016} estimator (the only aforementioned prior easily varied). We find that this has to be set to a 1\,per cent probability of the source being at the far distance in order to recover a value consistent with the near kinematic distance (3.6\,$\pm$\,0.7\,kpc).\footnote{See \url{http://bessel.vlbi-astrometry.org/bayesian}, where the default value that the source is at the far distance is 50\,per cent. Adjusting the weightings of the other three priors is possible, yet beyond the scope of this work.} Taking this distance estimate for GMF 38.1-32.4a (and Cloud F) places it in-between the near and far Sagittarius arms, as previously suggested by \citet{ragan_2014}. This would make this region of particular interest for further study, as one of the most massive and extended inter-arm star-forming complexes in the Milky Way.

\subsubsection{Could Cloud F be interacting with the supernova remnant W44?}

We note that the complication with the source distance may be caused by the higher than average uncertainty in the spiral arm models towards the 33$^{\circ}$\,$>$\,$l$\,$>$\,36$^{\circ}$ longitude region, due to the W44 supernova remnant, which has spread the gas over a large velocity range \citep{dame_1986, cardillo_2014}. A speculatory scenario could then be that the supernova remnant is directly influencing the gas within Cloud F, forcing it to a higher velocity than this predicted for the Sagittarius arm near-arm ($\sim$\,30\kms; \citealp{reid_2014}). Assuming that the cloud had an original velocity of $\sim$\,30\kms, the \citet{reid_2016} estimator places the source at a distance of 2.12\,$\pm$\,0.17\,kpc, which is in better agreement with the parallax distance from \cite{kurayama_2011}. This is, however, then significantly closer than the distance to W44 of $\sim$\,3.2\,kpc, determined from pulsar timing \citep{wolszczan_1991}. Furthermore, the structure of the W44, observed in the infrared, radio and x-ray, doesn't appear to extend high enough in galactic latitude to be interacting with the Cloud F (or GMF 38.1-32.4a; \citealp{castelletti_2007, castelletti_2011, cardillo_2014}). It is then not clear if it is possible that these two sources are interacting, nevertheless, such a scenario would further this cloud as an interesting source for future studies.

\section{Conclusions}\label{Conclusions}


In this work, we have identified the kinematic structures within a relatively quiescent massive IRDC: G034.43+00.24 (or Cloud F; e.g. \citealp{butler_2009}). To do so, we have acquired high-sensitivity, high spectral resolution maps of the \ceooz\ and \ntwohoz\ molecular line transitions taken with the IRAM-30m telescope. These lines were chosen as they are thought to trace the moderate to high-density gas within quiescent star-forming regions ($\sim$\,10$^{3-5}$cm$^{-3}$). Multiple velocity components are seen in the \ceooz\ spectra at the majority of positions throughout the cloud. To separate and link these into coherent structures, we use semi-automated Gaussian line fitting and hierarchical clustering ({\sc scouse} and {\sc acorns}; \citealp{henshaw_2016}, in prep). Compared to moment and channel map analysis, which are typically \change{used} for kinematic studies, the use of these algorithms removes much of the subjectivity in identifying velocity structures, allowing for a reliable investigation into their properties.

We find \change{four} distinct coherent velocity components within Cloud F in \ceooz\ emission, some of which are extended along the majority of the cloud ($\sim$\,4\,pc). We compare the properties of these to the velocity components identified within a similar IRDC, G035.39-00.33 (Cloud H; e.g. \citealp{butler_2009}). We find that these share many similar properties, such as the components appear to be very dense (approximately $>$\,10$^{4}$\,cm$^{-3}$, as inferred from the extended \ntwoh\ emitting area), the components have mildly supersonic velocity dispersions, the components have a similar separation in velocity, and there is a significant (positive) velocity difference between similar components identified in \ceo\ and \ntwoh emission. The latter two of these could hint at a common scenario of gentle filament merging, although this requires further investigation. 

We investigate the large-scale kinematic structure surrounding Cloud F, by using the lower density tracer \tcooz\ from the Galactic Ring Survey. Several very extended ($>$\,10\,pc) structures are identified throughout the GRS region, some of which may, in fact, be larger if not artificially split in velocity by optical depth effects. We find that the structures identified from the IRAM-30m observations are coincident with the central, brightest and most extended component in the GRS, suggesting that the IRDCs are the densest central parts of less dense, larger scale structures. We find that the structure identified here could be the Giant Molecular Filament 38.1-32.4a found by \citet{ragan_2014}, which when taking the kinematic source distance places it as an ``inter-arm cloud'' (``spur'' or ``feather'') residing in-between the near and far Sagittarius arm.

\section*{ACKNOWLEDGEMENTS}

A.T.B would like to acknowledge the funding provided by Liverpool John Moores University, the Max-Plank -Institute for Extraterrestrial Physics and the University of Leeds. P.C. acknowledges financial support from the European Research Council (ERC Advanced Grant; PALs 320620). Partial salary support for A.P. was provided by a Canadian Institute for Theoretical Astrophysics (CITA) National Fellowship. I.J.-S. acknowledges the financial support received from the STFC through an Ernest Rutherford Fellowship (proposal number ST/L004801/2). The research leading to these results has also received funding from the European Commission (FP/2007-2013) under grant agreement No 283393 (RadioNet3). This analysis has made use of the {\sc spectral cube} Version 0.4.0 package, and the {\sc acorns} package.

\bibliographystyle{mnras}
\bibliography{references}

\begin{thebibliography}{}
\makeatletter
\relax
\def\mn@urlcharsother{\let\do\@makeother \do\$\do\&\do\#\do\^\do\_\do\%\do\~}
\def\mn@doi{\begingroup\mn@urlcharsother \@ifnextchar [ {\mn@doi@}
  {\mn@doi@[]}}
\def\mn@doi@[#1]#2{\def\@tempa{#1}\ifx\@tempa\@empty \href
  {http://dx.doi.org/#2} {doi:#2}\else \href {http://dx.doi.org/#2} {#1}\fi
  \endgroup}
\def\mn@eprint#1#2{\mn@eprint@#1:#2::\@nil}
\def\mn@eprint@arXiv#1{\href {http://arxiv.org/abs/#1} {{\tt arXiv:#1}}}
\def\mn@eprint@dblp#1{\href {http://dblp.uni-trier.de/rec/bibtex/#1.xml}
  {dblp:#1}}
\def\mn@eprint@#1:#2:#3:#4\@nil{\def\@tempa {#1}\def\@tempb {#2}\def\@tempc
  {#3}\ifx \@tempc \@empty \let \@tempc \@tempb \let \@tempb \@tempa \fi \ifx
  \@tempb \@empty \def\@tempb {arXiv}\fi \@ifundefined
  {mn@eprint@\@tempb}{\@tempb:\@tempc}{\expandafter \expandafter \csname
  mn@eprint@\@tempb\endcsname \expandafter{\@tempc}}}

\bibitem[\protect\citeauthoryear{{Andr{\'e}}, {Belloche}, {Motte}  \&
  {Peretto}}{{Andr{\'e}} et~al.}{2007}]{andre_2007}
{Andr{\'e}} P.,  {Belloche} A.,  {Motte} F.,   {Peretto} N.,  2007, \mn@doi
  [\aap] {10.1051/0004-6361:20077422}, \href
  {http://cdsads.u-strasbg.fr/abs/2007A%26A...472..519A} {472, 519}

\bibitem[\protect\citeauthoryear{{Bailey}, {Basu}  \& {Caselli}}{{Bailey}
  et~al.}{2015}]{bailey_2015}
{Bailey} N.~D.,  {Basu} S.,   {Caselli} P.,  2015, \mn@doi [\apj]
  {10.1088/0004-637X/798/2/75}, \href
  {http://adsabs.harvard.edu/abs/2015ApJ...798...75B} {798, 75}

\bibitem[\protect\citeauthoryear{{Ballesteros-Paredes} \& {Mac
  Low}}{{Ballesteros-Paredes} \& {Mac Low}}{2002}]{ballesteros-paredes_2002}
{Ballesteros-Paredes} J.,  {Mac Low} M.-M.,  2002, \mn@doi [\apj]
  {10.1086/339624}, \href
  {http://ukads.nottingham.ac.uk/abs/2002ApJ...570..734B} {570, 734}

\bibitem[\protect\citeauthoryear{{Barnes}, {Kong}, {Tan}, {Henshaw}, {Caselli},
  {Jim{\'e}nez-Serra}  \& {Fontani}}{{Barnes} et~al.}{2016}]{barnes_2016}
{Barnes} A.~T.,  {Kong} S.,  {Tan} J.~C.,  {Henshaw} J.~D.,  {Caselli} P.,
  {Jim{\'e}nez-Serra} I.,   {Fontani} F.,  2016, \mn@doi [\mnras]
  {10.1093/mnras/stw403}, \href
  {http://adsabs.harvard.edu/abs/2016MNRAS.458.1990B} {458, 1990}

\bibitem[\protect\citeauthoryear{{Barnes}, {Longmore}, {Battersby}, {Bally},
  {Kruijssen}, {Henshaw}  \& {Walker}}{{Barnes} et~al.}{2017}]{barnes_2017}
{Barnes} A.~T.,  {Longmore} S.~N.,  {Battersby} C.,  {Bally} J.,  {Kruijssen}
  J.~M.~D.,  {Henshaw} J.~D.,   {Walker} D.~L.,  2017, \mn@doi [\mnras]
  {10.1093/mnras/stx941}, \href
  {http://adsabs.harvard.edu/abs/2017MNRAS.469.2263B} {469, 2263}

\bibitem[\protect\citeauthoryear{{Butler} \& {Tan}}{{Butler} \&
  {Tan}}{2009}]{butler_2009}
{Butler} M.~J.,  {Tan} J.~C.,  2009, \mn@doi [\apj]
  {10.1088/0004-637X/696/1/484}, \href
  {http://adsabs.harvard.edu/abs/2009ApJ...696..484B} {696, 484}

\bibitem[\protect\citeauthoryear{{Butler} \& {Tan}}{{Butler} \&
  {Tan}}{2012}]{butler_2012}
{Butler} M.~J.,  {Tan} J.~C.,  2012, \mn@doi [\apj]
  {10.1088/0004-637X/754/1/5}, \href
  {http://adsabs.harvard.edu/abs/2012ApJ...754....5B} {754, 5}

\bibitem[\protect\citeauthoryear{{Cardillo} et~al.,}{{Cardillo}
  et~al.}{2014}]{cardillo_2014}
{Cardillo} M.,  et~al., 2014, \mn@doi [\aap] {10.1051/0004-6361/201322685},
  \href {http://adsabs.harvard.edu/abs/2014A%26A...565A..74C} {565, A74}

\bibitem[\protect\citeauthoryear{{Carey}, {Clark}, {Egan}, {Price}, {Shipman}
  \& {Kuchar}}{{Carey} et~al.}{1998}]{carey_1998}
{Carey} S.~J.,  {Clark} F.~O.,  {Egan} M.~P.,  {Price} S.~D.,  {Shipman} R.~F.,
    {Kuchar} T.~A.,  1998, \mn@doi [\apj] {10.1086/306438}, \href
  {http://adsabs.harvard.edu/abs/1998ApJ...508..721C} {508, 721}

\bibitem[\protect\citeauthoryear{{Caselli}, {Myers}  \& {Thaddeus}}{{Caselli}
  et~al.}{1995}]{caselli_1995}
{Caselli} P.,  {Myers} P.~C.,   {Thaddeus} P.,  1995, \mn@doi [\apjl]
  {10.1086/309805}, \href {http://adsabs.harvard.edu/abs/1995ApJ...455L..77C}
  {455, L77+}

\bibitem[\protect\citeauthoryear{{Caselli}, {Walmsley}, {Zucconi}, {Tafalla},
  {Dore}  \& {Myers}}{{Caselli} et~al.}{2002a}]{caselli_2002a}
{Caselli} P.,  {Walmsley} C.~M.,  {Zucconi} A.,  {Tafalla} M.,  {Dore} L.,
  {Myers} P.~C.,  2002a, \mn@doi [\apj] {10.1086/324302}, \href
  {http://adsabs.harvard.edu/abs/2002ApJ...565..344C} {565, 344}

\bibitem[\protect\citeauthoryear{{Caselli}, {Benson}, {Myers}  \&
  {Tafalla}}{{Caselli} et~al.}{2002b}]{caselli_2002}
{Caselli} P.,  {Benson} P.~J.,  {Myers} P.~C.,   {Tafalla} M.,  2002b, \mn@doi
  [\apj] {10.1086/340195}, \href
  {http://adsabs.harvard.edu/abs/2002ApJ...572..238C} {572, 238}

\bibitem[\protect\citeauthoryear{{Castelletti}, {Dubner}, {Brogan}  \&
  {Kassim}}{{Castelletti} et~al.}{2007}]{castelletti_2007}
{Castelletti} G.,  {Dubner} G.,  {Brogan} C.,   {Kassim} N.~E.,  2007, \mn@doi
  [\aap] {10.1051/0004-6361:20077062}, \href
  {http://ukads.nottingham.ac.uk/abs/2007A%26A...471..537C} {471, 537}

\bibitem[\protect\citeauthoryear{{Castelletti}, {Dubner}, {Clarke}  \&
  {Kassim}}{{Castelletti} et~al.}{2011}]{castelletti_2011}
{Castelletti} G.,  {Dubner} G.,  {Clarke} T.,   {Kassim} N.~E.,  2011, \mn@doi
  [\aap] {10.1051/0004-6361/201016081}, \href
  {http://ukads.nottingham.ac.uk/abs/2011A%26A...534A..21C} {534, A21}

\bibitem[\protect\citeauthoryear{{Cazzoli}, {Puzzarini}  \&
  {Lapinov}}{{Cazzoli} et~al.}{2003}]{cazzoli_2003}
{Cazzoli} G.,  {Puzzarini} C.,   {Lapinov} A.~V.,  2003, \mn@doi [\apjl]
  {10.1086/377527}, \href {http://cdsads.u-strasbg.fr/abs/2003ApJ...592L..95C}
  {592, L95}

\bibitem[\protect\citeauthoryear{{Cazzoli}, {Cludi}, {Buffa}  \&
  {Puzzarini}}{{Cazzoli} et~al.}{2012}]{cazzoli_2012}
{Cazzoli} G.,  {Cludi} L.,  {Buffa} G.,   {Puzzarini} C.,  2012, \mn@doi
  [\apjs] {10.1088/0067-0049/203/1/11}, \href
  {http://ukads.nottingham.ac.uk/abs/2012ApJS..203...11C} {203, 11}

\bibitem[\protect\citeauthoryear{{Chambers}, {Jackson}, {Rathborne}  \&
  {Simon}}{{Chambers} et~al.}{2009}]{chambers_2009}
{Chambers} E.~T.,  {Jackson} J.~M.,  {Rathborne} J.~M.,   {Simon} R.,  2009,
  \mn@doi [\apjs] {10.1088/0067-0049/181/2/360}, \href
  {http://adsabs.harvard.edu/abs/2009ApJS..181..360C} {181, 360}

\bibitem[\protect\citeauthoryear{{Chira}, {Kainulainen},
  {Ib{\`a}{\~n}ez-Mej{\'{\i}}a}, {Henning}  \& {Mac Low}}{{Chira}
  et~al.}{2017}]{chira_2017}
{Chira} R.-A.,  {Kainulainen} J.,  {Ib{\`a}{\~n}ez-Mej{\'{\i}}a} J.~C.,
  {Henning} T.,   {Mac Low} M.-M.,  2017, preprint, \href
  {http://adsabs.harvard.edu/abs/2017arXiv171101417C} {} (\mn@eprint {arXiv}
  {1711.01417})

\bibitem[\protect\citeauthoryear{{Dame}, {Elmegreen}, {Cohen}  \&
  {Thaddeus}}{{Dame} et~al.}{1986}]{dame_1986}
{Dame} T.~M.,  {Elmegreen} B.~G.,  {Cohen} R.~S.,   {Thaddeus} P.,  1986,
  \mn@doi [\apj] {10.1086/164304}, \href
  {http://adsabs.harvard.edu/abs/1986ApJ...305..892D} {305, 892}

\bibitem[\protect\citeauthoryear{{Devine}, {Chandler}, {Brogan}, {Churchwell},
  {Indebetouw}, {Shirley}  \& {Borg}}{{Devine} et~al.}{2011}]{devine_2011}
{Devine} K.~E.,  {Chandler} C.~J.,  {Brogan} C.,  {Churchwell} E.,
  {Indebetouw} R.,  {Shirley} Y.,   {Borg} K.~J.,  2011, \mn@doi [\apj]
  {10.1088/0004-637X/733/1/44}, \href
  {http://cdsads.u-strasbg.fr/abs/2011ApJ...733...44D} {733, 44}

\bibitem[\protect\citeauthoryear{{Dirienzo}, {Brogan}, {Indebetouw},
  {Chandler}, {Friesen}  \& {Devine}}{{Dirienzo} et~al.}{2015}]{dirienzo_2015}
{Dirienzo} W.~J.,  {Brogan} C.,  {Indebetouw} R.,  {Chandler} C.~J.,  {Friesen}
  R.~K.,   {Devine} K.~E.,  2015, \mn@doi [\aj] {10.1088/0004-6256/150/5/159},
  \href {http://adsabs.harvard.edu/abs/2015AJ....150..159D} {150, 159}

\bibitem[\protect\citeauthoryear{{Egan}, {Shipman}, {Price}, {Carey}, {Clark}
  \& {Cohen}}{{Egan} et~al.}{1998}]{egan_1998}
{Egan} M.~P.,  {Shipman} R.~F.,  {Price} S.~D.,  {Carey} S.~J.,  {Clark} F.~O.,
    {Cohen} M.,  1998, \mn@doi [\apjl] {10.1086/311198}, \href
  {http://adsabs.harvard.edu/abs/1998ApJ...494L.199E} {494, L199+}

\bibitem[\protect\citeauthoryear{{Evans}}{{Evans}}{1999}]{evans_1999}
{Evans} II N.~J.,  1999, \mn@doi [\araa] {10.1146/annurev.astro.37.1.311},
  \href {http://adsabs.harvard.edu/abs/1999ARA%26A..37..311E} {37, 311}

\bibitem[\protect\citeauthoryear{{Fontani}, {Caselli}, {Crapsi}, {Cesaroni},
  {Molinari}, {Testi}  \& {Brand}}{{Fontani} et~al.}{2006}]{fontani_2006}
{Fontani} F.,  {Caselli} P.,  {Crapsi} A.,  {Cesaroni} R.,  {Molinari} S.,
  {Testi} L.,   {Brand} J.,  2006, \mn@doi [\aap] {10.1051/0004-6361:20066105},
  \href {http://adsabs.harvard.edu/abs/2006A%26A...460..709F} {460, 709}

\bibitem[\protect\citeauthoryear{{Fontani} et~al.,}{{Fontani}
  et~al.}{2011}]{fontani_2011}
{Fontani} F.,  et~al., 2011, \mn@doi [\aap] {10.1051/0004-6361/201116631},
  \href {http://adsabs.harvard.edu/abs/2011A%26A...529L...7F} {529, L7+}

\bibitem[\protect\citeauthoryear{{Foster}, {Stead}, {Benjamin}, {Hoare}  \&
  {Jackson}}{{Foster} et~al.}{2012}]{foster_2012}
{Foster} J.~B.,  {Stead} J.~J.,  {Benjamin} R.~A.,  {Hoare} M.~G.,   {Jackson}
  J.~M.,  2012, \mn@doi [\apj] {10.1088/0004-637X/751/2/157}, \href
  {http://ukads.nottingham.ac.uk/abs/2012ApJ...751..157F} {751, 157}

\bibitem[\protect\citeauthoryear{{Foster} et~al.,}{{Foster}
  et~al.}{2014}]{foster_2014}
{Foster} J.~B.,  et~al., 2014, \mn@doi [\apj] {10.1088/0004-637X/791/2/108},
  \href {http://ukads.nottingham.ac.uk/abs/2014ApJ...791..108F} {791, 108}

\bibitem[\protect\citeauthoryear{{Frerking}, {Langer}  \& {Wilson}}{{Frerking}
  et~al.}{1982}]{frerking_1982}
{Frerking} M.~A.,  {Langer} W.~D.,   {Wilson} R.~W.,  1982, \mn@doi [\apj]
  {10.1086/160451}, \href {http://adsabs.harvard.edu/abs/1982ApJ...262..590F}
  {262, 590}

\bibitem[\protect\citeauthoryear{{Friesen}, {Di Francesco}, {Myers},
  {Belloche}, {Shirley}, {Bourke}  \& {Andr{\'e}}}{{Friesen}
  et~al.}{2010}]{friesen_2010}
{Friesen} R.~K.,  {Di Francesco} J.,  {Myers} P.~C.,  {Belloche} A.,  {Shirley}
  Y.~L.,  {Bourke} T.~L.,   {Andr{\'e}} P.,  2010, \mn@doi [\apj]
  {10.1088/0004-637X/718/2/666}, \href
  {http://adsabs.harvard.edu/abs/2010ApJ...718..666F} {718, 666}

\bibitem[\protect\citeauthoryear{{Garay}, {Fa{\'u}ndez}, {Mardones},
  {Bronfman}, {Chini}  \& {Nyman}}{{Garay} et~al.}{2004}]{garay_2004}
{Garay} G.,  {Fa{\'u}ndez} S.,  {Mardones} D.,  {Bronfman} L.,  {Chini} R.,
  {Nyman} L.-{\AA}.,  2004, \mn@doi [\apj] {10.1086/421437}, \href
  {http://adsabs.harvard.edu/abs/2004ApJ...610..313G} {610, 313}

\bibitem[\protect\citeauthoryear{{Gerner}, {Shirley}, {Beuther}, {Semenov},
  {Linz}, {Abertsson}  \& {Henning}}{{Gerner} et~al.}{2015}]{gerner_2015}
{Gerner} T.,  {Shirley} Y.,  {Beuther} H.,  {Semenov} D.,  {Linz} H.,
  {Abertsson} T.,   {Henning} T.,  2015, preprint, \href
  {http://adsabs.harvard.edu/abs/2015arXiv150306594G} {} (\mn@eprint {arXiv}
  {1503.06594})

\bibitem[\protect\citeauthoryear{{Goodman}, {Benson}, {Fuller}  \&
  {Myers}}{{Goodman} et~al.}{1993}]{goodman_1993}
{Goodman} A.~A.,  {Benson} P.~J.,  {Fuller} G.~A.,   {Myers} P.~C.,  1993,
  \mn@doi [\apj] {10.1086/172465}, \href
  {http://adsabs.harvard.edu/abs/1993ApJ...406..528G} {406, 528}

\bibitem[\protect\citeauthoryear{{Hacar}, {Tafalla}, {Kauffmann}  \&
  {Kov{\'a}cs}}{{Hacar} et~al.}{2013}]{hacar_2013}
{Hacar} A.,  {Tafalla} M.,  {Kauffmann} J.,   {Kov{\'a}cs} A.,  2013, \mn@doi
  [\aap] {10.1051/0004-6361/201220090}, \href
  {http://adsabs.harvard.edu/abs/2013A%26A...554A..55H} {554, A55}

\bibitem[\protect\citeauthoryear{{Hacar}, {Kainulainen}, {Tafalla}, {Beuther}
  \& {Alves}}{{Hacar} et~al.}{2016}]{hacar_2016a}
{Hacar} A.,  {Kainulainen} J.,  {Tafalla} M.,  {Beuther} H.,   {Alves} J.,
  2016, \mn@doi [\aap] {10.1051/0004-6361/201526015}, \href
  {http://adsabs.harvard.edu/abs/2016A%26A...587A..97H} {587, A97}

\bibitem[\protect\citeauthoryear{{Hacar}, {Tafalla}  \& {Alves}}{{Hacar}
  et~al.}{2017}]{hacar_2017b}
{Hacar} A.,  {Tafalla} M.,   {Alves} J.,  2017, preprint, \href
  {http://adsabs.harvard.edu/abs/2017arXiv170307029H} {} (\mn@eprint {arXiv}
  {1703.07029})

\bibitem[\protect\citeauthoryear{{Henshaw}, {Caselli}, {Fontani},
  {Jim{\'e}nez-Serra}, {Tan}  \& {Hernandez}}{{Henshaw}
  et~al.}{2013}]{henshaw_2013}
{Henshaw} J.~D.,  {Caselli} P.,  {Fontani} F.,  {Jim{\'e}nez-Serra} I.,  {Tan}
  J.~C.,   {Hernandez} A.~K.,  2013, \mn@doi [\mnras] {10.1093/mnras/sts282},
  \href {http://adsabs.harvard.edu/abs/2013MNRAS.428.3425H} {428, 3425}

\bibitem[\protect\citeauthoryear{{Henshaw}, {Caselli}, {Fontani},
  {Jim{\'e}nez-Serra}  \& {Tan}}{{Henshaw} et~al.}{2014}]{henshaw_2014}
{Henshaw} J.~D.,  {Caselli} P.,  {Fontani} F.,  {Jim{\'e}nez-Serra} I.,   {Tan}
  J.~C.,  2014, \mn@doi [\mnras] {10.1093/mnras/stu446}, \href
  {http://adsabs.harvard.edu/abs/2014MNRAS.440.2860H} {440, 2860}

\bibitem[\protect\citeauthoryear{{Henshaw} et~al.,}{{Henshaw}
  et~al.}{2016}]{henshaw_2016}
{Henshaw} J.~D.,  et~al., 2016, \mn@doi [\mnras] {10.1093/mnras/stw121}, \href
  {http://adsabs.harvard.edu/abs/2016MNRAS.457.2675H} {457, 2675}

\bibitem[\protect\citeauthoryear{{Henshaw} et~al.,}{{Henshaw}
  et~al.}{2017}]{henshaw_2016d}
{Henshaw} J.~D.,  et~al., 2017, \mn@doi [\mnras] {10.1093/mnrasl/slw154}, \href
  {http://adsabs.harvard.edu/abs/2017MNRAS.464L..31H} {464, L31}

\bibitem[\protect\citeauthoryear{{Hernandez} \& {Tan}}{{Hernandez} \&
  {Tan}}{2011}]{hernandez_2011a}
{Hernandez} A.~K.,  {Tan} J.~C.,  2011, \mn@doi [\apj]
  {10.1088/0004-637X/730/1/44}, \href
  {http://cdsads.u-strasbg.fr/abs/2011ApJ...730...44H} {730, 44}

\bibitem[\protect\citeauthoryear{{Hernandez} \& {Tan}}{{Hernandez} \&
  {Tan}}{2015}]{hernandez_2015}
{Hernandez} A.~K.,  {Tan} J.~C.,  2015, \mn@doi [\apj]
  {10.1088/0004-637X/809/2/154}, \href
  {http://adsabs.harvard.edu/abs/2015ApJ...809..154H} {809, 154}

\bibitem[\protect\citeauthoryear{{Hernandez}, {Tan}, {Caselli}, {Butler},
  {Jim{\'e}nez-Serra}, {Fontani}  \& {Barnes}}{{Hernandez}
  et~al.}{2011}]{hernandez_2011}
{Hernandez} A.~K.,  {Tan} J.~C.,  {Caselli} P.,  {Butler} M.~J.,
  {Jim{\'e}nez-Serra} I.,  {Fontani} F.,   {Barnes} P.,  2011, \mn@doi [\apj]
  {10.1088/0004-637X/738/1/11}, \href
  {http://cdsads.u-strasbg.fr/abs/2011ApJ...738...11H} {738, 11}

\bibitem[\protect\citeauthoryear{{Hernandez}, {Tan}, {Kainulainen}, {Caselli},
  {Butler}, {Jim{\'e}nez-Serra}  \& {Fontani}}{{Hernandez}
  et~al.}{2012}]{hernandez_2012a}
{Hernandez} A.~K.,  {Tan} J.~C.,  {Kainulainen} J.,  {Caselli} P.,  {Butler}
  M.~J.,  {Jim{\'e}nez-Serra} I.,   {Fontani} F.,  2012, \mn@doi [\apjl]
  {10.1088/2041-8205/756/1/L13}, \href
  {http://adsabs.harvard.edu/abs/2012ApJ...756L..13H} {756, L13}

\bibitem[\protect\citeauthoryear{{Jackson} et~al.,}{{Jackson}
  et~al.}{2006}]{jackson_2006}
{Jackson} J.~M.,  et~al., 2006, \mn@doi [\apjs] {10.1086/500091}, \href
  {http://adsabs.harvard.edu/abs/2006ApJS..163..145J} {163, 145}

\bibitem[\protect\citeauthoryear{{Jim{\'e}nez-Serra}, {Caselli}, {Tan},
  {Hernandez}, {Fontani}, {Butler}  \& {van Loo}}{{Jim{\'e}nez-Serra}
  et~al.}{2010}]{jimnez-serra_2010}
{Jim{\'e}nez-Serra} I.,  {Caselli} P.,  {Tan} J.~C.,  {Hernandez} A.~K.,
  {Fontani} F.,  {Butler} M.~J.,   {van Loo} S.,  2010, \mn@doi [\mnras]
  {10.1111/j.1365-2966.2010.16698.x}, \href
  {http://adsabs.harvard.edu/abs/2010MNRAS.406..187J} {406, 187}

\bibitem[\protect\citeauthoryear{{Jim{\'e}nez-Serra}, {Caselli}, {Fontani},
  {Tan}, {Henshaw}, {Kainulainen}  \& {Hernandez}}{{Jim{\'e}nez-Serra}
  et~al.}{2014}]{jimnez-serra_2014}
{Jim{\'e}nez-Serra} I.,  {Caselli} P.,  {Fontani} F.,  {Tan} J.~C.,  {Henshaw}
  J.~D.,  {Kainulainen} J.,   {Hernandez} A.~K.,  2014, \mn@doi [\mnras]
  {10.1093/mnras/stu078}, \href
  {http://adsabs.harvard.edu/abs/2014MNRAS.439.1996J} {439, 1996}

\bibitem[\protect\citeauthoryear{{Kainulainen} \& {Tan}}{{Kainulainen} \&
  {Tan}}{2013}]{kainulainen_2013}
{Kainulainen} J.,  {Tan} J.~C.,  2013, \mn@doi [\aap]
  {10.1051/0004-6361/201219526}, \href
  {http://adsabs.harvard.edu/abs/2013A%26A...549A..53K} {549, A53}

\bibitem[\protect\citeauthoryear{{Kauffmann}, {Bertoldi}, {Bourke}, {Evans}  \&
  {Lee}}{{Kauffmann} et~al.}{2008}]{kauffmann_2008}
{Kauffmann} J.,  {Bertoldi} F.,  {Bourke} T.~L.,  {Evans} II N.~J.,   {Lee}
  C.~W.,  2008, \mn@doi [\aap] {10.1051/0004-6361:200809481}, \href
  {http://adsabs.harvard.edu/abs/2008A%26A...487..993K} {487, 993}

\bibitem[\protect\citeauthoryear{{Kirk}, {Myers}, {Bourke}, {Gutermuth},
  {Hedden}  \& {Wilson}}{{Kirk} et~al.}{2013}]{kirk_2013}
{Kirk} H.,  {Myers} P.~C.,  {Bourke} T.~L.,  {Gutermuth} R.~A.,  {Hedden} A.,
  {Wilson} G.~W.,  2013, \mn@doi [\apj] {10.1088/0004-637X/766/2/115}, \href
  {http://adsabs.harvard.edu/abs/2013ApJ...766..115K} {766, 115}

\bibitem[\protect\citeauthoryear{{Kong} et~al.,}{{Kong}
  et~al.}{2016}]{kong_2016}
{Kong} S.,  et~al., 2016, \mn@doi [\apj] {10.3847/0004-637X/821/2/94}, \href
  {http://adsabs.harvard.edu/abs/2016ApJ...821...94K} {821, 94}

\bibitem[\protect\citeauthoryear{{Kong}, {Tan}, {Caselli}, {Fontani}, {Liu}  \&
  {Butler}}{{Kong} et~al.}{2017}]{kong_2017}
{Kong} S.,  {Tan} J.~C.,  {Caselli} P.,  {Fontani} F.,  {Liu} M.,   {Butler}
  M.~J.,  2017, \mn@doi [\apj] {10.3847/1538-4357/834/2/193}, \href
  {http://adsabs.harvard.edu/abs/2017ApJ...834..193K} {834, 193}

\bibitem[\protect\citeauthoryear{{Kurayama}, {Nakagawa}, {Sawada-Satoh},
  {Sato}, {Honma}, {Sunada}, {Hirota}  \& {Imai}}{{Kurayama}
  et~al.}{2011}]{kurayama_2011}
{Kurayama} T.,  {Nakagawa} A.,  {Sawada-Satoh} S.,  {Sato} K.,  {Honma} M.,
  {Sunada} K.,  {Hirota} T.,   {Imai} H.,  2011, \mn@doi [\pasj]
  {10.1093/pasj/63.3.513}, \href
  {http://adsabs.harvard.edu/abs/2011PASJ...63..513K} {63, 513}

\bibitem[\protect\citeauthoryear{{Longmore} et~al.,}{{Longmore}
  et~al.}{2012}]{longmore_2012}
{Longmore} S.~N.,  et~al., 2012, \mn@doi [\apj] {10.1088/0004-637X/746/2/117},
  \href {http://adsabs.harvard.edu/abs/2012ApJ...746..117L} {746, 117}

\bibitem[\protect\citeauthoryear{{Lou} \& {Gao}}{{Lou} \&
  {Gao}}{2011}]{lou_2011}
{Lou} Y.-Q.,  {Gao} Y.,  2011, \mn@doi [\mnras]
  {10.1111/j.1365-2966.2010.18011.x}, \href
  {http://adsabs.harvard.edu/abs/2011MNRAS.412.1755L} {412, 1755}

\bibitem[\protect\citeauthoryear{{Miettinen}, {Hennemann}  \&
  {Linz}}{{Miettinen} et~al.}{2011}]{miettinen_2011}
{Miettinen} O.,  {Hennemann} M.,   {Linz} H.,  2011, \mn@doi [\aap]
  {10.1051/0004-6361/201117187}, \href
  {http://adsabs.harvard.edu/abs/2011A%26A...534A.134M} {534, A134}

\bibitem[\protect\citeauthoryear{{Miralles}, {Rodriguez}  \&
  {Scalise}}{{Miralles} et~al.}{1994}]{miralles_1994}
{Miralles} M.~P.,  {Rodriguez} L.~F.,   {Scalise} E.,  1994, \mn@doi [\apjs]
  {10.1086/191965}, \href {http://adsabs.harvard.edu/abs/1994ApJS...92..173P}
  {92, 173}

\bibitem[\protect\citeauthoryear{{Molinari}, {Brand}, {Cesaroni}  \&
  {Palla}}{{Molinari} et~al.}{1996}]{molinari_1996}
{Molinari} S.,  {Brand} J.,  {Cesaroni} R.,   {Palla} F.,  1996, \aap, \href
  {http://adsabs.harvard.edu/abs/1996A%26A...308..573M} {308, 573}

\bibitem[\protect\citeauthoryear{{Molinari} et~al.,}{{Molinari}
  et~al.}{2016}]{molinari_2016}
{Molinari} S.,  et~al., 2016, \mn@doi [\aap] {10.1051/0004-6361/201526380},
  \href {http://adsabs.harvard.edu/abs/2016A%26A...591A.149M} {591, A149}

\bibitem[\protect\citeauthoryear{{Myers}}{{Myers}}{1983}]{myers_1983}
{Myers} P.~C.,  1983, \mn@doi [\apj] {10.1086/161101}, \href
  {http://adsabs.harvard.edu/abs/1983ApJ...270..105M} {270, 105}

\bibitem[\protect\citeauthoryear{{Pagani}, {Daniel}  \& {Dubernet}}{{Pagani}
  et~al.}{2009}]{pagani_2009}
{Pagani} L.,  {Daniel} F.,   {Dubernet} M.,  2009, \mn@doi [\aap]
  {10.1051/0004-6361:200810570}, \href
  {http://adsabs.harvard.edu/abs/2009A%26A...494..719P} {494, 719}

\bibitem[\protect\citeauthoryear{{Perault} et~al.,}{{Perault}
  et~al.}{1996}]{perault_1996}
{Perault} M.,  et~al., 1996, \aap, \href
  {http://adsabs.harvard.edu/abs/1996A%26A...315L.165P} {315, L165}

\bibitem[\protect\citeauthoryear{{Peretto} et~al.,}{{Peretto}
  et~al.}{2010}]{peretto_2010}
{Peretto} N.,  et~al., 2010, \mn@doi [\aap] {10.1051/0004-6361/201014652},
  \href {http://adsabs.harvard.edu/abs/2010A%26A...518L..98P} {518, L98}

\bibitem[\protect\citeauthoryear{{Pillai}, {Wyrowski}, {Carey}  \&
  {Menten}}{{Pillai} et~al.}{2006}]{pillai_2006}
{Pillai} T.,  {Wyrowski} F.,  {Carey} S.~J.,   {Menten} K.~M.,  2006, \mn@doi
  [\aap] {10.1051/0004-6361:20054128}, \href
  {http://cdsads.u-strasbg.fr/abs/2006A%26A...450..569P} {450, 569}

\bibitem[\protect\citeauthoryear{{Pon}, {Caselli}, {Johnstone}, {Kaufman},
  {Butler}, {Fontani}, {Jim{\'e}nez-Serra}  \& {Tan}}{{Pon}
  et~al.}{2015}]{pon_2015}
{Pon} A.,  {Caselli} P.,  {Johnstone} D.,  {Kaufman} M.,  {Butler} M.~J.,
  {Fontani} F.,  {Jim{\'e}nez-Serra} I.,   {Tan} J.~C.,  2015, \mn@doi [\aap]
  {10.1051/0004-6361/201525681}, \href
  {http://adsabs.harvard.edu/abs/2015A%26A...577A..75P} {577, A75}

\bibitem[\protect\citeauthoryear{{Pon} et~al.,}{{Pon}
  et~al.}{2016a}]{pon_2016a}
{Pon} A.,  et~al., 2016a, \mn@doi [\aap] {10.1051/0004-6361/201527154}, \href
  {http://adsabs.harvard.edu/abs/2016A%26A...587A..96P} {587, A96}

\bibitem[\protect\citeauthoryear{{Pon} et~al.,}{{Pon}
  et~al.}{2016b}]{pon_2016b}
{Pon} A.,  et~al., 2016b, \mn@doi [\apj] {10.3847/0004-637X/827/2/107}, \href
  {http://adsabs.harvard.edu/abs/2016ApJ...827..107P} {827, 107}

\bibitem[\protect\citeauthoryear{{Ragan}, {Bergin}, {Plume}, {Gibson},
  {Wilner}, {O'Brien}  \& {Hails}}{{Ragan} et~al.}{2006}]{ragan_2006}
{Ragan} S.~E.,  {Bergin} E.~A.,  {Plume} R.,  {Gibson} D.~L.,  {Wilner} D.~J.,
  {O'Brien} S.,   {Hails} E.,  2006, \mn@doi [\apjs] {10.1086/506594}, \href
  {http://adsabs.harvard.edu/abs/2006ApJS..166..567R} {166, 567}

\bibitem[\protect\citeauthoryear{{Ragan}, {Bergin}  \& {Wilner}}{{Ragan}
  et~al.}{2011}]{ragan_2011}
{Ragan} S.~E.,  {Bergin} E.~A.,   {Wilner} D.,  2011, \mn@doi [\apj]
  {10.1088/0004-637X/736/2/163}, \href
  {http://cdsads.u-strasbg.fr/abs/2011ApJ...736..163R} {736, 163}

\bibitem[\protect\citeauthoryear{{Ragan}, {Henning}, {Tackenberg}, {Beuther},
  {Johnston}, {Kainulainen}  \& {Linz}}{{Ragan} et~al.}{2014}]{ragan_2014}
{Ragan} S.~E.,  {Henning} T.,  {Tackenberg} J.,  {Beuther} H.,  {Johnston}
  K.~G.,  {Kainulainen} J.,   {Linz} H.,  2014, \mn@doi [\aap]
  {10.1051/0004-6361/201423401}, \href
  {http://adsabs.harvard.edu/abs/2014A%26A...568A..73R} {568, A73}

\bibitem[\protect\citeauthoryear{{Ramesh}, {Bronfman}  \& {Deguchi}}{{Ramesh}
  et~al.}{1997}]{ramesh_1997}
{Ramesh} B.,  {Bronfman} L.,   {Deguchi} S.,  1997, \mn@doi [\pasj]
  {10.1093/pasj/49.3.307}, \href
  {http://ukads.nottingham.ac.uk/abs/1997PASJ...49..307R} {49, 307}

\bibitem[\protect\citeauthoryear{{Rathborne}, {Jackson}, {Chambers}, {Simon},
  {Shipman}  \& {Frieswijk}}{{Rathborne} et~al.}{2005}]{rathborne_2005}
{Rathborne} J.~M.,  {Jackson} J.~M.,  {Chambers} E.~T.,  {Simon} R.,  {Shipman}
  R.,   {Frieswijk} W.,  2005, \mn@doi [\apjl] {10.1086/491656}, \href
  {http://adsabs.harvard.edu/abs/2005ApJ...630L.181R} {630, L181}

\bibitem[\protect\citeauthoryear{{Rathborne}, {Jackson}  \&
  {Simon}}{{Rathborne} et~al.}{2006}]{rathborne_2006}
{Rathborne} J.~M.,  {Jackson} J.~M.,   {Simon} R.,  2006, \mn@doi [\apj]
  {10.1086/500423}, \href {http://adsabs.harvard.edu/abs/2006ApJ...641..389R}
  {641, 389}

\bibitem[\protect\citeauthoryear{{Rathborne}, {Jackson}, {Zhang}  \&
  {Simon}}{{Rathborne} et~al.}{2008}]{rathborne_2008b}
{Rathborne} J.~M.,  {Jackson} J.~M.,  {Zhang} Q.,   {Simon} R.,  2008, \mn@doi
  [\apj] {10.1086/592733}, \href
  {http://adsabs.harvard.edu/abs/2008ApJ...689.1141R} {689, 1141}

\bibitem[\protect\citeauthoryear{{Reid} et~al.,}{{Reid}
  et~al.}{2014}]{reid_2014}
{Reid} M.~J.,  et~al., 2014, \mn@doi [\apj] {10.1088/0004-637X/783/2/130},
  \href {http://adsabs.harvard.edu/abs/2014ApJ...783..130R} {783, 130}

\bibitem[\protect\citeauthoryear{{Reid}, {Dame}, {Menten}  \&
  {Brunthaler}}{{Reid} et~al.}{2016}]{reid_2016}
{Reid} M.~J.,  {Dame} T.~M.,  {Menten} K.~M.,   {Brunthaler} A.,  2016, \mn@doi
  [\apj] {10.3847/0004-637X/823/2/77}, \href
  {http://adsabs.harvard.edu/abs/2016ApJ...823...77R} {823, 77}

\bibitem[\protect\citeauthoryear{{Roman-Duval}, {Jackson}, {Heyer}, {Johnson},
  {Rathborne}, {Shah}  \& {Simon}}{{Roman-Duval}
  et~al.}{2009}]{roman-duval_2009}
{Roman-Duval} J.,  {Jackson} J.~M.,  {Heyer} M.,  {Johnson} A.,  {Rathborne}
  J.,  {Shah} R.,   {Simon} R.,  2009, \mn@doi [\apj]
  {10.1088/0004-637X/699/2/1153}, \href
  {http://adsabs.harvard.edu/abs/2009ApJ...699.1153R} {699, 1153}

\bibitem[\protect\citeauthoryear{{Rygl}, {Wyrowski}, {Schuller}  \&
  {Menten}}{{Rygl} et~al.}{2013}]{rygl_2013}
{Rygl} K.~L.~J.,  {Wyrowski} F.,  {Schuller} F.,   {Menten} K.~M.,  2013,
  \mn@doi [\aap] {10.1051/0004-6361/201219574}, \href
  {http://adsabs.harvard.edu/abs/2013A%26A...549A...5R} {549, A5}

\bibitem[\protect\citeauthoryear{{Sakai} et~al.,}{{Sakai}
  et~al.}{2013}]{sakai_2013}
{Sakai} T.,  et~al., 2013, \mn@doi [\apjl] {10.1088/2041-8205/775/1/L31}, \href
  {http://adsabs.harvard.edu/abs/2013ApJ...775L..31S} {775, L31}

\bibitem[\protect\citeauthoryear{{Sanhueza}, {Garay}, {Bronfman}, {Mardones},
  {May}  \& {Saito}}{{Sanhueza} et~al.}{2010}]{sanhueza_2010}
{Sanhueza} P.,  {Garay} G.,  {Bronfman} L.,  {Mardones} D.,  {May} J.,
  {Saito} M.,  2010, \mn@doi [\apj] {10.1088/0004-637X/715/1/18}, \href
  {http://adsabs.harvard.edu/abs/2010ApJ...715...18S} {715, 18}

\bibitem[\protect\citeauthoryear{{Shepherd}, {N{\"u}rnberger}  \&
  {Bronfman}}{{Shepherd} et~al.}{2004}]{shepherd_2004}
{Shepherd} D.~S.,  {N{\"u}rnberger} D.~E.~A.,   {Bronfman} L.,  2004, \mn@doi
  [\apj] {10.1086/381050}, \href
  {http://adsabs.harvard.edu/abs/2004ApJ...602..850S} {602, 850}

\bibitem[\protect\citeauthoryear{{Shepherd} et~al.,}{{Shepherd}
  et~al.}{2007}]{shepherd_2007}
{Shepherd} D.~S.,  et~al., 2007, \mn@doi [\apj] {10.1086/521331}, \href
  {http://adsabs.harvard.edu/abs/2007ApJ...669..464S} {669, 464}

\bibitem[\protect\citeauthoryear{{Simon}, {Jackson}, {Rathborne}  \&
  {Chambers}}{{Simon} et~al.}{2006a}]{simon_2006a}
{Simon} R.,  {Jackson} J.~M.,  {Rathborne} J.~M.,   {Chambers} E.~T.,  2006a,
  \mn@doi [\apj] {10.1086/499342}, \href
  {http://adsabs.harvard.edu/abs/2006ApJ...639..227S} {639, 227}

\bibitem[\protect\citeauthoryear{{Simon}, {Rathborne}, {Shah}, {Jackson}  \&
  {Chambers}}{{Simon} et~al.}{2006b}]{simon_2006b}
{Simon} R.,  {Rathborne} J.~M.,  {Shah} R.~Y.,  {Jackson} J.~M.,   {Chambers}
  E.~T.,  2006b, \mn@doi [\apj] {10.1086/508915}, \href
  {http://adsabs.harvard.edu/abs/2006ApJ...653.1325S} {653, 1325}

\bibitem[\protect\citeauthoryear{{Smith}, {Shetty}, {Beuther}, {Klessen}  \&
  {Bonnell}}{{Smith} et~al.}{2013}]{smith_2013}
{Smith} R.~J.,  {Shetty} R.,  {Beuther} H.,  {Klessen} R.~S.,   {Bonnell}
  I.~A.,  2013, \mn@doi [\apj] {10.1088/0004-637X/771/1/24}, \href
  {http://adsabs.harvard.edu/abs/2013ApJ...771...24S} {771, 24}

\bibitem[\protect\citeauthoryear{{Smith}, {Glover}, {Klessen}  \&
  {Fuller}}{{Smith} et~al.}{2016}]{smith_2016}
{Smith} R.~J.,  {Glover} S.~C.~O.,  {Klessen} R.~S.,   {Fuller} G.~A.,  2016,
  \mn@doi [\mnras] {10.1093/mnras/stv2559}, \href
  {http://adsabs.harvard.edu/abs/2016MNRAS.455.3640S} {455, 3640}

\bibitem[\protect\citeauthoryear{{Sokolov} et~al.,}{{Sokolov}
  et~al.}{2017}]{sokolov_2017}
{Sokolov} V.,  et~al., 2017, \mn@doi [\aap] {10.1051/0004-6361/201630350},
  \href {http://adsabs.harvard.edu/abs/2017A%26A...606A.133S} {606, A133}

\bibitem[\protect\citeauthoryear{{Tackenberg} et~al.,}{{Tackenberg}
  et~al.}{2014}]{tackenberg_2014}
{Tackenberg} J.,  et~al., 2014, \mn@doi [\aap] {10.1051/0004-6361/201321555},
  \href {http://adsabs.harvard.edu/abs/2014A%26A...565A.101T} {565, A101}

\bibitem[\protect\citeauthoryear{{Tan}, {Kong}, {Butler}, {Caselli}  \&
  {Fontani}}{{Tan} et~al.}{2013}]{tan_2013}
{Tan} J.~C.,  {Kong} S.,  {Butler} M.~J.,  {Caselli} P.,   {Fontani} F.,  2013,
  \mn@doi [\apj] {10.1088/0004-637X/779/2/96}, \href
  {http://adsabs.harvard.edu/abs/2013ApJ...779...96T} {779, 96}

\bibitem[\protect\citeauthoryear{{Vasyunina}, {Linz}, {Henning}, {Stecklum},
  {Klose}  \& {Nyman}}{{Vasyunina} et~al.}{2009}]{vasyunina_2009}
{Vasyunina} T.,  {Linz} H.,  {Henning} T.,  {Stecklum} B.,  {Klose} S.,
  {Nyman} L.-{\AA}.,  2009, \mn@doi [\aap] {10.1051/0004-6361/200811226}, \href
  {http://cdsads.u-strasbg.fr/abs/2009A%26A...499..149V} {499, 149}

\bibitem[\protect\citeauthoryear{{Walsh}, {Bertoldi}, {Burton}  \&
  {Nikola}}{{Walsh} et~al.}{2001}]{walsh_2001}
{Walsh} A.~J.,  {Bertoldi} F.,  {Burton} M.~G.,   {Nikola} T.,  2001, \mn@doi
  [\mnras] {10.1046/j.1365-8711.2001.04352.x}, \href
  {http://adsabs.harvard.edu/abs/2001MNRAS.326...36W} {326, 36}

\bibitem[\protect\citeauthoryear{{Wang}, {Zhang}, {Rathborne}, {Jackson}  \&
  {Wu}}{{Wang} et~al.}{2006}]{wang_2006}
{Wang} Y.,  {Zhang} Q.,  {Rathborne} J.~M.,  {Jackson} J.,   {Wu} Y.,  2006,
  \mn@doi [\apjl] {10.1086/508939}, \href
  {http://adsabs.harvard.edu/abs/2006ApJ...651L.125W} {651, L125}

\bibitem[\protect\citeauthoryear{{Wilson} \& {Matteucci}}{{Wilson} \&
  {Matteucci}}{1992}]{wilson_1992}
{Wilson} T.~L.,  {Matteucci} F.,  1992, \mn@doi [\aapr] {10.1007/BF00873568},
  \href {http://adsabs.harvard.edu/abs/1992A%26ARv...4....1W} {4, 1}

\bibitem[\protect\citeauthoryear{{Wilson} \& {Rood}}{{Wilson} \&
  {Rood}}{1994}]{wilson_1994}
{Wilson} T.~L.,  {Rood} R.,  1994, \mn@doi [\araa]
  {10.1146/annurev.aa.32.090194.001203}, \href
  {http://cdsads.u-strasbg.fr/abs/1994ARA%26A..32..191W} {32, 191}

\bibitem[\protect\citeauthoryear{{Wolszczan}, {Cordes}  \& {Dewey}}{{Wolszczan}
  et~al.}{1991}]{wolszczan_1991}
{Wolszczan} A.,  {Cordes} J.~M.,   {Dewey} R.~J.,  1991, \mn@doi [\apjl]
  {10.1086/186033}, \href {http://adsabs.harvard.edu/abs/1991ApJ...372L..99W}
  {372, L99}

\bibitem[\protect\citeauthoryear{{Xu}, {Li}, {Zhang}, {Liu}, {Wang}, {Ning}  \&
  {Ju}}{{Xu} et~al.}{2016}]{xu_j_2016}
{Xu} J.-L.,  {Li} D.,  {Zhang} C.-P.,  {Liu} X.-L.,  {Wang} J.-J.,  {Ning}
  C.-C.,   {Ju} B.-G.,  2016, \mn@doi [\apj] {10.3847/0004-637X/819/2/117},
  \href {http://adsabs.harvard.edu/abs/2016ApJ...819..117X} {819, 117}

\bibitem[\protect\citeauthoryear{{Yanagida} et~al.,}{{Yanagida}
  et~al.}{2014}]{yanagida_2014}
{Yanagida} T.,  et~al., 2014, \mn@doi [\apjl] {10.1088/2041-8205/794/1/L10},
  \href {http://adsabs.harvard.edu/abs/2014ApJ...794L..10Y} {794, L10}

\bibitem[\protect\citeauthoryear{{Zamora-Avil{\'e}s}, {Ballesteros-Paredes}  \&
  {Hartmann}}{{Zamora-Avil{\'e}s} et~al.}{2017}]{zamora-aviles_2017}
{Zamora-Avil{\'e}s} M.,  {Ballesteros-Paredes} J.,   {Hartmann} L.~W.,  2017,
  preprint, \href {http://adsabs.harvard.edu/abs/2017arXiv170801669Z} {}
  (\mn@eprint {arXiv} {1708.01669})

\bibitem[\protect\citeauthoryear{{Zhang}, {Yuan}, {Li}, {Zhou}  \&
  {Wang}}{{Zhang} et~al.}{2017}]{zhang_2017}
{Zhang} C.-P.,  {Yuan} J.-H.,  {Li} G.-X.,  {Zhou} J.-J.,   {Wang} J.-J.,
  2017, \mn@doi [\aap] {10.1051/0004-6361/201629771}, \href
  {http://adsabs.harvard.edu/abs/2017A%26A...598A..76Z} {598, A76}

\bibitem[\protect\citeauthoryear{{Zucker}, {Battersby}  \& {Goodman}}{{Zucker}
  et~al.}{2015}]{zucker_2015}
{Zucker} C.,  {Battersby} C.,   {Goodman} A.,  2015, \mn@doi [\apj]
  {10.1088/0004-637X/815/1/23}, \href
  {http://adsabs.harvard.edu/abs/2015ApJ...815...23Z} {815, 23}

\makeatother
\end{thebibliography}

\appendix

\section{The kinematic structure of Cloud F}\label{Appendix D}

The kinematic structure of Cloud F is discussed in this section with reference to previous work on this source. For reference, the \citet{rathborne_2006} and \citet{butler_2012} core regions mentioned in the follow section are labeled on Figure\,\ref{cloudf_msd}. We also give a note on the physical interpretation of structures identified in molecular line observations. 

\subsection{The F1/MM8 region}

Towards the south-west of the mapped region, the F1/MM8 core region, we find two distinct velocity components: F$_{\rm PPV2}$ and F$_{\rm PPV4}$, both seen in the \ceooz\ emission. As shown in Figure\,\ref{spec_plot_cloudf} (right panel), the higher velocity of these components shows a factor of two narrower line width, with respect to the mean value of this component (line width towards this region and mean width of the F$_{\rm PPV4}$ component are $\sim$\,1\,\kms\ and $\sim$1.8\,\kms, respectively). 

A similar double-peaked line profile may, however, be produced as a result of optically depth. If the emission were optically thick, it would be self-absorbed at the mean centroid velocity of the region (as traced by optically thin emission). Unfortunately, significant \ntwohoz\ emission is not observed towards this region, however, another high-density tracing, optically thin molecular line transition, N$_2$D$^+$\,($3-2$) has been shown to have emission at velocities coincident with only the lower velocity component \citep{pon_2016a, tan_2013}. This is not expected if these components were produced by optical depth, as instead, the optically thin emission would have a centroid velocity at the centre of these components (see section\,\ref{grs_comparison} for a discussion). 

The depletion of CO-bearing molecules, such as \ceo, onto the dust grains within dense, cold environments, may also artificially produce multiple velocity components. However, a dip in the emission profile of the CO-bearing molecule emission (e.g. \ceo) would typically be seen at the centroid velocity of the emission from a non-CO bearing molecules (such as N$_2$D$^+$), which we again do not see \citep{tan_2013}. Furthermore, we calculate the level of CO depletion throughout Cloud F and find an average value towards this region of 1.3\,$\pm$\,0.1 (no depletion of CO would be represented by a value of unity), which we believe is not significant enough to cause this effect (see Appendix\,\ref{Appendix B}). We, therefore, find that two distinct velocity components with different line profiles are indeed present along the line of sight towards this region, pointing to different internal conditions within these components. 

Possible formation scenarios for the interesting structure observed towards this region have been discussed in a series of papers by \citet{pon_2015, pon_2016a, pon_2016b}. These authors identify both narrow and broad velocity components in JCMT mid-J transition $^{13}$CO observations (with a spatial resolution of $\sim$\,11\arcsec) towards this region, showing similar centroid velocities to the F$_{\rm PPV2}$ and F$_{\rm PPV4}$ components identified here. \citet{pon_2016a} suggest that protostars associated with the 24\,\micron\ source just to the north of the F1 core have created a wind-blown bubble, where the broader, lower velocity component ($\sim$\,56\kms, F$_{\rm PPV2}$ here) traces the compressed gas within the bubble wall. This lower velocity component is also seen in high-J CO transitions (e.g. $J=8\rightarrow7$ upwards; \citealp{pon_2015}). Comparison between the PDR models and the high-J transition emission shows that there may be a hot gas component (of around $\sim$\,100\,K) within this region, which could have been formed by the dissipation of turbulence as this component is compressed within the shell \citep{pon_2015, pon_2016b}. The origin of the narrow velocity component F$_{\rm PPV4}$ is, however, still unknown.

\subsection{The F4/MM3 region}

Initial molecular line studies towards the F4/MM3 region showed it to be cold, dense and quiescent, hence an ideal region to study the initial stages of massive star formation \citep{garay_2004}. Several more recent studies have, however, found signs that protostars could be already present within this region (e.g. \citealp{foster_2014}). For example, there is a clear point source towards this region in the {\it Spitzer} and {\it Herschel} images, which can be seen as a negative mass surface density values in Figure\,\ref{moment_maps_cloudf}. Indeed, \citet{chambers_2009} found two sources within the MM3 region which could be classified as a ``green fuzzie'' from their excess of 4.5\micron\ emission, a signpost of heated dust by embedded protostars, and Cosentino et al. (accepted) have found bright and broad SiO emission toward this core, indicative of an outflow (also see \citealp{sakai_2013, yanagida_2014}). These authors also detect water and methanol maser emission towards both of these sources, suggesting that massive protostars may be present within this region (e.g. \citealp{walsh_2001}). \cite{wang_2006} identified that the water maser emission towards this core has a single component, which is red-shifted by $\sim$\,20\kms\ with respect to the molecular gas at $\sim$\,55\kms, suggesting that the embedded protostar(s) within this region have already begun to influence the kinematics of the surrounding gas. \citet{sanhueza_2010} found that the optically thick emission from CO ($3-2$) towards these regions have blue- and red-shifted lobes of $\pm$\,15\kms\ around the mean centroid velocity of optically thin lines, such as \ceo\ and CS (similarly broad profiles were found by \citealp{rathborne_2005}). These authors suggest that such profiles are the result of a molecular outflow with a total mass of $\sim$\,40\sol. Using the lower mass limit of the embedded sources within this region from \citet{shepherd_2007},\footnote{The total mass of the sources with SED fits towards this region -- IDs ``25'', ``26'', ``27'', ``28'', ``29'' and ``30'' -- is $\sim$\,10\sol. We note that the uncertainty on this value, and on the total embedded stellar mass estimate, could be larger than a factor of three (see \citealp{barnes_2017} for a discussion of the uncertainties present when determining embedded stellar masses).} and extrapolated with a Kroupa IMF, we estimate that the total mass of protostars within this region is comparable to the mass of the molecular outflow. 

Towards the F4/MM3 region, we find a relatively simple velocity structure of only a single velocity component, F$_{\rm PPV4}$. However, in light of the above discussion, it is possible that this kinematic structure has been influenced by protostellar feedback and/or is causing the star formation within this region. Indeed, we find a systematic offset between the \ntwoh\ emission towards higher velocities, with respect to the \ceo\ emission (of $\sim$\,0.2\,\kms; see section\,\ref{shift}). This red-shift is not, however, as large as the red-shift lobe of the optically thick CO emission from \citet{sanhueza_2010} or the water maser emission from \cite{wang_2006}. Interestingly, recent high-resolution \ntwod\,($3-2$) ALMA observations towards the F4 region show emission at a velocity of 57.10\,$\pm$\,0.05\,\kms\ \citep{kong_2017}, which is offset from the \ntwohoz\ and \ceo\ emission by $>$\,0.8\,\kms\ for the same position (58.67\,$\pm$\,0.07\,kms and 57.90\,$\pm$\,0.02\,kms, respectively). Moreover, despite the evidence for an outflow within the region, observations of optically think molecular lines (HCO$^{+}$, HCN; \citealp{zhang_2017}) towards this region show asymmetric line profile characteristic of infall motions \citep{evans_1999}. This suggests that the active star-forming region within this northern portion of Cloud F, F2/MM3, is currently accreting material from the gas reservoir of the cloud.

\subsection{The MM1 and MM2 regions}

We observe the most complex spectra towards the south of the cloud, which at some positions show three velocity components along the line of sight: F$_{\rm PPV1}$, F$_{\rm PPV2}$ and F$_{\rm PPV4}$. This region is referred to as the MM1 region \citep{rathborne_2005, rathborne_2006}, and is located approximately 40\arcsec (or $\sim$\,0.75\,pc, assuming a source distance of 3.7\,kpc; \citealp{simon_2006b}), from the MM2 region, also known as the IRAS 18507+0121 (refer to Figure\,\ref{cloudf_msd}). The MM1 region is thought to host a young, embedded protostar, with a spectral type B2 \citep{shepherd_2004, rathborne_2008b, shepherd_2007}, whereas the MM2 region is thought to be more massive and evolved, harbouring a B0.5 class star surrounded by an ultracompact HII region \citep{miralles_1994, molinari_1996}. These sources are thought to have a combined mass of $\sim$\,50\sol, and are driving a molecular outflow of $\sim$\,100\sol \citep{shepherd_2007}. When extrapolated using a Kroupa IMF the total embedded stellar mass within these regions is $\sim$\,200\sol. As with the MM3 region, despite the on-going star formation within MM1 and MM2, there is evidence to show large-scale infall motions towards these regions \citep{ramesh_1997, sanhueza_2010, zhang_2017}. Indeed, the spectral profiles of the optically thick \tco\ emission towards these regions show asymmetric profiles with enhanced blue-shifted peaks (see section\,\ref{grs_comparison}). Again, this is suggesting that this active star-forming region is accreting material over scales of up to 2\,pc, given the approximate extent of the double-peaked, blue-shifted profile seen in the \tco\ emission (see Figure\,\ref{spec_plot_cloudf_grs}).

\section{A note on the physical interpretation of velocity components}\label{Appendix E}

In this section, we would like to briefly mention the relation between the observed position-position-velocity space and the physical position-position-position space. As throughout this work, we make the assumption that the identified velocity components correspond directly to physical structures within the cloud, however, this has recently been suggested to have several caveats within low-mass star-forming regions (e.g. \citealp{ballesteros-paredes_2002, smith_2016}). For example, \citet{zamora-aviles_2017}, used observational techniques to analyse a molecular cloud within a three dimensional, magnetohydrodynamic simulation and found that unassociated density enhancements can artificially appear as single coherent structures when superposed along the line-of-sight, particularly towards low-density regions. It is not clear, however, how these simulations apply to the massive star-forming regions, which typically have higher densities than their lower mass counterparts. In this work, we have identified velocity component(s) which are coherent across several parsecs within both moderate and higher density molecular line tracers. It would seem unlikely that these extended structures could be produced by the superposition of low-density random fluctuations. 
\section{Reduction and analysis of the IRAM-30m observations towards Cloud H (G035.39-00.33)}\label{Appendix A}

One of the aims of this work is a comparison of the kinematic structure of Cloud F (G034.43+00.24) and Cloud H (G035.39-00.33; \citealp{butler_2009}). The kinematic analysis of the \ceooz\ and \ntwohoz\ IRAM-30m observation has already been conducted by \citet{henshaw_2013}, yet we choose to analyse these data using the same techniques as Cloud F, such that no artificial differences are introduced into the comparison. 

\subsection{Observations}\label{sec:observations,appendix}

\begin{figure}
\centering
\includegraphics[trim = 3mm 3mm 3mm 3mm, clip,angle=0,width=1\columnwidth]{\dir 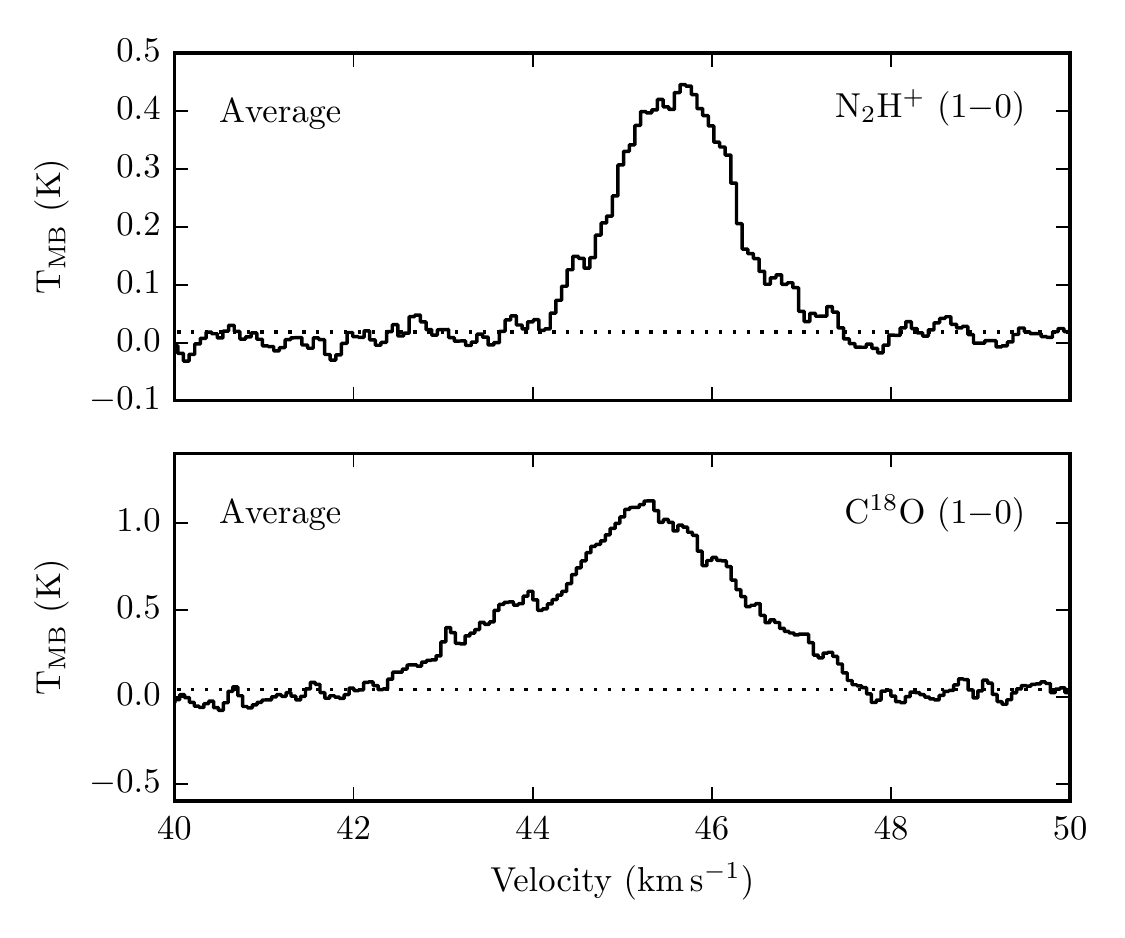}  \vspace{-3mm}
\caption{Shown are the average spectrum of the \ntwohoz\ transition (upper) and the \ceooz\ transition (lower) across the mapped region of Cloud H. The horizontal dotted line represents the {\it rms} level of $\sim$\,0.02\,K and $\sim$\,0.04\,K for \ntwohoz\ and \ceooz, respectively.} 
\label{spec_ave_plot_cloudf}
\end{figure}

Details of the \ceooz\ and \ntwohoz\ observations towards Cloud H are presented in Table 1 of \citet{henshaw_2013}. Here, we smooth these observations to an angular resolution of 36\arcsec, with a pixel spacing of 18\arcsec, such that they have a spatial resolution of $\sim$\,0.5\,pc at the source distance of 2.9\,kpc \citep{simon_2006b}. This was done to match the spatial resolution of the Cloud F observations, the results of which can be found in section\,\ref{comparison_to_cloudh}.

\subsection{Gaussian fitting and hierarchical clustering}\label{subsection:spectral line fitting and linking cloud h}

\begin{figure*}
\centering
\includegraphics[trim = 4.9cm 10mm 6.3cm 9mm, clip,angle=0,width=1\columnwidth]{\dir 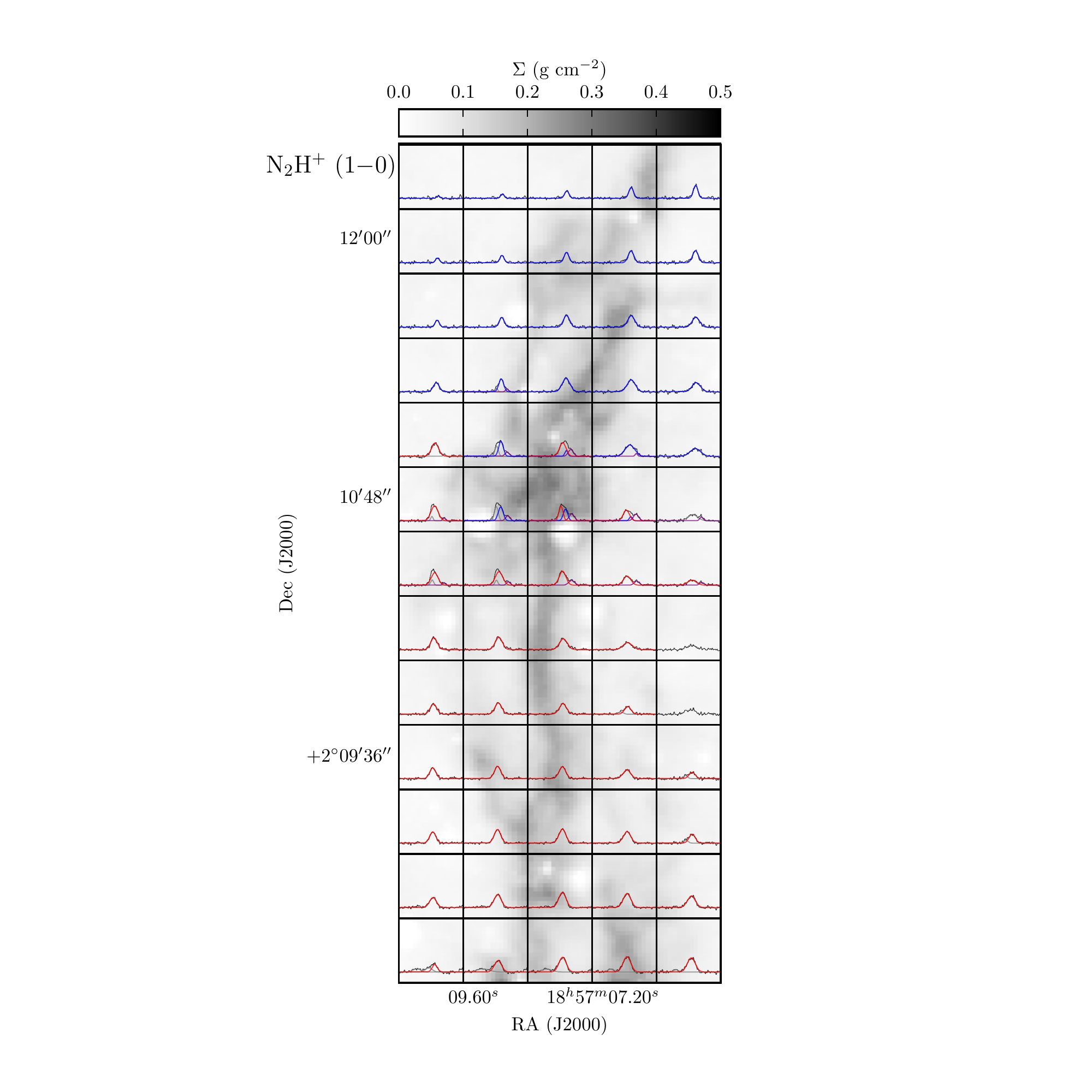}
\hspace{3mm}
\includegraphics[trim = 4.9cm 10mm 6.3cm 9mm, clip,angle=0,width=1\columnwidth]{\dir 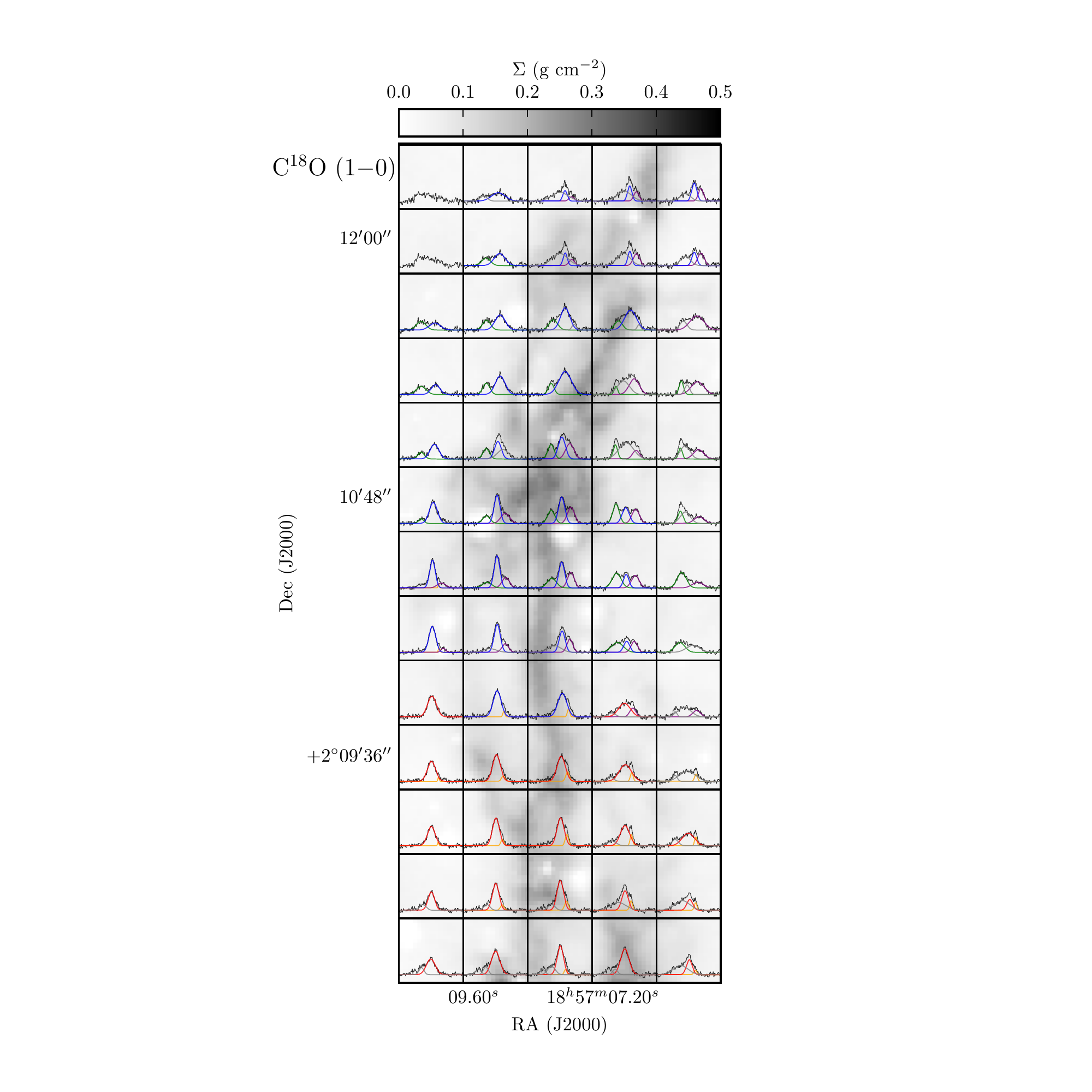}
\caption{The upper two panels show the average spectrum of the \ntwohoz\ transition (left) and the \ceooz\ transition (right) across the mapped region of Cloud H. The horizontal dotted line represents the {\it rms} level of $\sim$\,0.03\,K and $\sim$\,0.07\,K for \ntwohoz\ and \ceooz, respectively. The lower two panels show the spectra at each pixel position across the cloud (shown in black). The velocity ranges are 40 to 50\,\kms, and the intensity ranges are -0.5 to 2.5\,K for \ntwoh and -0.5 to 3.5\,K for \ceo. Overlaid on each spectrum are the results of the line fitting ({\sc scouse}) and clustering ({\sc acorns}) routines, which are discussed in section\,\ref{subsection:spectral line fitting and linking cloud h}. The colours of these profiles represent the various velocity component associations \change{given in Table\,\ref{velocity component parameters}}. The background greyscale is the mass surface density map \citep{kainulainen_2013}.}
\label{spec_plot_cloudh}.
\end{figure*}

\begin{figure*}
\centering
\includegraphics[trim =  46mm 33mm 48mm 25mm, clip,angle=0, clip,angle=0,width=1\columnwidth]{\dir 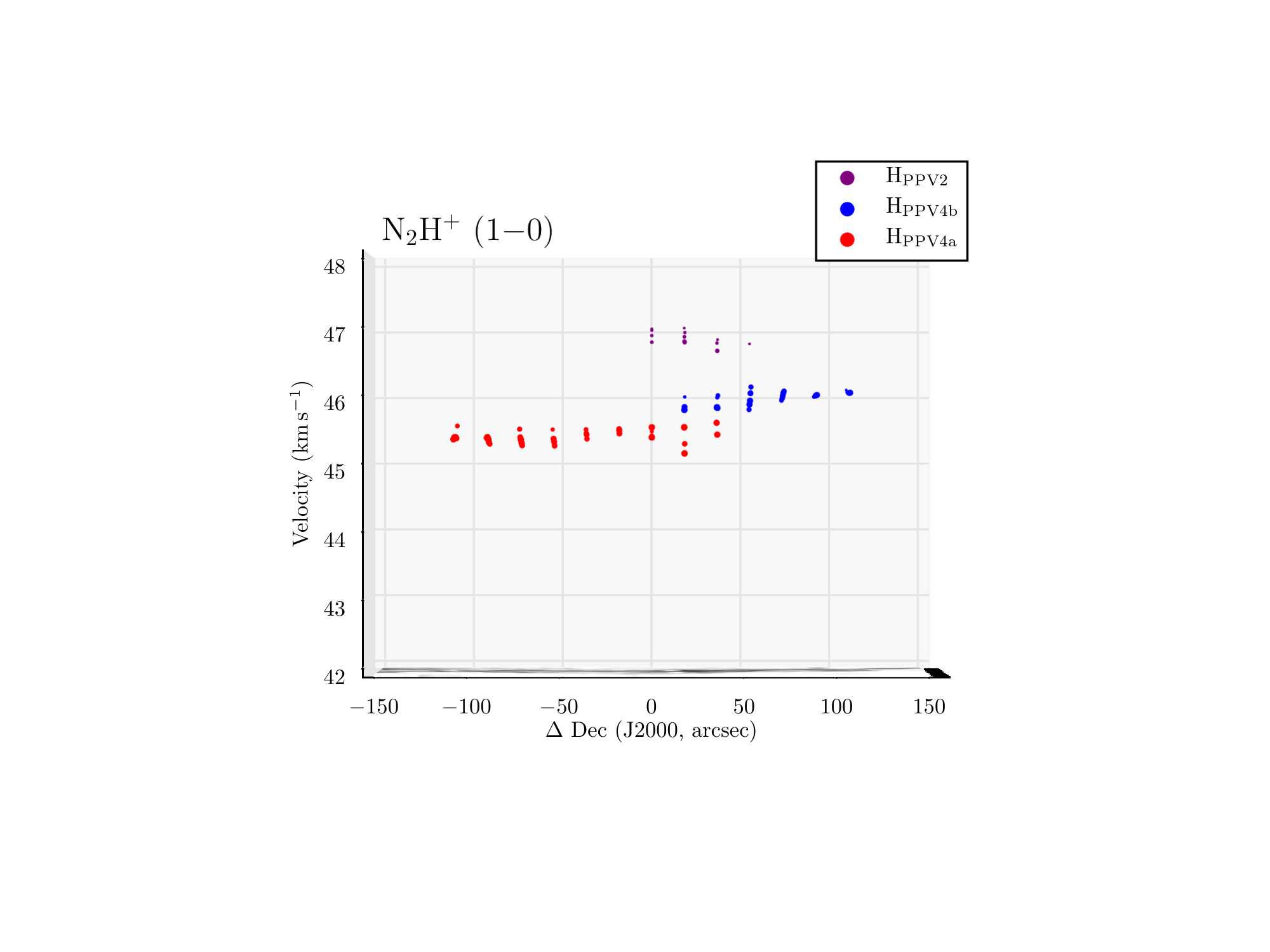} 
\includegraphics[trim =  46mm 33mm 48mm 25mm, clip,angle=0, clip,angle=0,width=1\columnwidth]{\dir 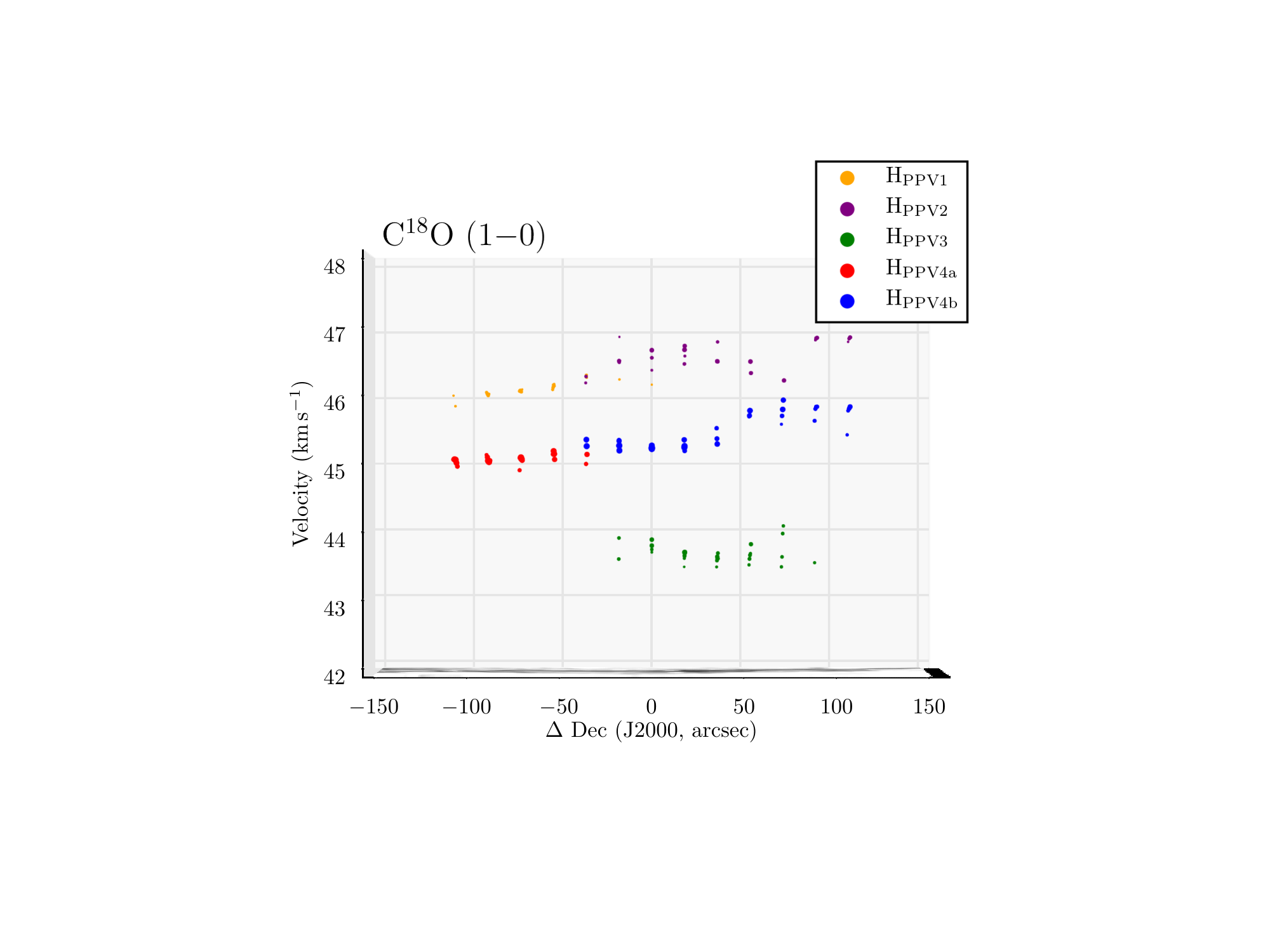}

\includegraphics[trim = 86mm 33mm 30mm 34mm, clip, angle=0, height=0.7\columnwidth]{\dir 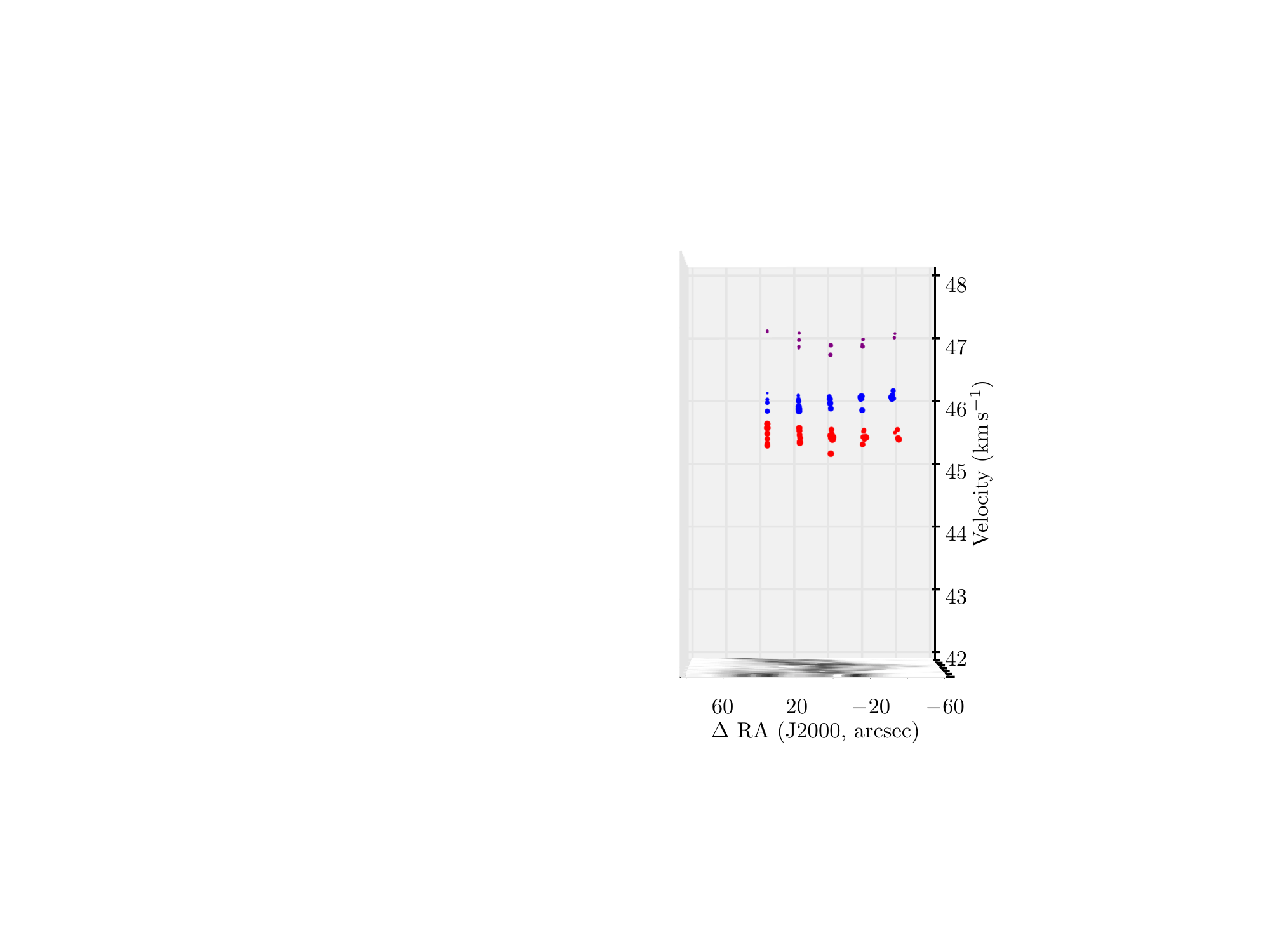} 
\includegraphics[trim = 86mm 33mm 30mm 34mm, clip, angle=0, height=0.7\columnwidth]{\dir 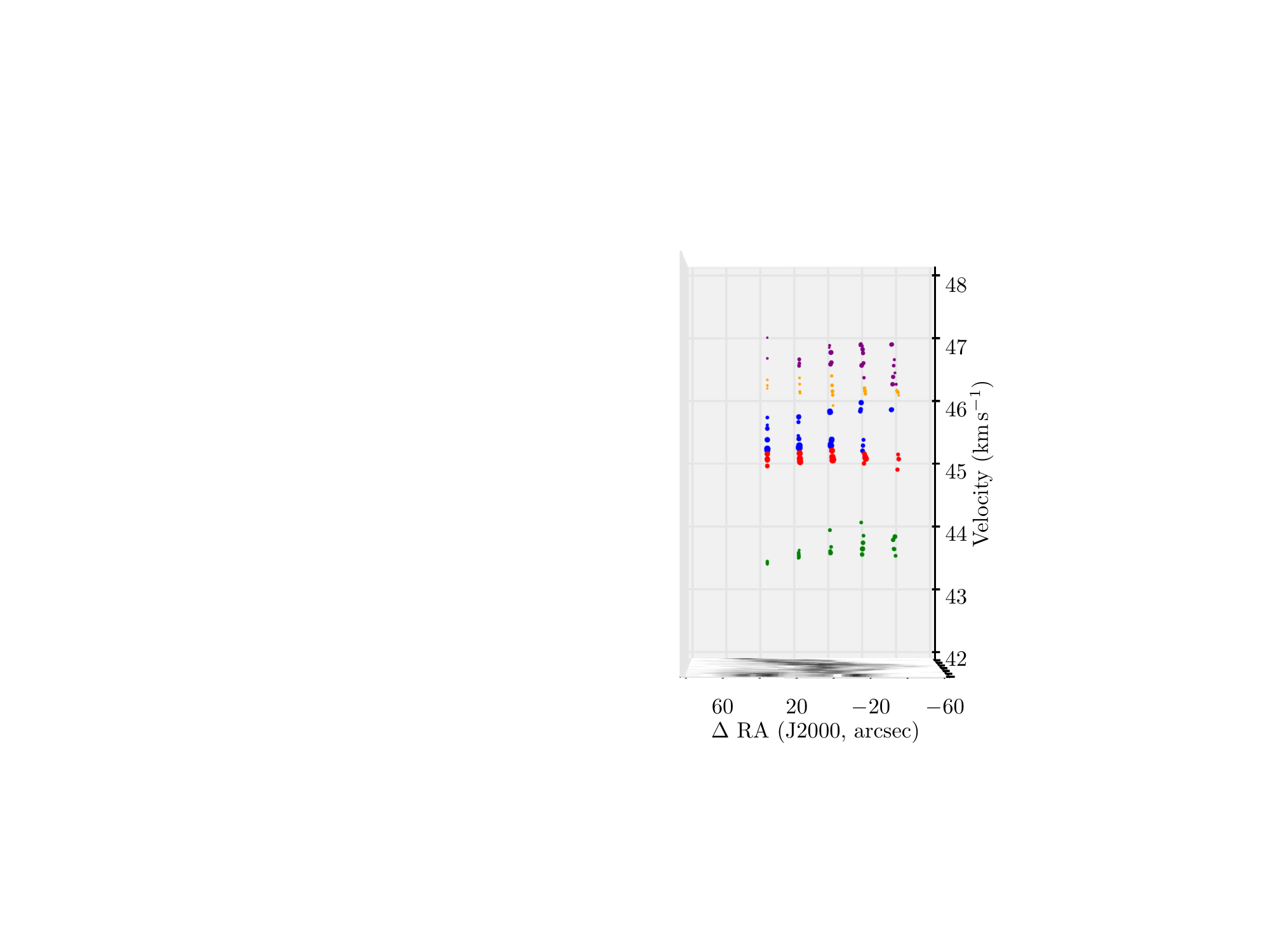}

\includegraphics[trim = 53mm 12mm 25mm 42mm, clip,angle=0, clip,angle=0,width=1.03\columnwidth]{\dir 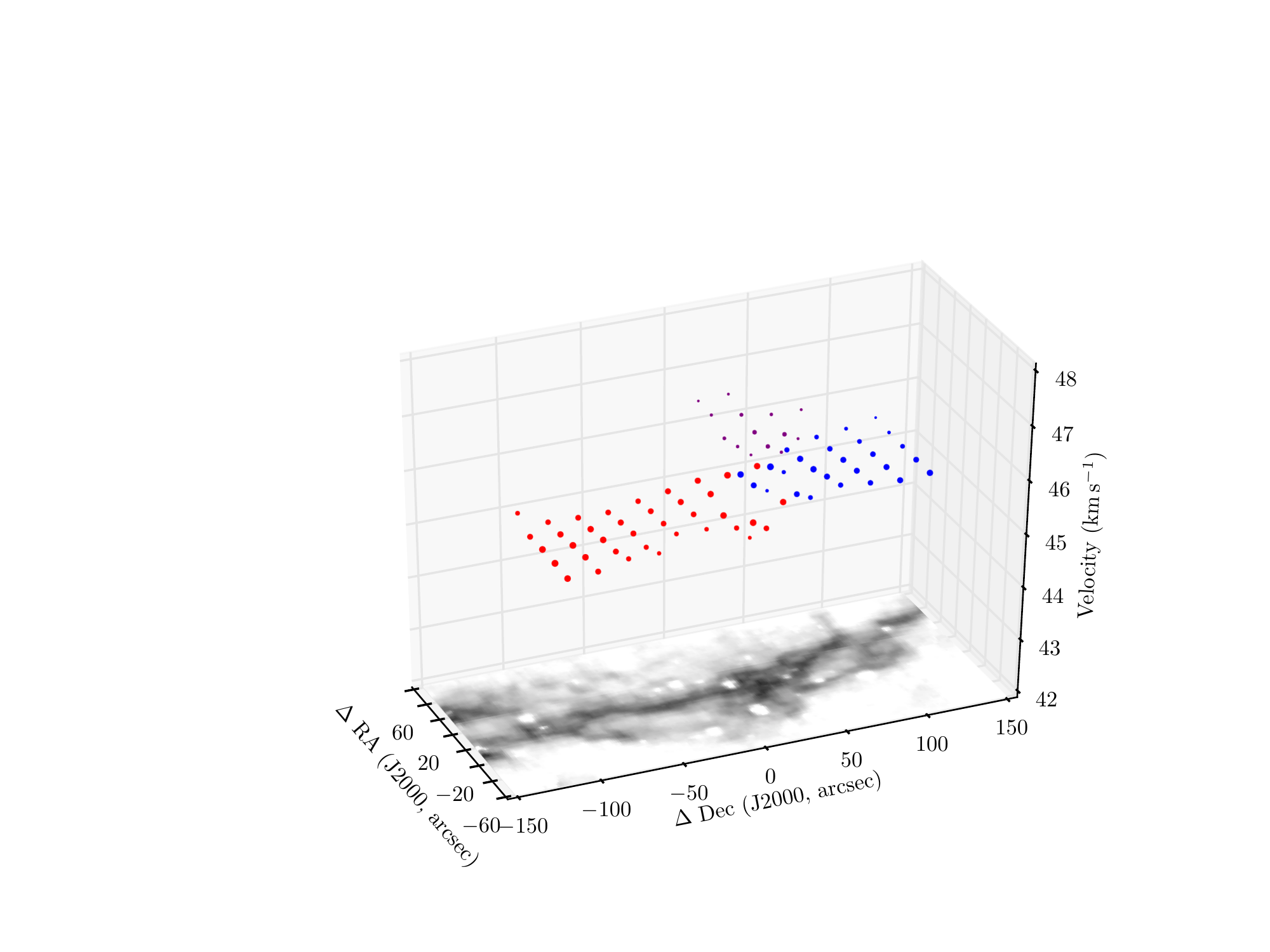} 
\includegraphics[trim = 53mm 12mm 25mm 42mm, clip,angle=0, clip,angle=0,width=1.03\columnwidth]{\dir 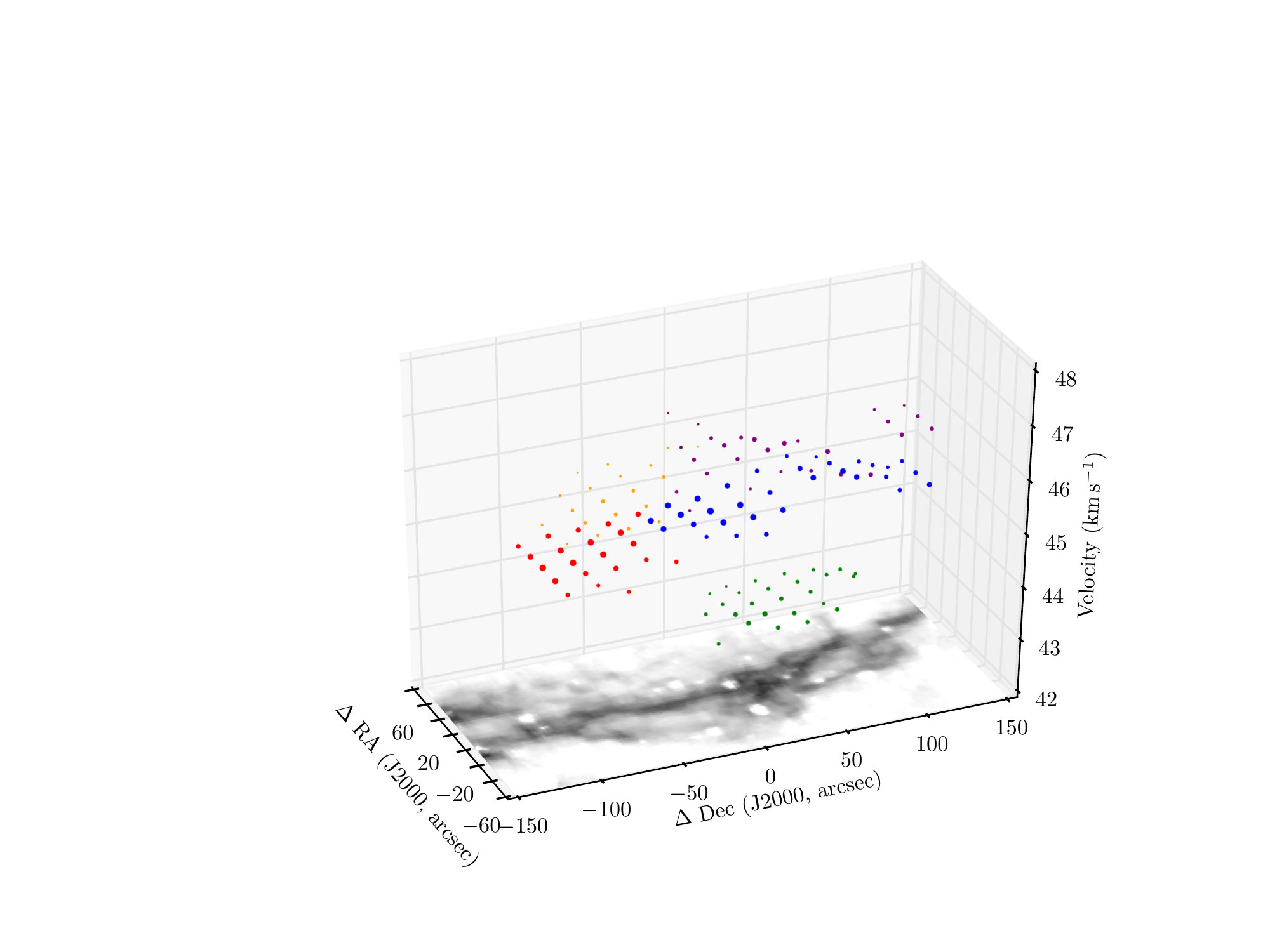}
\caption{Displayed in each panel is the position-position-velocity diagram of Cloud H, shown at various viewing angles. The left and right panels show \ntwohoz\ and \ceooz\ results, respectively. The colour of each point represents the association to one of the distinct coherent velocity components identified using the clustering algorithm {\sc acorns} (Henshaw et al. in prep), H$_{\rm PPV1}$ in orange, H$_{\rm PPV2}$ in purple, H$_{\rm PPV3}$ in green, and H$_{\rm PPV4a}$ in red and H$_{\rm PPV4b}$ in blue. The size of each point represents its relative peak intensity. The mass surface density map of \citet{kainulainen_2013} is shown on the base of each plot. Note, the coordinate offsets of these plots are relative to the centre of the mapped region: RA\,(J2000)\,=\,18$^h$57$^m$08$^s$, Dec\,(J2000)\,=\,02$^{\circ}$10$'$12${''}$ ({\it l} = 35.512$^{\circ}$, {\it b} = -0.277$^{\circ}$).}
\label{ppv_plots_cloudh}
\end{figure*}

Figure\,\ref{spec_ave_plot_cloudf} shows the average spectrum for the \ntwohoz\ hyperfine component and \ceooz\ transitions across Cloud H. We find that the majority of the emission above the {\it rms} levels is between approximately $43 - 57$\,\kms. Figure\,\ref{spec_plot_cloudh} shows the spectra at each pixel position across the cloud, plotted using the same velocity range as the average spectra. As with Cloud F, several spectra across the mapped region, particularly in \ceooz, appear to have more than one peak, and show that multiple velocity components are present along the line of sight.

We use the Gaussian profile fitting algorithm {\sc scouse} to separate the velocity components within the spectra, then the hierarchical clustering routine {\sc acorns} to identify the coherent velocity structures across the cloud. The same spectral averaging area (SAA) radius and threshold values in {\sc scouse} that were used for Cloud F were used for Cloud H. These gave reasonable fits, and $<10$ per cent of the data had to be checked and re-fitted. The same input parameters as used in Cloud F were in {\sc acorns} for the identification of the initial hierarchy. The parameter space survey of the relaxation factors, however, showed that for values of 0.5, 1.25, 1.0, for the peak intensity, centroid velocity and line width, respectively, were required to identify the most robust structures.\footnote{We note that, choosing the same relaxation factors as Cloud F did not significantly change the main structures in Cloud H, rather these cause {\sc acorns} to identify additional structures using the lower peak intensity positions throughout the cloud.} The results of the Gaussian fitting and structure finding routines are presented in Figures\,\ref{spec_plot_cloudh} and \ref{ppv_plots_cloudh}. 

We identify five structures in the \ceooz\ emission, defined as H$_{\rm PPV1}$, H$_{\rm PPV2}$, H$_{\rm PPV3}$, and H$_{\rm PPV4a}$ and H$_{\rm PPV4b}$ (shown in orange, purple, green, red and blue), three of which are also identified in the \ntwohoz\ emission. We choose to define H$_{\rm PPV4a}$ and H$_{\rm PPV4b}$ in this way, as, although they have been defined as separate structures (see \citealp{henshaw_2014}), previous single dish studies have defined them as one \citep{henshaw_2013, jimnez-serra_2014}. The basic properties are given in Table\,\ref{velocity component parameters}.    
\section{CO depletion within Cloud F}\label{Appendix B}


The depletion of CO onto dust-grain surfaces occurs in the coolest, densest regions of IRDCs, and therefore it is a signpost for material at the earliest phases of star formation, away from the effects of stellar feedback. This could artificially give the appearance of multiple velocity components, as the emission at the centroid velocity of the source could be reduced, mimicking optical depth effects. Here we calculate the column density and abundance of \ceo\ within Cloud F, which are used to estimate the average CO depletion. 

The column density, $N$(\ceo), is calculated following the procedure outlined by \citet{caselli_2002a}, assuming the \ceooz\ emission is optically thin. The column density of hydrogen is given as $N$(H$_2$)\,=\,$\Sigma$\,/\,$\mu_p$\,$ m_{\rm H}$, where $\mu_p$\,=\,2.33 a.m.u is the mean mass per particle, $m_{\rm H}$ is the mass of hydrogen, and $\Sigma$ is the mass surface density taken from \citet{kainulainen_2013}. The abundance of \ceo\ with respect to H$_2$ is calculated as $X$(\ceo)\,=\,$N$(\ceo)\,/\,$N$(H$_2$). To determine the abundance of CO we use the oxygen isotope ratio $^{16}$O\,/\,$^{18}$O\,=\,58.5\,$d_{\rm GC}$\,+\,37.1~$\sim$~372 \citep{wilson_1994}, given the Galactocentric distance of Cloud F, $d_{\rm GC}$\,$\sim$\,5.7\,kpc (assuming source distance of 3.7\,kpc and a distance to the galactic centre of $\sim$\,8.3\,kpc; \citealp{simon_2006b, reid_2014}). We find an average CO abundance across the cloud of X(CO)\,=\,1.3\,$\times$\,10$^{-4}$. Comparing this measured CO abundance to the reference (or ``expected'') value, X$^{ref}$(CO), gives the CO depletion factor, $f_{\rm D}$\,=\,$X^{\rm ref}$(CO)\,/\,X(CO). Using the abundance gradients in the Galactic Disk from \citealp{wilson_1992} and the Solar neighbourhood abundance of CO from \citet{frerking_1982}, \citet{fontani_2006} find that the reference abundance of CO is given by $X^{\rm ref}$(CO)\,=\,9.5\,$\times$\,10$^{-5}$\,exp(1.105\,-\,0.13\,$d_{\rm GC}$), which for the Galactocentric distance of Cloud F is $\sim$\,1.4\,$\times$\,10$^{-4}$. We find that the average CO depletion factor across Cloud F is $\sim$\,1.2, which peaks with a value of $\sim$\,2.1 towards the MM3 core region (see Figure\,\ref{cloudf_msd}). These values are in the range previously calculated by \citet{hernandez_2011a} using \tco\ emission towards Cloud F, albeit these authors used a slightly higher value of the reference abundance of $\sim$\,2\,$\times$\,10$^{-4}$. We note \citet{pon_2016b} find higher CO depletion values than observed in this work ($f_{\rm D}$\,=\,$5-9$), using higher resolution, higher CO J-transition observations with the James Clerk Maxwell Telescope. These observations are, however, more sensitive to the higher density gas, where CO is expected to be more depleted. 

In summary, here we have shown that Cloud F contains only a moderate level of CO depletion. Therefore, artificially split line profiles are not expected to contaminate the velocity component analysis from the \ceooz\ emission.
\section{Analysis of the GRS observations towards Cloud F (G0.34.43+00.24)}\label{Appendix C}

\subsection{Observations}\label{sec:observations,appendix}

Observations covering a large scale of Cloud F have been taken as part of the Galactic Ring Survey \citep[GRS][]{jackson_2006}. These \tcooz\ observations have an angular resolution of $\sim$\,44\arcsec\ and a spectral resolution of $\sim$\,0.2\kms -- factors of $\sim$\,1.5 and $\sim$\,3 larger than the (smoothed) IRAM-30m observations. The data are publicly available from \url{https://www.bu.edu/galacticring/new_data.html}, from which we take the data cube over the region $34<l<36^{\circ}$, $|b|<1^{\circ}$, $0< \upsilon< 100$\,\kms. This $2^{\circ} \times 2^{\circ}$ image is significantly larger than required, hence we trim the image to a $\sim$\,1300\arcsec $\times$ 300\arcsec\ region covering the filamentary structure identified in the mass surface density map shown in Figure\,\ref{cloudf_msd}.    

\subsection{Gaussian fitting and hierarchical clustering}

\begin{figure*}
\centering
\includegraphics[trim = 0mm 0mm 0mm 0mm, clip,angle=0,width=1\columnwidth]{\dir 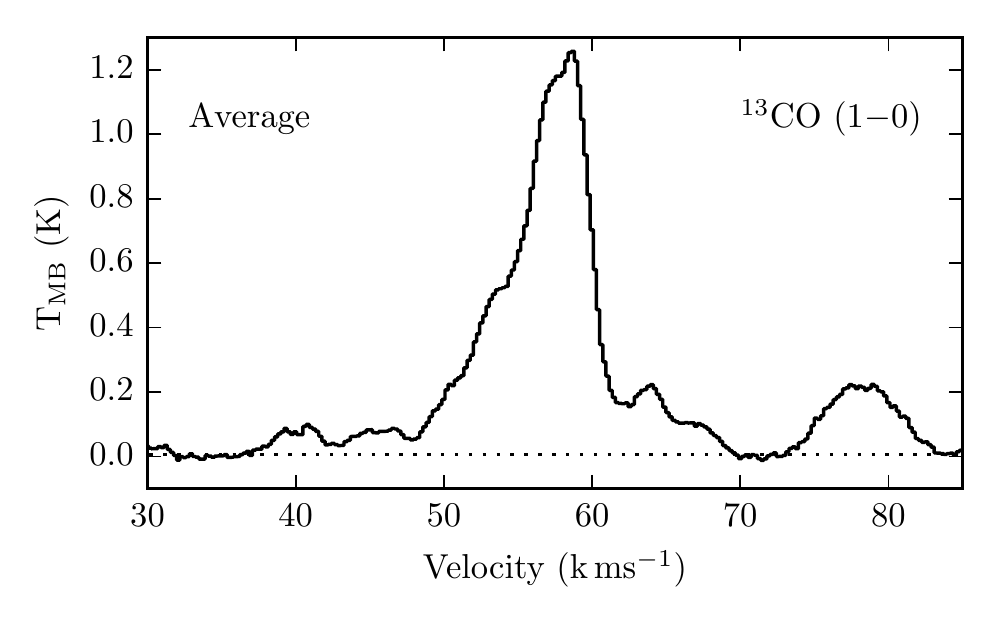}
\includegraphics[trim = 90mm 18mm 95mm 26mm, clip,angle=0, height=0.94\textheight]{\dir 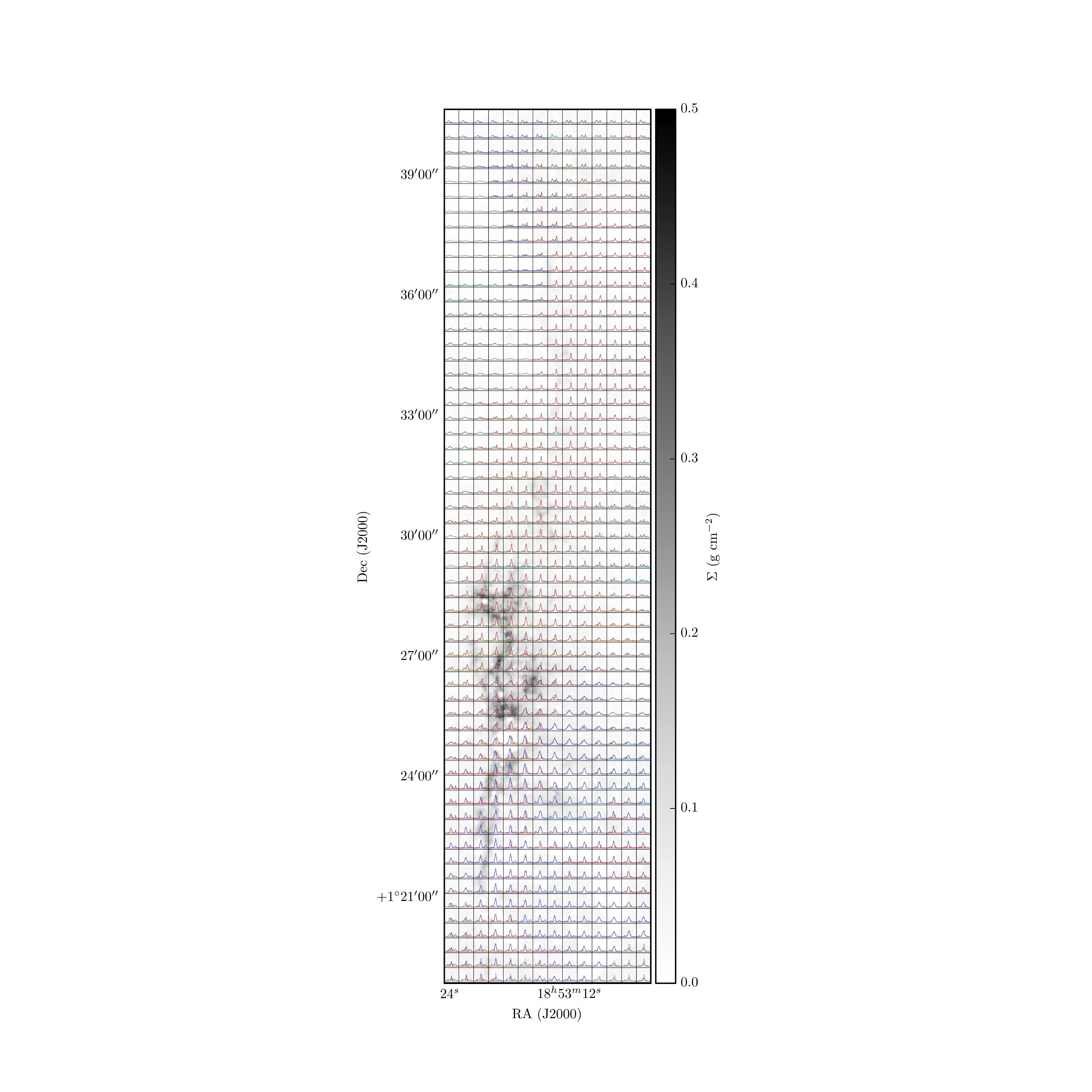}\vspace{-3mm}
\caption{The left panel shows the average spectrum of the \tcooz\ transition from the GRS. The horizontal dotted line represents the {\it rms} detection thresholds of $\sim$\,0.006\,K. The right panel shows the spectra at each pixel position across Cloud F, overlaid with coloured profiles of the various velocity components. The velocity ranges are 45 to 70\,\kms, and the intensity ranges are -0.5 to 5.0\,K. The background greyscale is the mass surface density map \citep{kainulainen_2013}.} 
\label{spec_plot_cloudf_grs}
\end{figure*}

\begin{table*}
\caption{Parameters of the velocity components identified in the GRS observations towards Clouds F (F$_{\rm GRS}$). Shown is the molecule used to identify the components (for consistency with Table\,\ref{velocity component parameters}), and for each component: the name, the total number of points, the average centroid velocity, the average line width, the velocity gradient and the angle of this gradient with respect to East of North. When the uncertainty on the velocity gradient is larger than or equal to the calculated velocity gradient, the velocity gradient angle is unconstrained, and therefore not shown.}
\centering
\begin{tabular}{ccccccccc}
\hline
Line &  Component & $\#$ points & Centroid velocity & Line width & Velocity gradient  & Gradient angle \\ 
& & & ($V_0$) \kms &  ($\Delta \upsilon$)  \kms & ($\nabla v$) \kms\,pc$^{-1}$ & ($\theta_{\nabla v}$) degrees \\
\hline
\tcooz \\
& F$_{\rm GRS1}$ & 297 & 56.80  \,$\pm$\, 0.39 & 3.69  \,$\pm$\, 1.17 & 0.02  \,$\pm$\, 0.01 & -80.85  \,$\pm$\, 9.57 \\
& F$_{\rm GRS2}$ & 498 & 58.91  \,$\pm$\, 0.71 & 2.11  \,$\pm$\, 0.56 & 0.14  \,$\pm$\, 0.03 & 82.09  \,$\pm$\, 2.55 \\
& F$_{\rm GRS3}$ & 357 & 53.58  \,$\pm$\, 1.05 & 3.89  \,$\pm$\, 1.27 & 0.42  \,$\pm$\, 0.07 & -86.67  \,$\pm$\, 1.02 \\
& F$_{\rm GRS4}$ & 60 & 67.44  \,$\pm$\, 0.38 & 2.11  \,$\pm$\, 0.35 & 0.20  \,$\pm$\, 0.07 & 75.62  \,$\pm$\,6.64 \\
& F$_{\rm GRS5}$ & 94 & 64.31  \,$\pm$\, 0.80 & 3.03  \,$\pm$\, 0.75 & 0.38  \,$\pm$\, 0.09 & 82.56  \,$\pm$\, 1.91 \\
& F$_{\rm GRS6}$ & 9 & 46.69  \,$\pm$\, 0.21 & 1.86  \,$\pm$\, 0.52 & 0.51  \,$\pm$\, 0.11 & -58.46  \,$\pm$\, 7.23 \\
& F$_{\rm GRS7}$ & 22 & 44.32  \,$\pm$\, 0.18 & 2.63  \,$\pm$\, 0.81 & 0.11  \,$\pm$\, 0.05 & 83.59  \,$\pm$\, 4.60 \\
& F$_{\rm GRS8}$ & 92 & 50.08  \,$\pm$\, 0.28 & 2.28  \,$\pm$\, 0.72 & 0.0  \,$\pm$\, 0.0 & \dots \\
& F$_{\rm GRS9}$ & 21 & 54.73  \,$\pm$\, 0.17 & 0.92  \,$\pm$\, 0.19 & 0.38  \,$\pm$\, 0.07 & -70.60  \,$\pm$\, 2.56 \\
& F$_{\rm GRS10}$ & 23 & 64.11  \,$\pm$\, 0.12 & 1.62  \,$\pm$\, 0.26 & 0.03  \,$\pm$\, 0.06 &  \dots \\
& F$_{\rm GRS11}$ & 9 & 55.10  \,$\pm$\, 0.57 & 7.24  \,$\pm$\, 0.63 & 0.29  \,$\pm$\, 0.30 & 83.89  \,$\pm$\, 9.12 \\
& F$_{\rm GRS12}$ & 9 & 54.82  \,$\pm$\, 0.10 & 2.60  \,$\pm$\, 0.19 & 0.18  \,$\pm$\, 0.15 & 78.89 \,$\pm$\, 13.21 \\
& F$_{\rm GRS13}$ & 9 & 57.71  \,$\pm$\, 0.12 & 1.54  \,$\pm$\, 0.27 & 0.26  \,$\pm$\, 0.07 & -82.11  \,$\pm$\, 2.39 \\
& F$_{\rm GRS14}$ & 13 & 58.83  \,$\pm$\, 0.23 & 1.38  \,$\pm$\, 0.88 & 0.29  \,$\pm$\, 0.13 & 86.05  \,$\pm$\, 2.99 \\
& F$_{\rm GRS15}$ & 16 & 45.79  \,$\pm$\, 0.23 & 2.16  \,$\pm$\, 0.23 & 0.41  \,$\pm$\, 0.08 & -76.43  \,$\pm$\, 2.62 \\
& F$_{\rm GRS16}$ & 45 & 62.78  \,$\pm$\, 0.31 & 3.30  \,$\pm$\, 0.50 & 0.18  \,$\pm$\, 0.03 & -63.14  \,$\pm$\, 5.15 \\
& F$_{\rm GRS17}$ & 40 & 41.01  \,$\pm$\, 0.18 & 1.30  \,$\pm$\, 0.34 & 0.03  \,$\pm$\, 0.02 & 85.75  \,$\pm$\, 9.67 \\
& F$_{\rm GRS18}$ & 51 & 57.30  \,$\pm$\, 0.17 & 2.21  \,$\pm$\, 0.63 & 0.05  \,$\pm$\, 0.04 & -82.96  \,$\pm$\, 10.16 \\
& F$_{\rm GRS19}$ & 77 & 51.80  \,$\pm$\, 1.27 & 3.40  \,$\pm$\, 1.35 & 0.09  \,$\pm$\, 0.10 & -82.55  \,$\pm$\, 10.78 \\
& F$_{\rm GRS20}$ & 29 & 60.11  \,$\pm$\, 0.20 & 1.66  \,$\pm$\, 0.37 & 0.05  \,$\pm$\, 0.05 & 82.00  \,$\pm$\, 11.72 \\
\hline
\end{tabular}

\label{grs velocity component parameters}
\end{table*}

Shown in Figure\,\ref{spec_plot_cloudf_grs} are the average spectrum and the spectrum at each position across the map. As with the IRAM-30m observations of this cloud, these spectra are complex, showing multiple velocity components for the majority of positions. We use the {\sc scouse} and {\sc acorns} algorithms to separate and identify the coherent velocity structures across the cloud. We used the same threshold value in {\sc scouse} as for the IRAM-30m observations. Given the larger number of pixels present in this dataset compared to the IRAM-30m observations, we choose a larger SAA in {\sc scouse} of $\sim$\,145\arcsec (i.e. each SAA contained 40 positions, given the pixel spacing of $\sim$\,22\arcsec). Nevertheless, this only resulted in a still manageable $\sim$\,20\,per cent of the positions requiring manual inspection. The same input parameters used for the IRAM-30m observations were used in {\sc acorns} for the identification of the initial hierarchy. The parameter space survey of the relaxation values showed that for values of 2.5, 1.75, 0.75, for the peak intensity, centroid velocity and line width, respectively, were required to identify the most robust structures. The results of these analyses are shown in Figures\,\ref{spec_plot_cloudf_grs} and \ref{ppv_plots_cloudf-grs}. Twenty distinct velocity components are identified here, for which the basic properties are presented in Table\,\ref{grs velocity component parameters}. Given that the observation used to identify these components are different in spatial resolution, angular resolution and extent of the IRAM-30m observations, we choose to differentiate these by using a different, ``F$_{\rm GRS}$'', nomenclature.

\label{lastpage}
\end{document}